\documentclass[fleqn,usenatbib]{mnras}

\usepackage{newtxtext,newtxmath}

\usepackage{comment}
\usepackage{lscape}
\usepackage{lscape}
\usepackage{rotating}
\usepackage{adjustbox}

\usepackage[T1]{fontenc}
\usepackage{ae,aecompl}

\usepackage{graphicx}	
\usepackage{amsmath}	
\usepackage{amssymb}	
\usepackage{gensymb}
\usepackage{xcolor}
\PassOptionsToPackage{hyphens}{url}\usepackage{hyperref}

\newcommand{\qtir}{q$_{\rm TIR}$}

\defcitealias{ref:D.Molnar2018}{M18}
\defcitealias{ref:I.Delvecchio2021}{D21}

\title[$z<2.5$ star-forming galaxy properties with SKA]{Looking ahead to the sky with the Square Kilometre Array: simulating flux densities \& resolved radio morphologies of $0<z<2.5$ star-forming galaxies}

\author[R. T. Coogan]{Rosemary T. Coogan$^{1,2,3}$\thanks{E-mail: rcoogan.astrophysics@gmail.com},
Mark T. Sargent$^{1,4}$,
Anna Cibinel$^{1}$,
Isabella Prandoni$^{5}$,
\newauthor Anna Bonaldi$^{6}$,
Emanuele Daddi$^{3}$,
Maximilien Franco$^{7,8}$
\\
$^{1}$Astronomy Centre, Department of Physics and Astronomy, University of Sussex, Brighton BN1 9QH, UK\\
$^{2}$Max-Planck-Institut f\"ur extraterrestrische Physik (MPE), Giessenbachstr. 1, D-85748 Garching, Germany\\
$^{3}$CEA, IRFU, DAp, AIM, Universit\'{e} Paris-Saclay, Universit\'{e} Paris Cit\'{e}, Sorbonne Paris Cit\'{e}, CNRS, 91191 Gif-sur-Yvette, France\\
$^{4}$International Space Science Institute (ISSI), Hallerstrasse 6, CH-3012 Bern, Switzerland\\
$^{5}$INAF-Istituto di Radioastronomia, via Gobetti 101, Bologna 40129, Italy\\
$^{6}$SKA Observatory, Jodrell Bank, Lower Whitington, Macclesfield SK11 9DL, UK\\
$^{7}$Centre for Astrophysics Research, School of Physics, Engineering and Computer Science, University of Hertfordshire, College Lane, Hatfield AL10 9AB, UK\\
$^{8}$Department of Astronomy, The University of Texas at Austin, 2515 Speedway Blvd Stop C1400, Austin, TX 78712, USA }

\date{Accepted 2023 July 12. Received 2023 July 12; in original form 2022 August 01}

\pubyear{2023}

\begin{document}
\label{firstpage}
\pagerange{\pageref{firstpage}--\pageref{lastpage}}
\maketitle

\begin{abstract}
SKA-MID surveys will be the first in the radio domain to achieve clearly sub-arcsecond resolution at high sensitivity over large areas, opening new science applications for galaxy evolution. To investigate the potential of these surveys, we create simulated SKA-MID images of a $\sim$0.04~deg$^{2}$ region of GOODS-North, constructed using multi-band HST imaging of 1723 real galaxies containing significant substructure at $0<z<2.5$. We create images at the proposed depths of the band 2 wide, deep and ultradeep reference surveys (RMS = 1.0~$\mu$Jy, 0.2~$\mu$Jy and 0.05~$\mu$Jy over 1000~deg$^{2}$, 10-30~deg$^{2}$ and 1~deg$^{2}$ respectively), using the telescope response of SKA-MID at 0.6" resolution. We quantify the star-formation rate - stellar mass space the surveys will probe, and asses to which stellar masses they will be complete. We measure galaxy flux density, half-light radius ($R_{50}$), concentration, Gini (distribution of flux), second-order moment of the brightest pixels ($M_{20}$) and asymmetry before and after simulation with the SKA response, to perform input-output tests as a function of depth, separating the effects of convolution and noise. We find that the recovery of Gini and asymmetry is more dependent on survey depth than for $R_{50}$, concentration and $M_{20}$. We also assess the relative ranking of parameters before and after observation with SKA-MID. $R_{50}$ best retains its ranking, whilst asymmetries are poorly recovered. We confirm that the wide tier will be suited to the study of highly star-forming galaxies across different environments, whilst the ultradeep tier will enable detailed morphological analysis to lower SFRs.
\end{abstract}




\section{Introduction}
\label{sec:intro}

One of the observatories at the forefront to probe galaxy evolution over the next few decades is the Square Kilometer Array (SKA\footnote{\url{http://skao.int}}). The SKA will be the world's largest radio observatory, with mid- and low-frequency telescopes situated in South Africa and Australia respectively, and provide data to address several outstanding research questions (see e.g. \citealt{ref:C.Carilli2004b, ref:A.Taylor2008, ref:R.Braun2015}). One of the focuses of extragalactic survey science with the SKA will be to study galaxy growth and AGN activity through cosmic time via radio continuum emission (e.g. \citealt{ref:M.Jarvis2004, ref:J.Afonso2015, ref:R.Deane2015, ref:C.Power2015, ref:V.Smolcic2015, ref:J.Wagg2015, ref:A.Iqbal2017}). Radio continuum emission is also a good tracer of high-redshift structures such as proto(clusters), due to the large overdensity of star-formation and Active Galactic Nucleus (AGN) activity that are present in these environments, giving rise to radio continuum emission in the frequency range of the SKA (e.g. \citealt{ref:N.Hatch2011, ref:E.Daddi2017, ref:C.Krishnan2017}). The sensitivity and field of view of the SKA therefore make this telescope a highly effective tool for the detection of large numbers of star-forming galaxies and AGNs across a wide range of environments.

This paper focuses on the galaxy evolution science enabled by SKA-MID, which will operate in the frequency range $\sim$350~MHz -15~GHz. SKA-MID will be comprised of 64$\times$13.5~m dishes from its precursor telescope MeerKAT, combined with 133 SKA 15~m dishes. With a planned maximum baseline of 150~km the telescope array will achieve angular resolutions between 0.04"-0.70" in this frequency range. This high angular resolution enables a number of scientific applications (e.g. \citealt{ref:L.Godfrey2012}), and also allows the identification of individual galaxies in crowded regions such as high-redshift overdensities. SKA-MID is expected \citep{ref:R.Braun2017} to reach sensitivities approximately an order of magnitude above that of the VLA at 1.4~GHz. The survey speed of the SKA will be approximately 100$\times$ that of facilities such as LOFAR and the VLA, due to the combination of the high sensitivity with a large field of view ($\sim$0.5~deg$^{2}$ at 1.4~GHz) compared to many of its precursors and pathfinders. With the SKA it will therefore be possible to trace the star-formation history of the Universe with unprecedented statistics (see Section~\ref{sec:surveys_SKA}). Precursor telescopes in this frequency range are already operational, such as the Australian SKA Pathfinder (ASKAP) and MeerKAT, which will perform significant radio continuum and spectral surveys before SKA-MID comes online \citep{ref:A.Duffy2012, ref:R.Norris2013, ref:R.Norris2021, ref:F.Camilo2018, ref:T.Reynolds2019, ref:A.Matthews2021}.

As the radio astronomy community prepares for the SKA, we need to gain a fuller understanding of the  ability of SKA-MID to both detect galaxies and resolve their morphological properties. This will depend on the angular resolution and sensitivity of the observations, as well as factors such as the redshift and intrinsic radio brightness of a galaxy. Since early planning for the SKA, considerable effort has been put into predicting observable properties of the radio sky. Publicly available resources for this include the T-RECS model source catalogue of \citet[see below]{ref:A.Bonaldi2019}, as well as the Square Kilometre Array Design Studies (SKADS) Simulated Skies (S$^{3}$, \citealt{ref:R.Wilman2008, ref:D.Obreschkow2009}). Such catalogues can be used -- alongside other models or extrapolations of current observations -- to predict, e.g., radio luminosity functions and AGN/galaxy number counts in the sensitivity regime of the SKA (e.g. \citealt{ref:A.Hopkins2000, ref:R.Windhorst2003, ref:C.Jackson2004, ref:M.Jarvis2004, ref:I.Whittam2017, ref:A.Bonaldi2019, ref:E.Zackrisson2019}). They can also be used as model inputs for realistic simulations of SKA maps. In recent years several teams have produced mock SKA (or pathfinder/precursor) sky images, ranging from cosmology to galaxy morphology and cluster studies (e.g. \citealt{ref:A.Willis2008, ref:C.Ferrari2015, ref:S.Makhathini2015, ref:A.Bonalidi2016, ref:F.Loi2019}). These simulations can be performed using a range of different software tools, such as \texttt{Miriad} \citep{ref:Sault1995}, the \texttt{Faraday} package \citep{ref:M.Murgia2004}, \texttt{MeqTrees} \citep{ref:J.Noordan2010} and OSKAR\footnote{\url{https://github.com/OxfordSKA/OSKAR}}. Not all of these tools are specific to the SKA, but they allow the user to define array configuration, frequency, and observing times for specific science observations. For instance, image simulations can be used to assess the ability of SKA to detect and characterise subsets of the galaxy population, for planned surveys such as those discussed in Section~\ref{sec:surveys_SKA}.

A recent related effort was SKA Science Data Challenge \#1 (SDC1), for which SKA-MID continuum images were produced using a custom \texttt{Miriad} pipeline and the T-RECS model source catalogue of \citet[see also \citealt{ref:J.Leahy1996}, \citealt{ref:R.Battye2020}]{ref:A.Bonaldi2019}, at several different exposure times and frequencies\footnote{\url{https://www.skao.int/en/464/ska-science-data-challenge-1}}. SDC1 aimed to explore how the recoverability of galaxy parameters depends on data analysis strategy, by asking the community to attempt to recover source properties in an unknown model SKA-MID sky \citep{ref:A.Bonaldi2021}. However, the T-RECS catalogue sources in SDC1 were implemented with smoothly varying radio brightness profiles. They did not involve examples of the clumpy or irregular morphologies that can be observed in some galaxies.

In this paper, we expand on this previous work by constructing mock radio sky images using resolved observations of {\it real} galaxies in the GOODS-North (GOODS-N) field, to create simulated SKA-MID maps of realistic star-formation distributions within galaxies across a range of redshifts (Section~\ref{sec:SKA_sim}). We produce these simulated images at depths corresponding to the SKA reference continuum surveys summarised in table 1 of \citet{ref:PrandoniSeymour2015}. Our aim is to answer the following questions: (i) Which parameter space (in terms of star-formation rate (SFR), stellar mass and redshift) is accessible via these SKA-MID reference continuum surveys?, (ii) How well can optical indicators of galaxy structure be measured in different depth tiers of SKA-MID continuum surveys, and therefore input into classification schemes based in multi-dimensional parameter space (e.g. CAS-like schemes)?, (iii) Which structural indicators are the most (and least) reliable across different survey depths?

The outline of the paper is as follows. In Section~\ref{sec:surveys_SKA}, we introduce the tiered continuum surveys proposed by the SKA Extragalactic Continuum Science Working Group (SWG) and outline our methodology for relating observed radio flux densities and galaxy SFRs (Section~\ref{sec:method_SKA}). In addition, we then characterise the parameter space that these surveys will probe, based on general relations, {such as the Main Sequence of star-forming galaxies (MS; e.g. \citealt{ref:D.Elbaz2007, ref:K.Noeske2007, ref:G.Rodighiero2011}), and the assumption of point-source like flux distributions (Section~\ref{sec:surveyresults}). In Section~\ref{sec:SKA_sim} we describe our SKA image simulation and underlying galaxy sample, where we depart from the assumption of point sources and explore the effect of both telescope resolution and surface brightness sensitivity using real galaxies. The associated results follow in Section~\ref{sec:results}. We discuss and summarise the work in Sections~\ref{sec:discussion} and \ref{sec:concs} respectively. We use a $\Lambda$CDM cosmology with H$_{0}$=70kms$^{-1}$Mpc$^{-1}$, $\Omega_{M}$=0.3 and $\Omega_{\Lambda}$=0.7, and adopt a \citet{ref:G.Chabrier2003} initial mass function (IMF).

\section{Continuum galaxy and AGN co-evolution surveys with the SKA}
\label{sec:surveys_SKA}

\subsection{Tiered continuum surveys}
\label{sec:tiered_desc}

Three reference surveys have been proposed for SKA-MID by the Extragalactic Continuum SWG, for the purpose of galaxy evolution and AGN-galaxy co-evolution studies \citep{ref:PrandoniSeymour2015}. Two of these -- in band 2 (0.95-1.76~GHz) and band 5b (8.3-15~GHz), respectively \citep{ref:M.Caiazzo2017} -- are `tiered' surveys, with deep observations over a restricted area, and shallower observations covering a much larger part of the sky.

In this paper, we discuss the band 2 ($\sim$1.4~GHz) tiered reference surveys, which would provide resolved radio imaging of a larger number of galaxies and out to higher redshift than ever before. The survey tiers have RMS depths of 0.05~$\mu$Jy, 0.2~$\mu$Jy and 1.0~$\mu$Jy over regions of size 1~deg$^{2}$, 10-30~deg$^{2}$ and 1000~deg$^{2}$ respectively, requiring observing times of the order of a few hours (wide), $\sim$100~hours (deep) and $\sim$1.5k~hours (ultradeep). Compared to pre-SKA surveys the ultradeep tier would be 10$\times$ deeper than the e-MERLIN Galaxy Evolution Survey (eMERGE, \citealt{ref:T.Muxlow2020}), and cover a 16$\times$ larger area. The deep tier would be 2-3$\times$ deeper than eMERGE, but over a few hundred times larger area. At a similar sensitivity as the MeerKAT International GHz Tiered Extragalactic Exploration (MIGHTEE) survey, the wide tier would cover a 50$\times$ larger area and achieve a 16$\times$ higher angular resolution. Although high angular resolution can currently be achieved at higher frequencies with the VLA, for example, the spectral slope of the radio spectra means that flux densities at $\nu>$1.4~GHz are intrinsically lower than at 1.4~GHz, and therefore become increasingly difficult to detect with increasing redshift. On the other hand, lower frequencies (e.g. those of band 1 SKA surveys), won’t have sufficient angular resolution to resolve high-z galaxies.

Based on the areas and continuum sensitivities of these surveys, the global physical properties of galaxies that each tier will detect has been outlined previously using simple relations and sensitivity arguments (e.g. \citealt{ref:PrandoniSeymour2015, ref:M.Jarvis2015}). In this paper we carry out a more detailed assessment of what SKA will be able to achieve. Specifically, we will investigate where detectable galaxies reside at different redshifts with respect to the Main Sequence of star-forming galaxies. To this end, we consider different plausible scenarios for relating radio luminosity and SFR (and vice-versa) in Section~\ref{sec:method_SKA}. We implement these relations to infer the minimal detectable SFR for each band 2 survey tier in Section~\ref{sec:surveyresults}, and to create simulated SKA images of star-forming galaxies in Section \ref{sec:SKA_sim}. In Section~\ref{sec:results} we examine how well galaxy flux densities and morphological properties can be recovered.

\begin{figure*}
\begin{minipage}{\textwidth}
\centering
\includegraphics[width=0.49\textwidth, clip, trim=0cm 0cm 0cm 0cm]{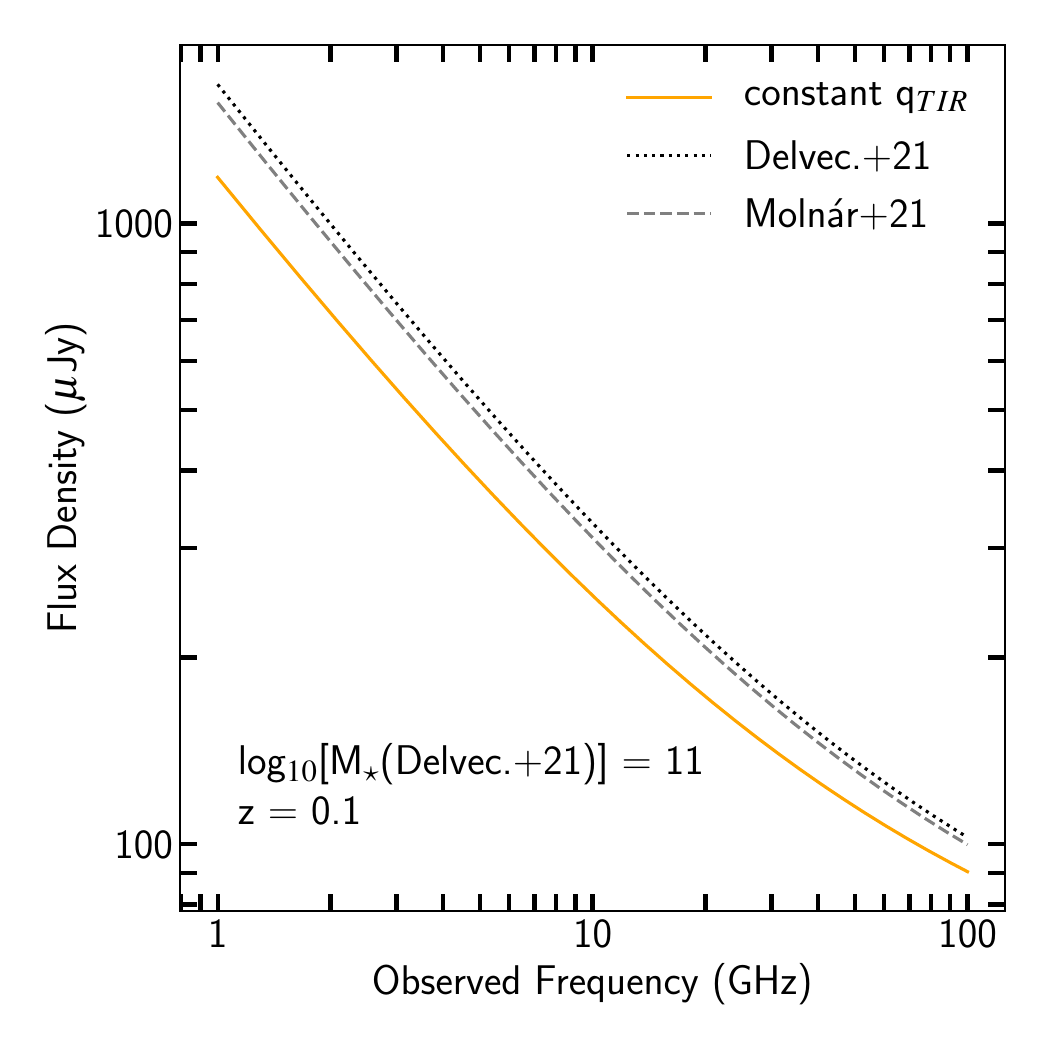}\hfill
\includegraphics[width=0.49\textwidth, clip, trim=0cm 0cm 0cm 0cm]{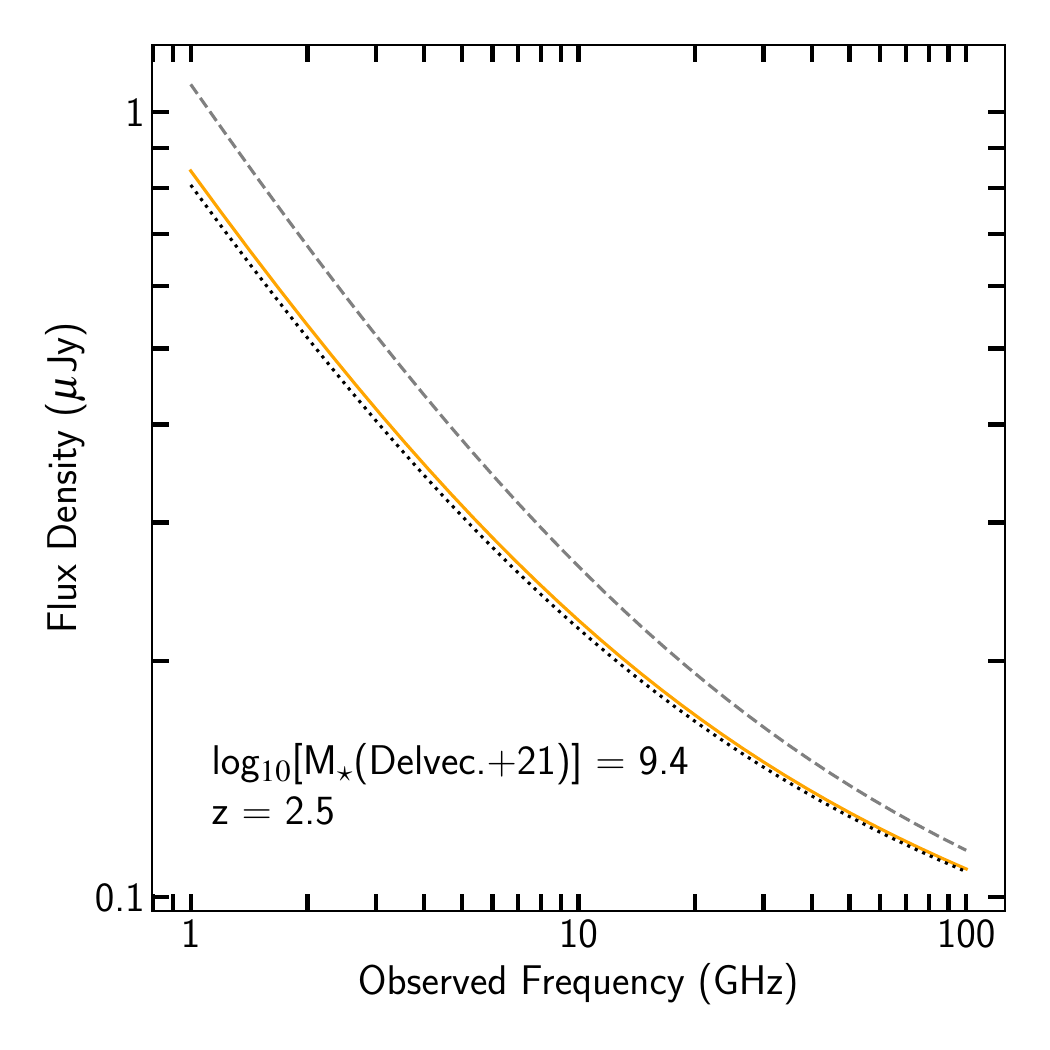}\hfill
\caption{Observed radio SEDs for an SFR\,=\,10~$M_{\odot}$yr$^{-1}$ galaxy at $z$\,=\,0.1 ({\it left}) and $z$\,=\,2.5 ({\it right}), using the radio SFR calibrations outlined in Section~\ref{sec:diff_radio_methods}. For the mass-dependent \citet{ref:I.Delvecchio2021} recipe, the stellar mass scale is chosen to correspond to a galaxy on the MS (as parametrised in \citealt{ref:M.Sargent2014}) with SFR~=~10~$M_{\odot}$yr$^{-1}$ at each redshift.}
\label{fig:radiomethods}
\end{minipage}
\end{figure*}

\subsection{Total radio luminosity and galaxy star-formation rate}
\label{sec:method_SKA}

Two mechanisms generate radio continuum emission from star-forming galaxies: synchrotron and free-free emission. At rest-frame frequencies $\lesssim$30~GHz the dominant emission mechanism is non-thermal synchrotron radiation \citep{ref:J.Condon1992}, produced by electrons being accelerated within a galaxy's magnetic field. In purely star-forming galaxies supernova shocks are responsible for this acceleration, such that synchrotron radiation is an effective indicator of their SFR \citep{ref:E.Berezhko2004, ref:M.Longair2011}. The synchrotron component of a star-forming galaxy's spectral energy distribution (SED) typically has a spectral index\footnote{S$_{\nu} \propto \nu^{\alpha}$, where S$_{\nu}$ is flux density, $\nu$ the frequency. The exponent $\alpha$ of the frequency is the spectral index.} $\alpha$ around -0.7 to -0.8 (\citealt{ref:U.Lisenfeld2000, ref:A.Kimball2008, ref:J.Delhaize2017, ref:V.Smolcic2017a}). In galaxies containing an AGN, electrons are also accelerated by the powerful magnetic fields around the jet of the AGN. While separating a galaxy's radio emission into purely star-forming and AGN components is an important topic, here we only consider star-forming galaxies that do not contain an AGN.

Radio emission in star-forming galaxies also includes so-called `free-free' emission. This arises when an electron comes into close proximity to a positive ion, e.g., in a HII region. The electron scatters off the ion but is not captured, and emits thermal bremsstrahlung radio emission in the interaction process. The contribution to the SED from free-free emission has a shallower spectral slope than synchrotron radiation, approximately $\beta$=-0.1 \citep{ref:J.Condon1992}. This emission is only significant in comparison to the synchrotron radiation at relatively high frequencies ($\sim$30-200~GHz rest-frame).

Using the SEDS of purely star-forming galaxies, both the free-free and synchrotron component of the radio SED have been calibrated as an indicator of star-formation rate (see below). Where ultraviolet emission is heavily attenuated, radio emission allows us to trace the obscured star-formation in galaxies, as radio emission is not absorbed by dust grains in the interstellar medium (ISM). At longer wavelengths, although single-dish telescopes such as the \textit{Herschel} Space Observatory and the \textit{James Clerk Maxwell} Telescope can probe dust continuum emission close to its infrared (IR) peak, their low angular resolution often leads to source blending. The ability to measure and resolve star-formation on small angular scales is, however, crucial to understanding galaxy evolution at high redshift, particularly across a wide range of galaxy morphologies and environments. Using radio observations rather than IR data can therefore be particularly advantageous, as interferometers like the SKA produce high angular resolution imaging with minimal source confusion.

We now outline how we can quantiatively relate synchrotron and free-free luminosities, and their respective contributions to the observed flux density, to a galaxy's SFR.

\subsubsection{Free-free radiation} 
\label{sec:ff}

We take the galaxy free-free luminosity to relate to the star-formation rate as in \citet{ref:E.Murphy2012}:
\begin{equation}
L_{\nu} = \frac{SFR}{4.3 \times 10^{-28} \times (\frac{T_{e}}{10^{4}})^{-0.45} \times \nu^{0.1}}
\label{eqn:free-free}
\end{equation}
where $L_{\nu}$ is the free-free luminosity in erg.s$^{-1}$ at frequency $\nu$ (in GHz), and T$_{e}$ is the electron temperature, which we set to T$_{e}$=10$^{4}$~K following standard practice. Having calculated the rest-frame 1.4~GHz luminosity arising from free-free emission using Equation~\ref{eqn:free-free}, we then use the free-free spectral slope $\beta$=-0.1 to infer the free-free luminosity and flux density at the frequency of interest (in this case 1.4~GHz in the observer frame).

\subsubsection{Synchrotron radiation} 
In order to relate SFR and total 1.4~GHz luminosity we use the infrared-radio correlation (IRRC) and its \qtir\ parameter \citep{ref:T.deJong1985, ref:G.Helou1985}, which represents the logarithmic ratio of the total infrared (8-1000$\micron$; $L_{\rm IR}$) and 1.4~GHz radio ($L_{\rm 1.4GHz}$) continuum luminosities. We convert between galaxy SFR and $L_{\rm IR}$ following \citet{ref:R.Kennicutt1998a}. The relationship between total radio luminosity and SFR (in units of $M_{\odot}$yr$^{-1}$) then can be written as:
\begin{equation}
L_{\rm 1.4GHz} = \frac{SFR}{10^{-24} \times 10^{q_{\rm TIR}}}
\label{eqn:irrc}
\end{equation}
where $L_{\rm 1.4GHz}$ is the derived spectral luminosity in units of W/Hz. Here we assume that at high luminosities $L_{\rm IR}$ is a good proxy for the total SFR, as the dust-obscured SFR is dominant in this regime \citep{ref:S.Heinis2014, ref:M.Pannella2015}.

Having calculated the total rest-frame 1.4~GHz luminosity from Equation~\ref{eqn:irrc}, we then subtract the free-free contribution calculated in Section~\ref{sec:ff}, in order to isolate the synchrotron contribution. We then use the spectral slope $\alpha$=-0.8 to infer the synchrotron luminosity and flux at 1.4~GHz in the observer frame. We note that inverse Compton scattering of cosmic ray electrons in galaxies off cosmic microwave background photons can potentially lead to synchrotron dimming. However, this effect is not thought to be significant at $z<3$ (e.g. \citealt{ref:C.Carilli2001, ref:E.Murphy2009, ref:B.Lacki2010}).

Although initially calibrated in the local Universe (e.g. \citealt{ref:T.deJong1985, ref:G.Helou1985, ref:E.Bell2003}), advances in the sensitivity of IR-radio observatories have recently allowed the exploration of the \qtir\ parameter at higher redshifts and across a wide range of stellar masses (e.g. \citealt{ref:J.Delhaize2017, ref:D.Molnar2018, ref:I.Delvecchio2021}). However, these studies have not yet reached a clear consensus on the value of \qtir\ beyond the local Universe. Several work by, e.g., \citet{ref:B.Magnelli2015}, \citet{ref:CalistroRivera2017} and \citet{ref:J.Delhaize2017} conclude that there is a smooth, decreasing trend of \qtir\ with redshift. On the other hand, \citet{ref:D.Molnar2018}  find that for disk-like galaxies at $z<1.5$ (which are expected to have the most straightforward correlation between IR and radio luminosity, with both IR and radio emission being driven by star-formation alone) \qtir\ is redshift-invariant, and consistent with values measured in the local Universe (in agreement with e.g. \citealt{ref:M.Garrett2002, ref:E.Ibar2008, ref:M.Jarvis2010, ref:M.Sargent2010, ref:D.Smith2014}). Recent work by \citet{ref:I.Delvecchio2021} finds evidence for a predominantly mass-dependent \qtir\ that, at fixed mass, shows little variation with redshift. Finally, \citet[]{ref:D.Molnar2021} find that \qtir\ depends on rest-frame 1.4~GHz luminosity itself, and thus define a relationship directly from $L_{\rm 1.4GHz}$ to SFR (their equation 22). Such a non-linear IRRC or mass-dependent \qtir\ could also play some role in explaining the redshift-dependence of \qtir\ reported by, e.g., \citet{ref:B.Magnelli2015}, \citet{ref:CalistroRivera2017} and \citet{ref:J.Delhaize2017}.

In order to assess the systematics arising from the different scenarios for relating radio luminosity and SFR, we therefore define three possible methods for deriving \qtir:
\begin{enumerate}
\item constant \qtir\ from \citet{ref:E.Bell2003} and \citet{ref:D.Molnar2018}
\item \qtir($M_{\star}$,z) as in \citet{ref:I.Delvecchio2021}
\item non-linear relation between SFR and $L_{\rm 1.4GHz}$, as in \citet{ref:D.Molnar2021}
\end{enumerate}

\begin{figure*}
\centering
\includegraphics[width=0.5\textwidth, clip, trim=0cm 1cm 0cm 0cm]{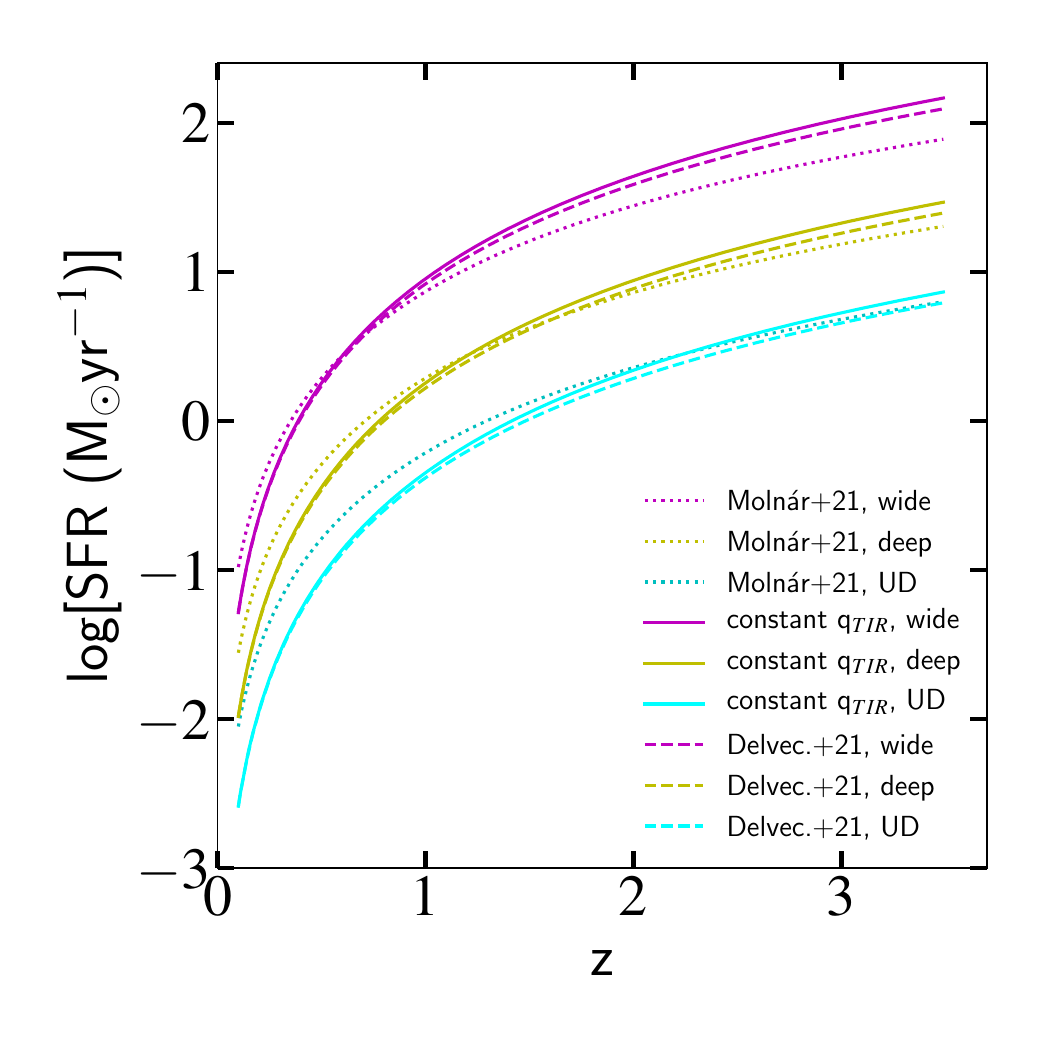}\hfill
\includegraphics[width=0.465\textwidth, clip, trim=0cm 1cm 0cm 0cm]{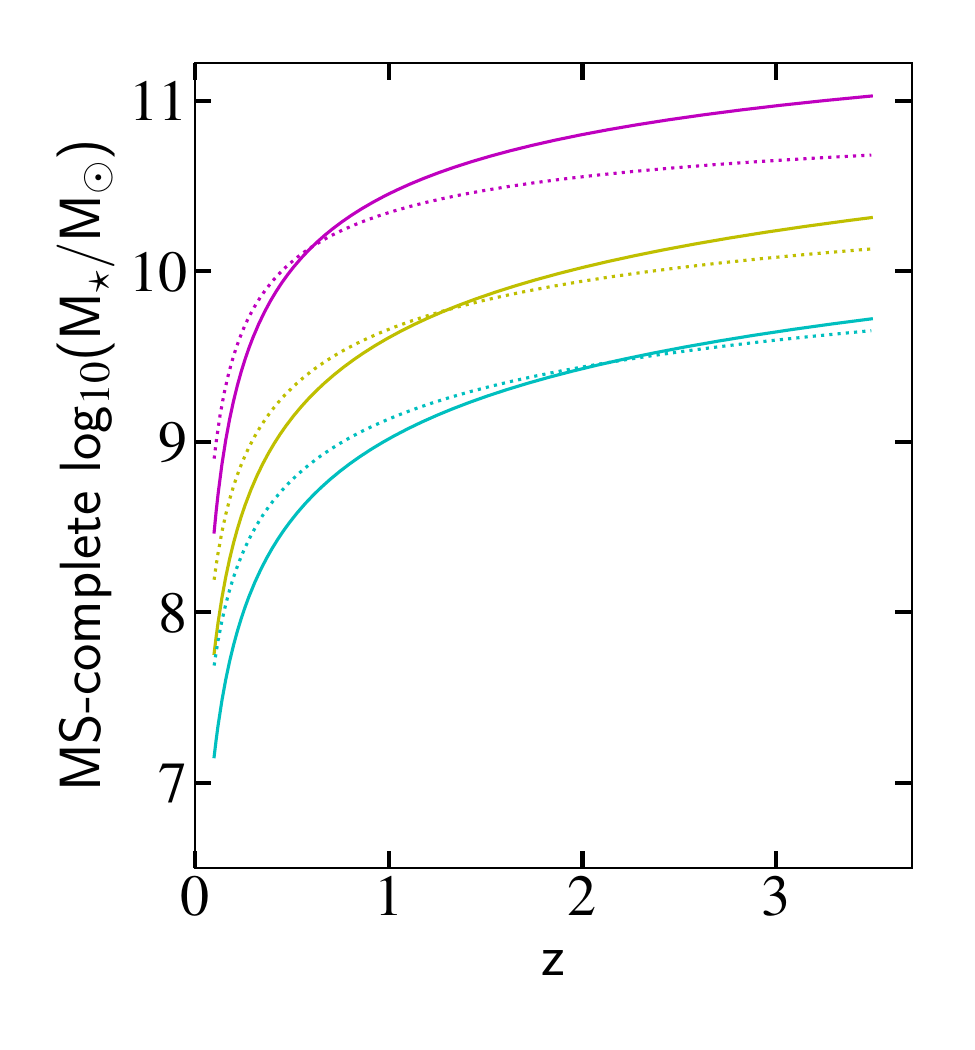}\hfill
\caption{{\it Left}: Minimum detectable SFRs (5$\sigma$) as a function of redshift, for an SKA-MID band 2 reference survey (see Sect. \ref{sec:tiered_desc}) with ultradeep (UD, cyan), deep (yellow) and wide (magenta) tiers. For the \citet{ref:I.Delvecchio2021} curves (dashes), we have assumed a stellar mass of $M_{\star}$=10$^{10}$~$M_{\odot}$. {\it Right}: the minimum stellar mass to which the same surveys will be able to detect Main Sequence complete samples, for the cases of a constant \qtir\ (solid lines) and \citet{ref:D.Molnar2021} (dotted lines).}
\label{fig:SFRlimits}
\end{figure*}

\subsubsection{Total radio luminosity}
\label{sec:diff_radio_methods}
The total radio luminosity is the sum of the synchrotron and free-free emission. We show in Fig.~\ref{fig:radiomethods} the total observed radio flux density derived from each method at $z=0.1$ and $z=2.5$, for a galaxy with SFR=10~$M_{\odot}$yr$^{-1}$. At $z=0.1$, the \citet{ref:I.Delvecchio2021} and \citet{ref:D.Molnar2021} \qtir\ derivations lead to similar results for total radio flux density, with the lowest flux densities arising from a constant \qtir.
However, when we move to $z=2.5$, the flux densities derived using the luminosity-dependent \qtir\ \citep{ref:D.Molnar2021} are raised above those derived using a constant \qtir\ (e.g. \citealt{ref:E.Bell2003}) and a \qtir($M_{\star}$,z) \citep{ref:I.Delvecchio2021}. However, we remind the reader that as the \citet{ref:D.Molnar2021} \qtir\ is SFR-dependent, the relationship between these curves will also change with galaxy SFR (see Fig.~\ref{fig:SFRlimits}).

\subsection{Detectable star-formation rates of SKA reference surveys}
\label{sec:surveyresults}

To assess the physical properties of the galaxies that will be detectable in the tiered band 2 continuum reference surveys of Sect. \ref{sec:tiered_desc}, we take the proposed survey RMS noise levels, and use the framework described in Section~\ref{sec:method_SKA} to calculate the SFRs that would give rise to the corresponding flux densities. In doing so, we assume unresolved radio sources. We show in the left panel of Fig.~\ref{fig:SFRlimits} the minimum galaxy-integrated SFRs that will be detectable by SKA-MID at the $\geq$5$\sigma$ level, for the three survey tiers. Note that, for resolved galaxies with realistic source structure and close to the lowest detectable SFRs, these surveys will be affected by incompleteness to a degree which varies with survey depth and source redshift. A quantitative discussion of this effect is included in Section~\ref{sec:SKA_sim}.

At $z=0.1$, the SKA will be able to probe galaxies with SFRs down to small values, $\sim$0.003-0.1~$M_{\odot}$yr$^{-1}$. At $z=2.5$, the minimum detectable SFR (5$\sigma$) is between $\sim$3-3.5~$M_{\odot}$yr$^{-1}$ in the ultradeep tier, 11-14~$M_{\odot}$yr$^{-1}$ in the deep tier and $\sim$40-70~$M_{\odot}$yr$^{-1}$ in the wide tier, respectively. These ranges quoted highlight the uncertainties arising from different radio SFR calibrations at high redshift. In the right panel of Fig.~\ref{fig:SFRlimits}, we show the `Main Sequence completeness' of these surveys. In order to be conservative in our predictions across a range of redshifts, we use the case of a constant \qtir\ in order to derive our stellar masses from SFR. We additionally include the case of \citet{ref:D.Molnar2021}, which represents the least conservative case at high redshift (Fig.~\ref{fig:radiomethods}). For each tier of the survey, we show the stellar masses down to which SKA-MID will be able to detect galaxies on the MS, by taking the most conservative estimate (largest mass) from three MS relations for each stellar mass \citep[see also Fig.~\ref{fig:MS_surveys}]{ref:M.Sargent2014, ref:J.Speagle2014, ref:C.Schreiber2015}. Similarly to the SFR limits themselves, the 5$\sigma$ MS mass limits increase with redshift, such that the different survey tiers will probe the MS population from $M_{\star}\sim$10$^{7.2}$-10$^{8.9}$~$M_{\odot}$ at $z=0.1$, to $M_{\star}\sim$10$^{9.5}$-10$^{11}$~$M_{\odot}$  at $z=3$ (with the range reflecting the different survey depths and the methods used for conversion between SFR and observed flux density).

\begin{figure*}
\begin{minipage}{0.95\textwidth}
\centering
\includegraphics[width=0.43\textwidth, clip, trim=0cm 0cm 0cm 0cm]{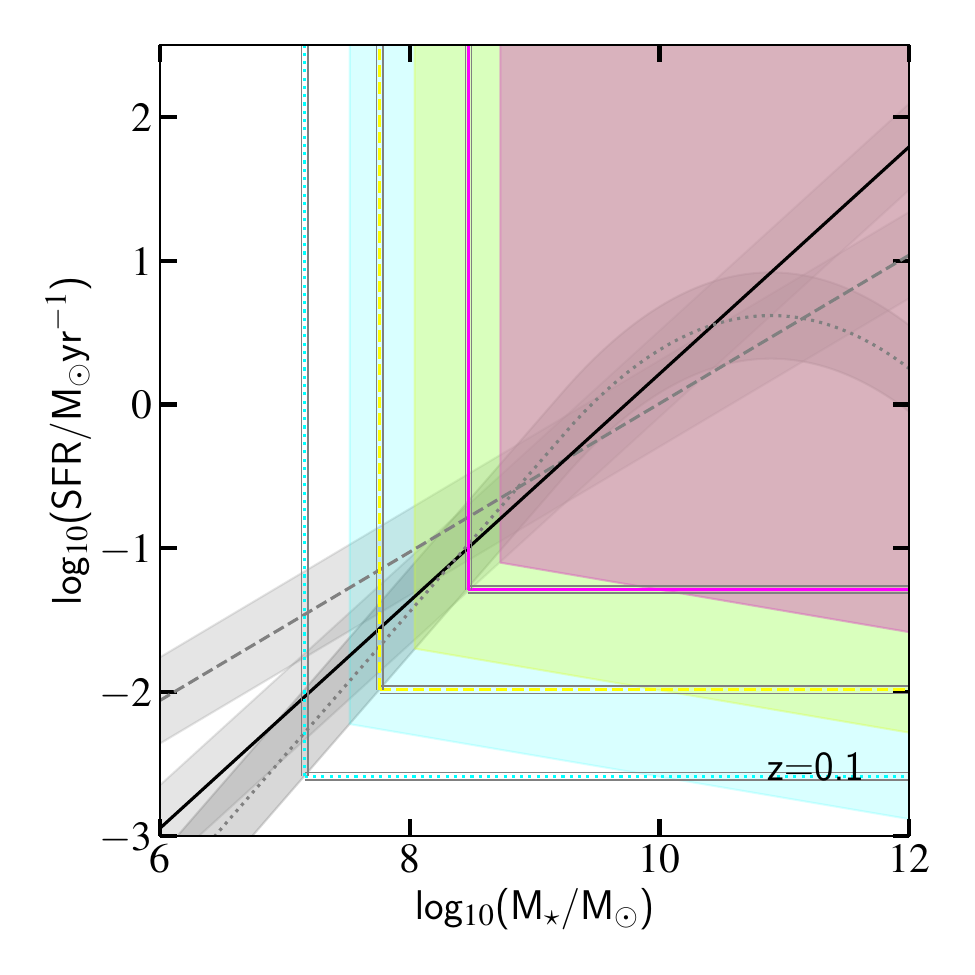}\hfill
\includegraphics[width=0.43\textwidth, clip, trim=0cm 0cm 0cm 0cm]{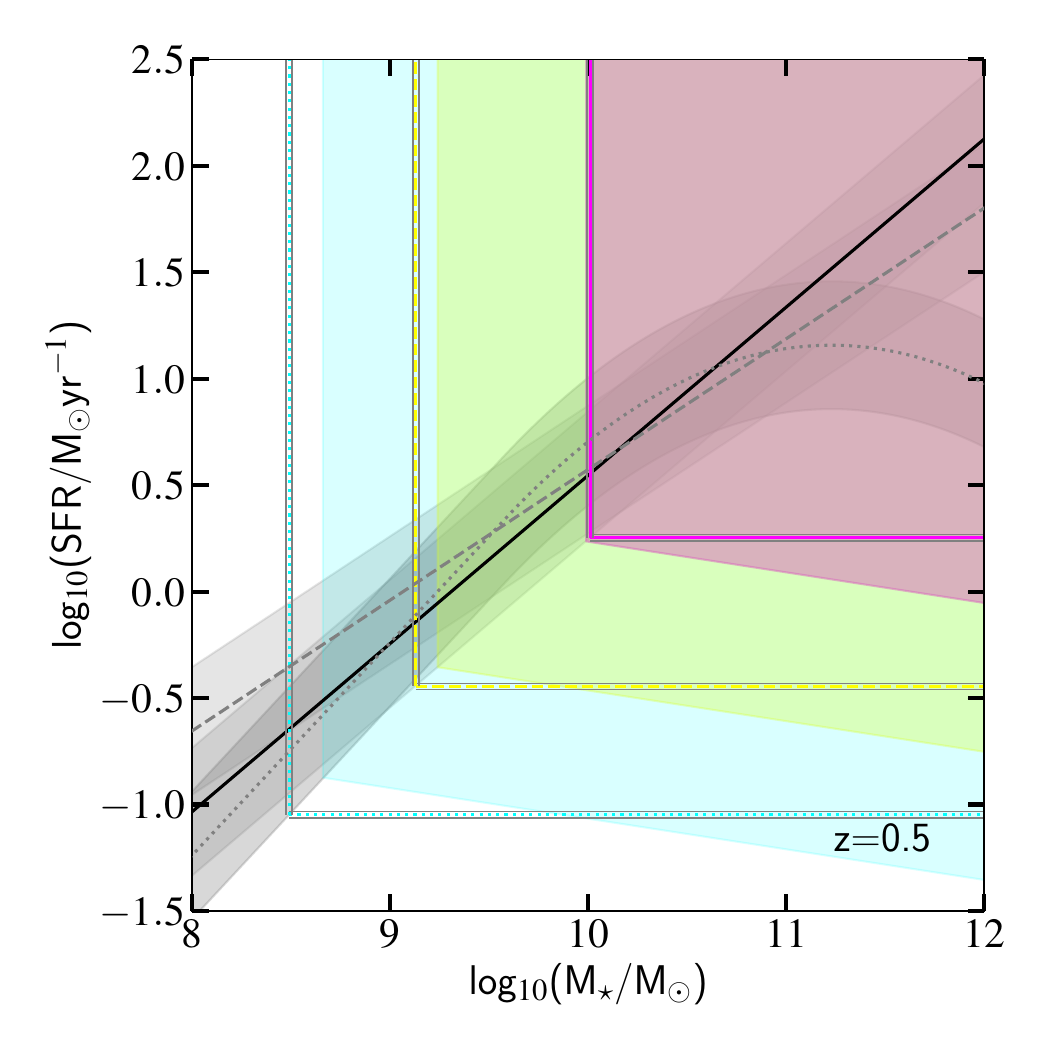}\hfill
\end{minipage}
\begin{minipage}{0.95\textwidth}
\includegraphics[width=0.43\textwidth, clip, trim=0cm 0cm 0cm 0cm]{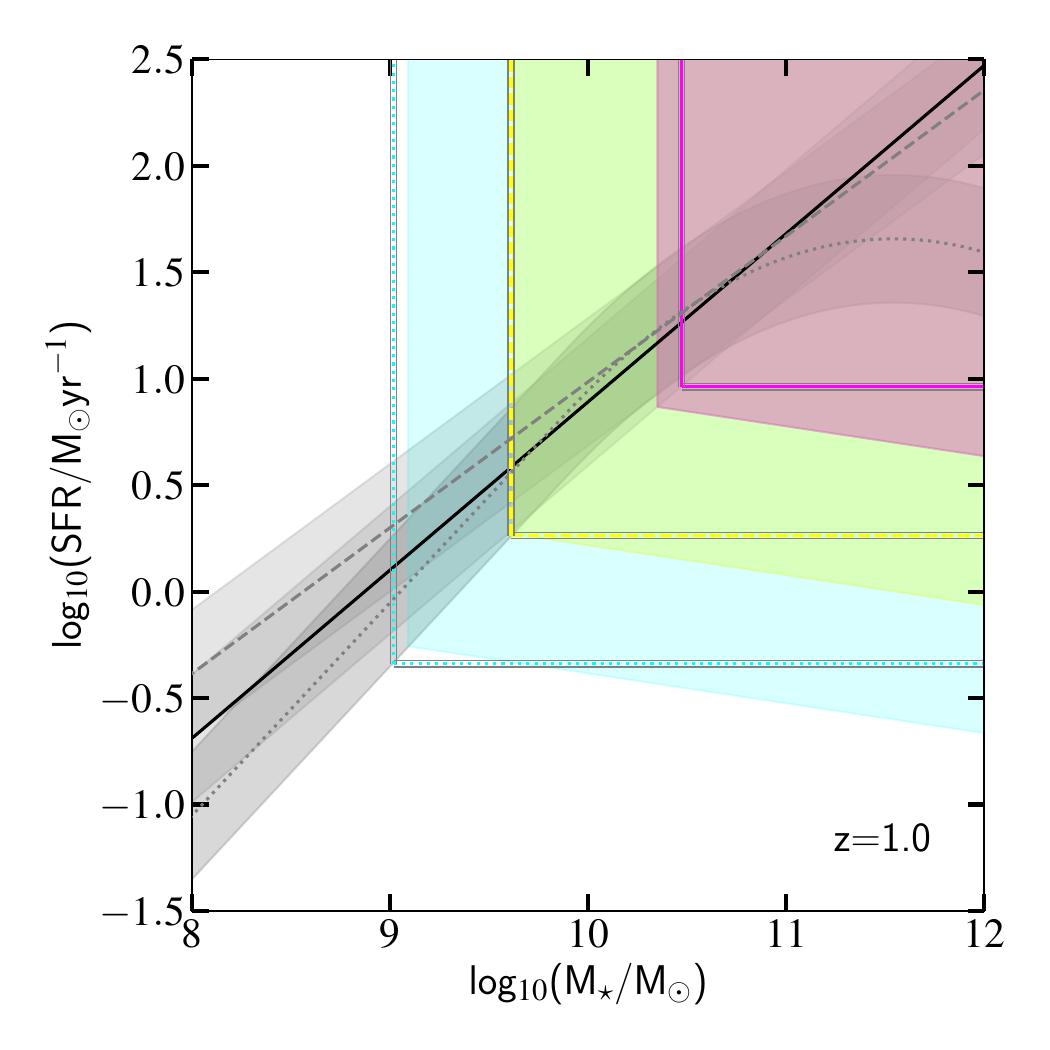}\hfill
\includegraphics[width=0.43\textwidth, clip, trim=0cm 0cm 0cm 0cm]{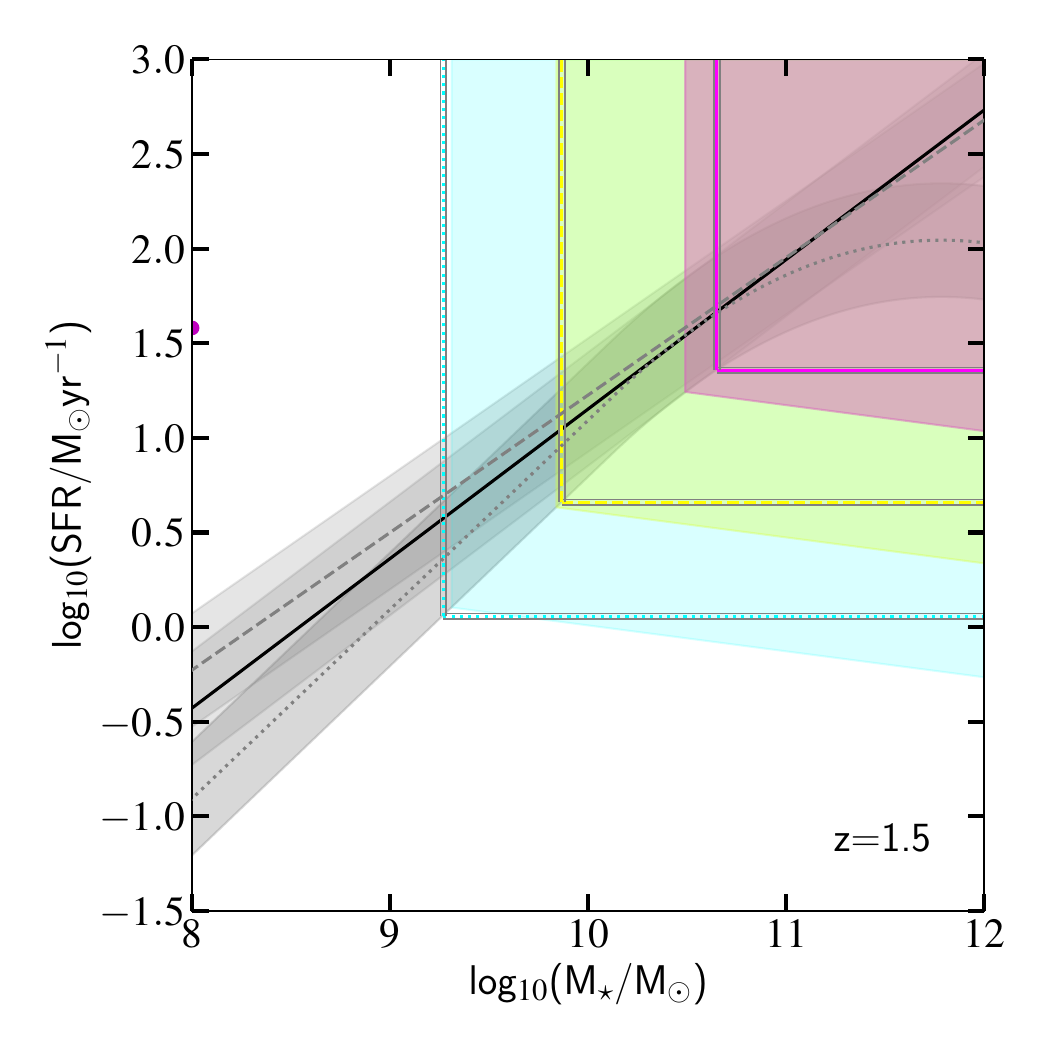}\hfill
\end{minipage}
\begin{minipage}{0.95\textwidth}
\includegraphics[width=0.43\textwidth, clip, trim=0cm 0cm 0cm 0cm]{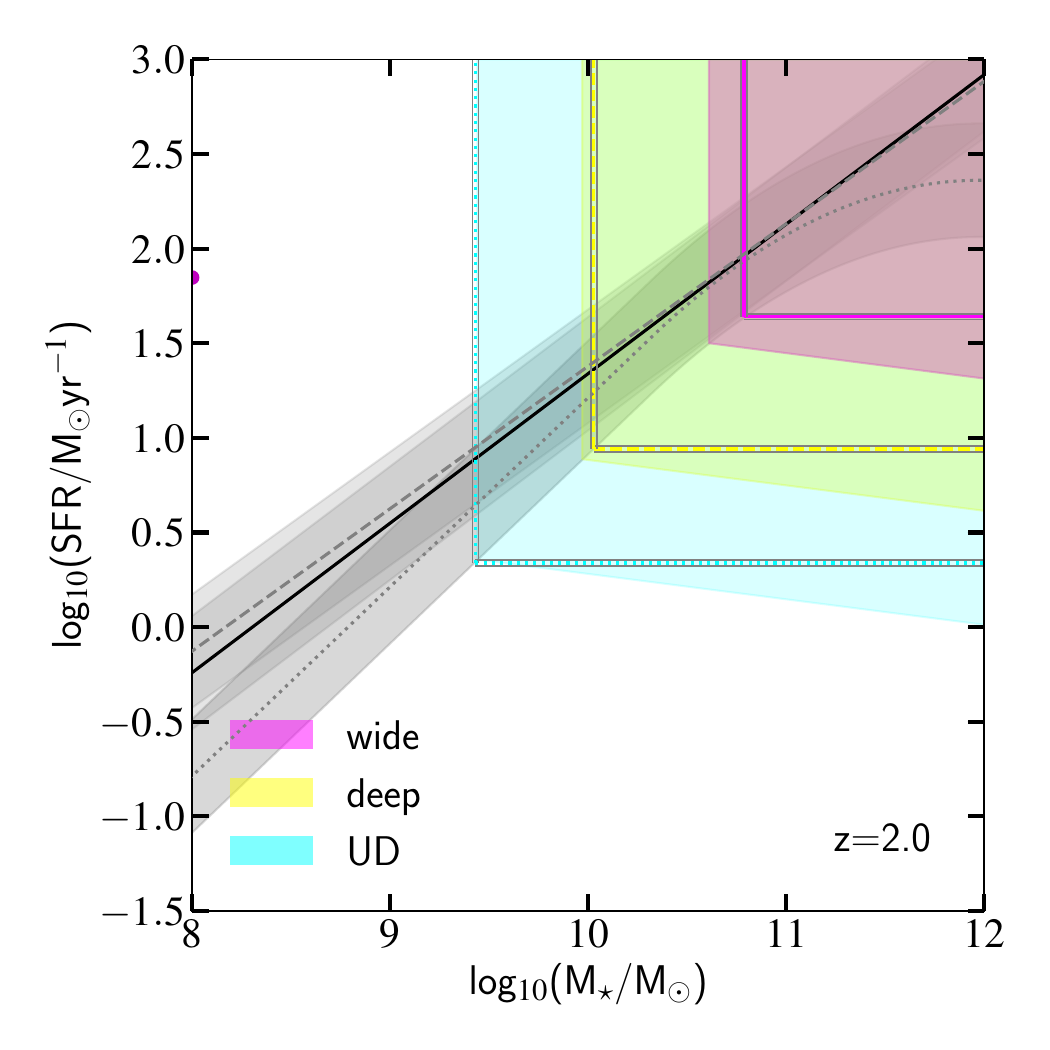}\hfill
\includegraphics[width=0.43\textwidth, clip, trim=0cm 0cm 0cm 0cm]{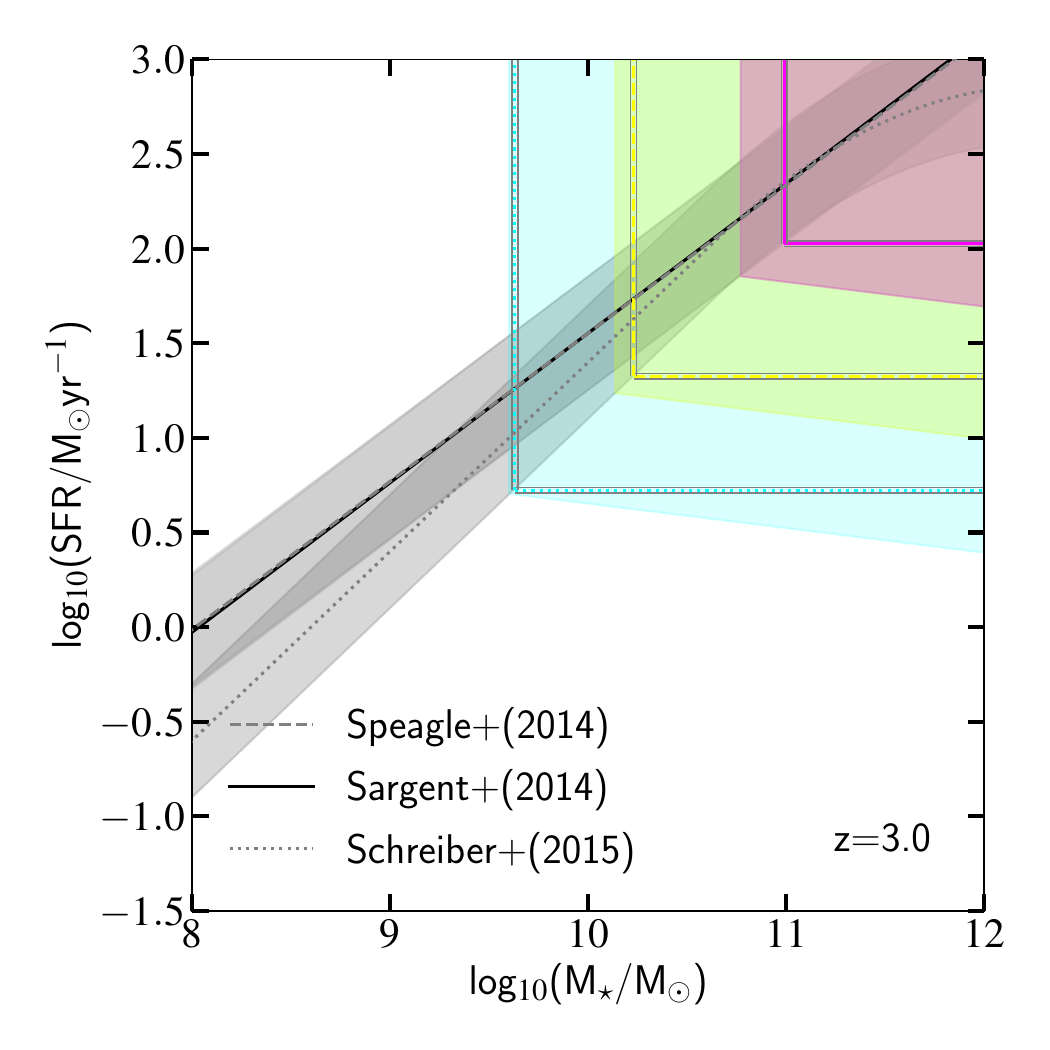}\hfill
\end{minipage}
\caption{Regions on the MS accessible to the SKA-MID band 2 continuum reference surveys, at the given redshifts. Three Main Sequence loci and their 0.3~dex scatter are shown. The coloured regions (cyan: ultradeep, yellow: deep, magenta: wide) and the corresponding coloured lines (dotted: ultradeep, dashed: deep, solid: wide) indicate areas in the SFR-$M_{\star}$ plane where mass-complete samples of MS galaxies can be selected in the different survey tiers, for \citet{ref:I.Delvecchio2021} and constant \qtir\ respectively, according to the 5$\sigma$ minimum SFRs shown in Fig.~\ref{fig:SFRlimits}.}
\label{fig:MS_surveys}
\end{figure*}

In order to illustrate the accessible regions of SFR-stellar mass space, we plot the SFR detection limits of the different survey tiers in relation to the locus of the MS in Fig.~\ref{fig:MS_surveys}. We show redshifts spanning $z=0.1$ to $z=3.0$, with three different MS determinations and their scatter. In each panel, we colour the regions of the MS that will be detectable by the different tiers of the band 2 surveys (with coloured shading and coloured lines for the $M_{\star}$-dependent \qtir\ from \citet{ref:I.Delvecchio2021} and a redshift-invariant, constant \qtir\ respectively). The horizontal boundaries (cyan, yellow and magenta) correspond to the minimum detectable 5$\sigma$ SFRs shown in Fig.~\ref{fig:SFRlimits}, assuming point sources. We draw our vertical survey boundaries where these minimum SFRs intersect with the lowest edge of the 1$\sigma$ scatter from the Main Sequence curves at that redshift. The vertical boundaries this represent a conservative stellar mass limit down to which  the survey tiers will be ``MS-complete".

Across the wide redshift range 0.1\,$<z<$\,3 shown in Fig.~\ref{fig:MS_surveys}, the MS mass-completeness limit changes by $\sim$2 orders of magnitude for the \citet{ref:I.Delvecchio2021} case, and around 2.5 orders of magnitude for a constant \qtir. At $z=0.1$, the band 2 wide tier will be MS complete to $M_{\star}$$\sim$10$^{8.5}$~$M_{\odot}$ (magenta) for a constant \qtir. The deep (yellow) and ultradeep (cyan) surveys will probe to even lower stellar masses at this redshift, detecting complete samples down to $M_{\star}$$\sim$10$^{7.8}$~$M_{\odot}$ and $M_{\star}$$\sim$10$^{7.2}$~$M_{\odot}$ respectively for unresolved sources, again for the case of a constant \qtir. Interestingly, the view in Fig.~\ref{fig:MS_surveys} highlights the stellar-mass dependency of the \citet{ref:I.Delvecchio2021} \qtir, as the horizontal boundaries slope downward towards higher stellar masses, meaning lower SFRs would be detectable at these higher masses.

Towards the beginning of the peak epoch of the cosmic star-formation history, at $z=3$, the band 2 surveys will be complete for galaxies on the MS down to $M_{\star}$=10$^{10.7-11.0}$~$M_{\odot}$, $M_{\star}$=10$^{10.1-10.2}$~$M_{\odot}$, and $\sim$$M_{\star}$$=$10$^{9.6}$~$M_{\odot}$, for the wide, deep, and ultradeep surveys respectively (here the ranges quoted reflect the difference between the different \qtir\ cases.) Fig.~\ref{fig:MS_surveys} also highlights that the SKA band 2 continuum surveys could detect galaxies with SFRs significantly below the MS, particularly in the ultradeep and deep survey regions covering 1-30~deg$^{2}$ respectively.

\begin{figure}
\centering
\includegraphics[width=0.45\textwidth, clip, trim=0.5cm 0.7cm 0.5cm 0cm]{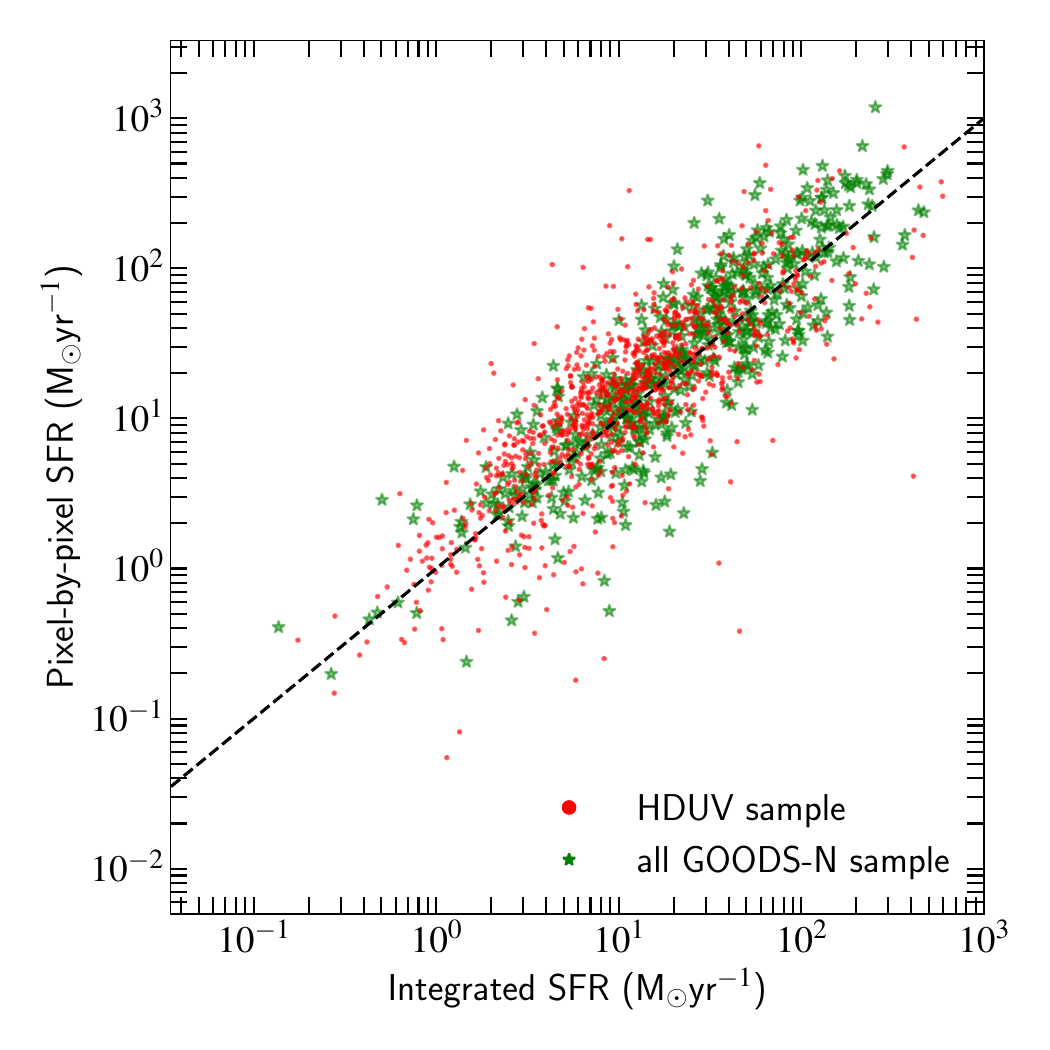}\hfill
\caption{Integrated star-formation rates vs. pixel-by-pixel SFR sums, for our galaxy sample. The galaxies are coloured according to their selection technique, as given in the legend. The dashed black line is the 1:1 relation.}
\label{fig:SFR_corrections}
\end{figure}

\begin{figure*}
\begin{minipage}{\textwidth}
\includegraphics[width=0.3328\textwidth, clip, trim=0cm 0cm .75cm 0cm]{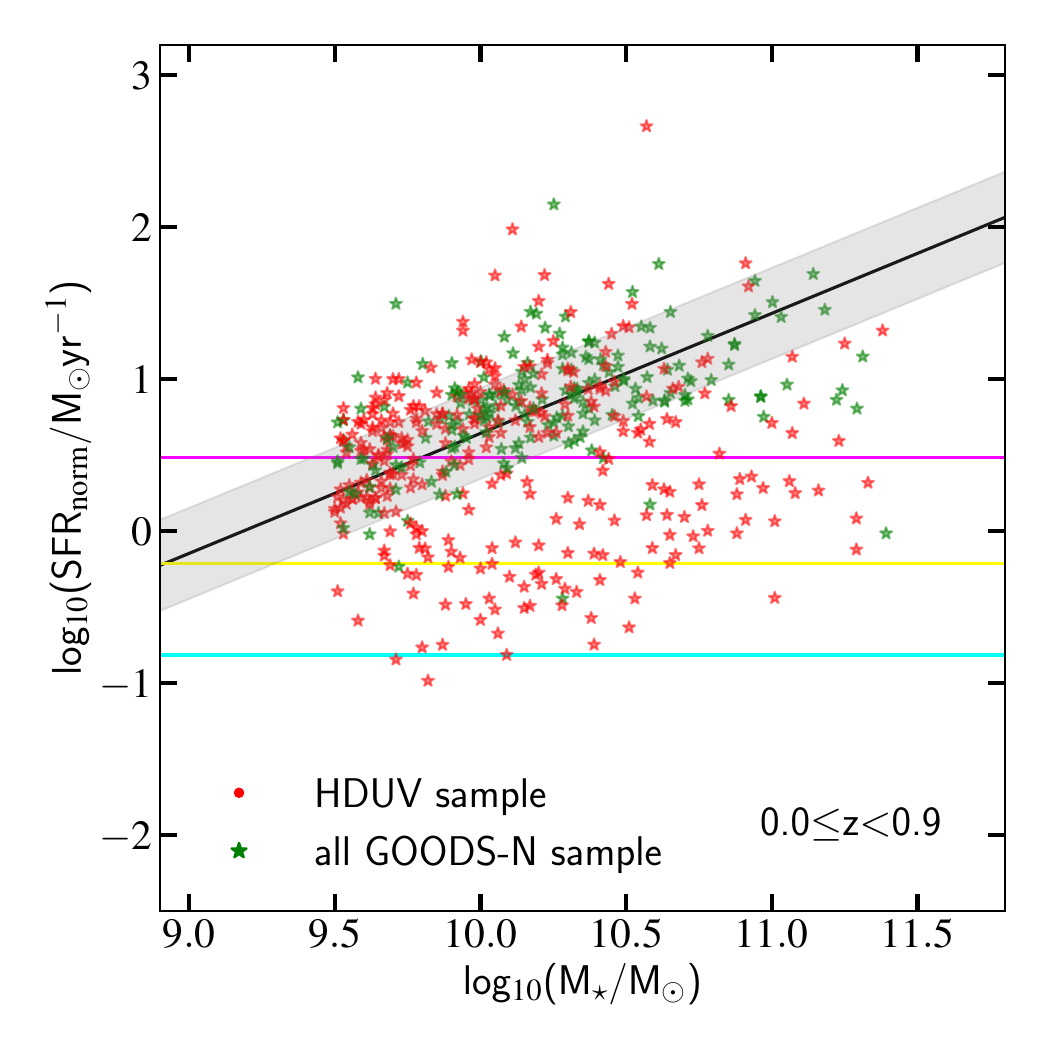}
\includegraphics[width=0.28\textwidth, clip, trim=2.7cm 0cm .75cm 0cm]{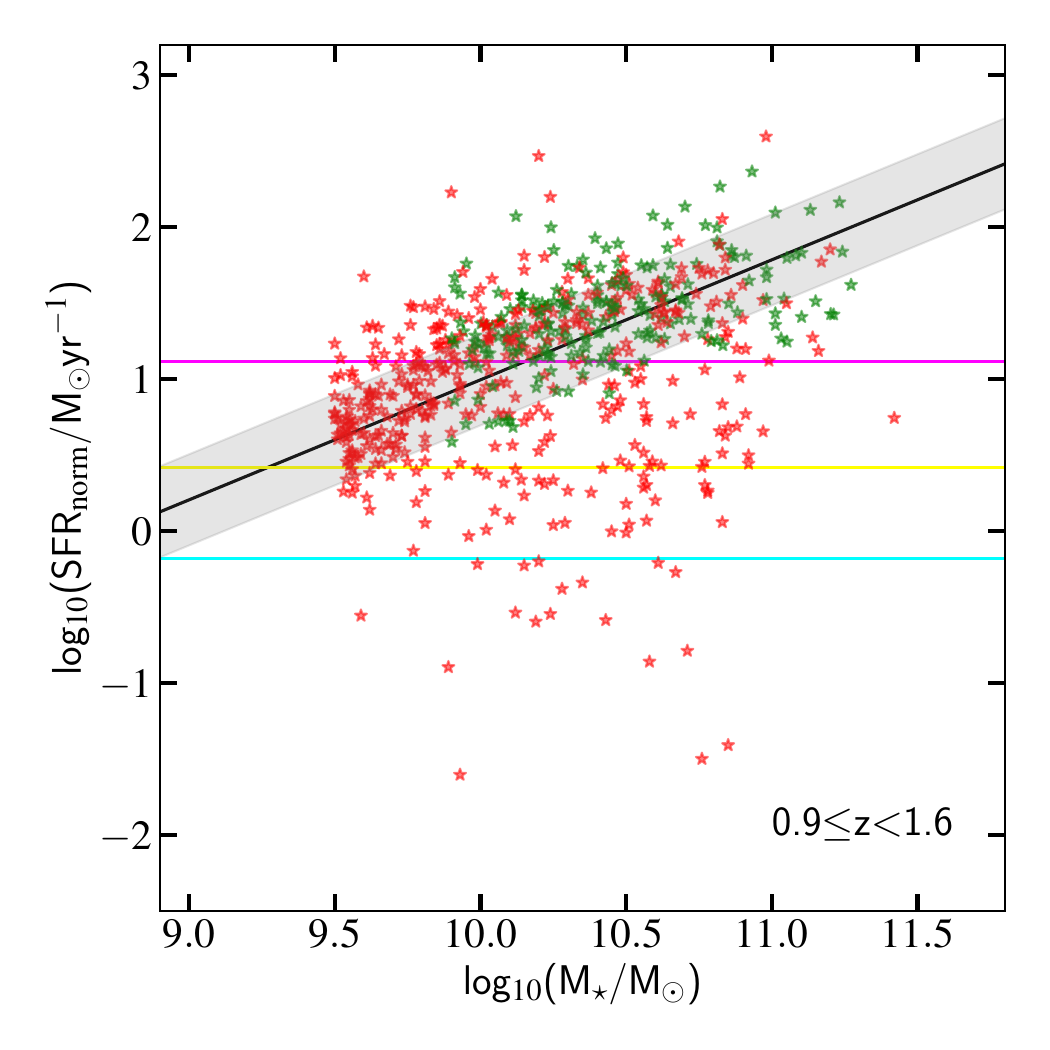}
\includegraphics[width=0.28\textwidth, clip, trim=2.7cm 0cm .75cm 0cm]{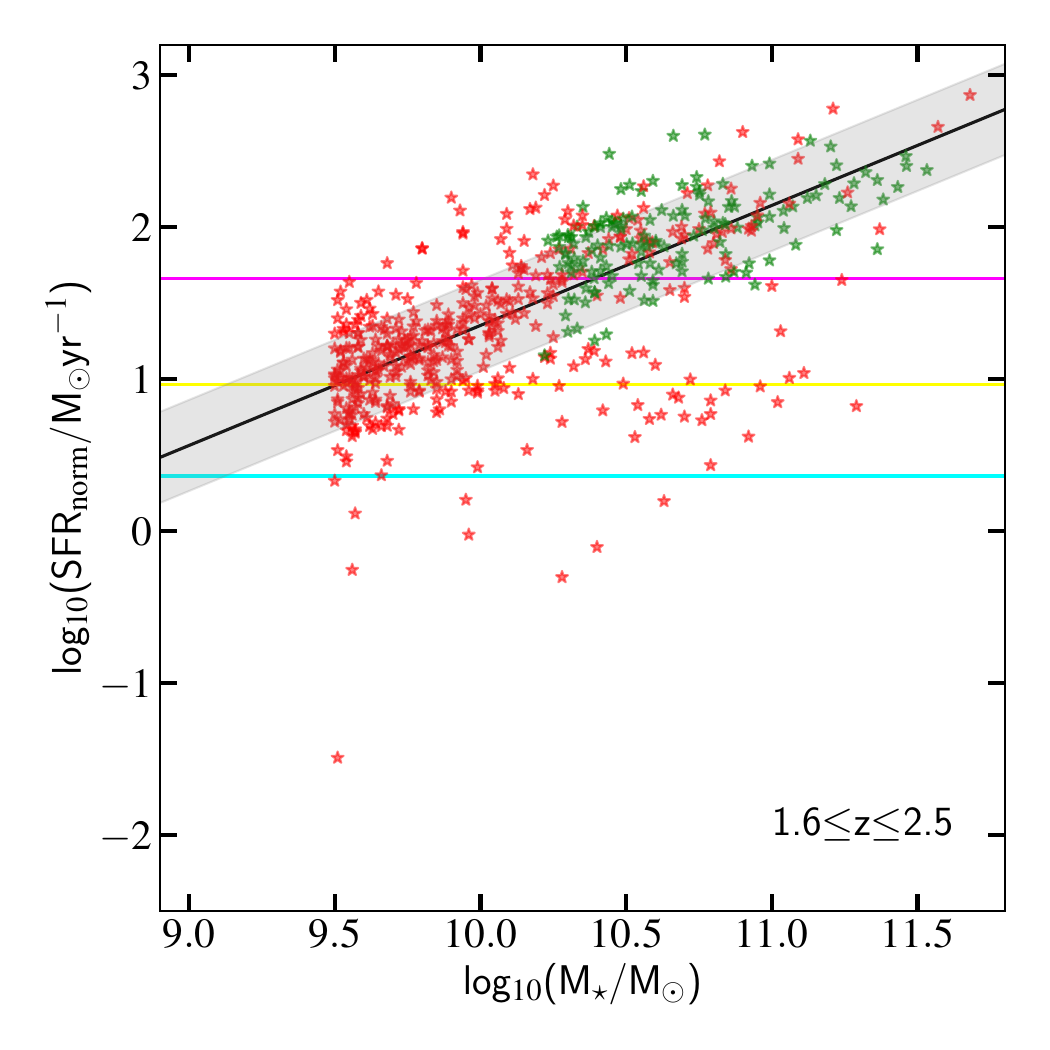}
\end{minipage}
\caption[{Position of our sample on the Main Sequence of star-formation.}]{The position of our galaxy sample on the Main Sequence of star-formation. Three redshift bins are shown, as indicated in the lower right of each panel. The MS of \citet{ref:M.Sargent2014} at the median sample redshift in each bin is shown by the solid black line, with the grey shaded area indicating a 0.3~dex scatter. The SFRs of the galaxies have been normalised in order to lie at their intrinsic $\Delta{\rm MS}$, with respect to the MS shown. The galaxies are coloured by sample, where red is the mass-complete sample (HDUV area) and green is the mass-complete MS \& SB galaxy sample (all GOODS-N). The horizontal lines show the 5$\sigma$ SFR limits that will be probed by the SKA surveys (magenta: wide, yellow: deep, cyan: UD, taken from Fig.~\ref{fig:SFRlimits}, at the median redshift of each panel), for a constant \qtir.}
\label{fig:Cibinel_MS}
\end{figure*}

\section{SKA simulations}
\label{sec:SKA_sim}

In this section, we create simulated images of the SKA sky, using previous multiwavelength, resolved observations of galaxies in the GOODS-N field. We use these previous observations to construct resolved radio flux distributions of the galaxies, as well as to derive galaxy morphological parameters relating to their concentration and flux distribution. Having convolved our model galaxies flux density distributions with the anticipated observational response of SKA-MID, we can test our ability to accurately recover these parameters with SKA in band 2 ($\sim$1.4\,GHz). In Sections~\ref{sec:sample_SKA_sample} to \ref{sec:sample_rad_SKA}, we describe our input galaxy sample, and the derivation of the resolved radio flux maps. We outline the simulation process in Section~\ref{sec:SKA_sims_method}. In Section~\ref{sec:results}, we present the results of our simulations.

\subsection{Sample selection}
\label{sec:sample_SKA_sample}

As input for our image simulations, we combine two galaxy samples from the GOODS-N field. The first is a mass-complete sample of 1174 galaxies at log$_{\rm 10}$($M_{\star}$)$>$9.5 in the central part of the GOODS-N field covered by the Hubble Deep UV Legacy Survey field (HDUV, 56.5~arcmin$^{2}$, \citealt{ref:P.Oesch2018}), which is at least >90\% complete at all $z<2.5$ for stellar masses $>$4.5$\times$10$^{9}$ $M_{\odot}$ \citep{ref:T.Tal2014}. These galaxies have SFRs both above and on the MS of star-formation, with some also falling into the region below the MS where they might be considered quiescent (see Fig.~\ref{fig:Cibinel_MS}).

The second sample is a mass complete sample of 965 MS and starburst galaxies (i.e. excluding quiescent galaxies), covering the full GOODS-N field and with completeness limit evolving from log$_{\rm 10}$($M_{\star}$/$M_{\rm \odot}$)=9.4 to $\sim$10.3 from low-redshift to z$\sim$3 (\citealt{ref:A.Cibinel2019}, see their fig. 2). Approximately 40\% of the galaxies in this 2nd sample are also part of the mass-complete sample located in the HDUV region of GOODS-N with log$_{\rm 10}$($M_{\star}$/$M_{\rm \odot}$)$>$9.5 as described above, leaving 549 unique galaxies in this second sample. In order to homogenise the redshift and mass coverage of our two samples for the subsequent statistical analyses, we limit this sample of MS and starburst galaxies from \citet{ref:A.Cibinel2019} to objects at $z<2.5$ and log$_{\rm 10}$($M_{\star}$/$M_{\rm \odot}$)$>$9.5\footnote{Galaxies in the \citet{ref:A.Cibinel2019} catalogue that do not fall in this stellar mass or redshift range are nevertheless included in our simulated radio image of the GOODS-N field, to create as realistic a mock sky as possible. Although they are present when cross-matching between catalogue and sky, they are not taken into account for any of the statistical analysis in this paper}. We will refer to the two above samples as ``mass-complete sample (HDUV area)'' and ``mass-complete MS \& SB galaxy sample (all GOODS-N)'' respectively.

Stellar masses were derived from multiwavelength SED fitting using the `Fitting and Assessment of Synthetic Templates' tool \texttt{FAST}, \citealt{ref:M.Kriek2009}). Redshifts were derived either from \texttt{EAZY} spectral energy distribution modelling \citep{ref:G.Brammer2008}, 3DHST spectroscopy \citep{ref:R.Skelton2014, ref:I.Momcheva2016}, or the spectroscopic redshift compilation in \citet{ref:D.Liu2018}, where available. All of our total 1723 galaxies have reliable photometry, as defined by the 3DHST flag ``use\_phot=1''.

Finally, we note that AGN were removed from our galaxy samples by \citet[$\sim$7\% of the total sample]{ref:A.Cibinel2019}. They considered AGN as those sources with either X-ray luminosity $L_{0.5-7 \rm keV}\geq$3$\times$10$^{42}$erg s$^{-1}$ or a radio flux excess indicative of radio-loud AGN activity, as detailed in \citet{ref:D.Liu2018}. Additionally, our simulation analysis is based on the assessment of input vs. output parameters, and is therefore not affected by the mechanism giving rise to the emission itself, in the case that our sample still contains AGN with low levels of radio emission.

\subsection{Star-formation rates of the GOODS-North sample}
\label{sec:sample_SKA_SFRs}

\subsubsection{Construction of resolved star-formation maps}

As our galaxies lie in a region of the sky that has previously been observed by many different instruments, large multiwavelength (MWL) auxiliary datasets exist for the derivation of resolved galaxy properties. As described in \citet{ref:A.Cibinel2015}, all HST WFC3 and ACS images of our galaxy sample were convolved to the same, HST H-band resolution before deriving resolved SFR maps. The adaptive smoothing routine \texttt{ADAPTSMOOTH} \citep{ref:S.Zibetti2009} was applied to stacks of HST WFC/ACS images to ensure homogenisation of the signal-to-noise-ratio (SNR) for SED fitting with \texttt{LePhare} \citep{ref:S.Arnouts1999, ref:O.Ilberts2006}. Resolved star-formation rate maps were then constructed for each galaxy using pixel-by-pixel SED fitting of the available HST photometry, using bands B$_{435}$, V$_{606}$, i$_{775}$, i$_{814}$, z$_{850}$, J$_{125}$ and J$_{140}$, as well as H-band and Y$_{105}$ images. For galaxies in the HDUV field, F275W and F336W data were also available. To correct for internal dust extinction, the dust reddening parameter was allowed to vary over a large range for the SED fits (see \citealt{ref:A.Cibinel2015, ref:A.Cibinel2019} for further details). Specifically, a \citet{ref:D.Calzetti2000} law was used with E(B-V) ranging between 0 and 0.9 mag, where the upper bound translates to A$_{FUV}$$\sim$9 and is suitable also for the SB regime \citep{ref:G.Meurer1999, ref:R.Overzier2011, ref:R.Nordon2013}. While it is possible that the SFRs of the most extreme obscured regions (A$_{FUV}$$\gg$10) could be underestimated, comparison with the galaxy-integrated SFR measurements discussed in Section \ref{sec:int_SFRs} suggests that the resolved SED fitting does well at recovering the overall attenuation on galaxy scales.
In constructing the SFR maps, the median SFRs from the \texttt{LePhare} outputs was assigned to the individual resolution elements. If the adaptive smoothing required averaging over $>$5 pixels in order to reach SNR=5 (as may be the case towards the edges of a galaxy), the resolved SFR maps were truncated at this point. If no such truncation occurred, SFR maps extend out to a maximum of one Kron radius ($R_{\rm Kron}$, \citealt{ref:R.Kron1980}), and a value of zero is imposed beyond this radius. The median major-axis $R_{\rm Kron}$ of our sample is 13.8~kpc (proper). The pixel scale of all resolved SFR maps is 0.06"/pixel.

\subsubsection{Integrated star-formation rates}
\label{sec:int_SFRs}

In order to discuss the global star-forming properties of our galaxies, we now demonstrate the consistency between our resolved SFRs and the integrated values present in the literature. For our mass-complete MS \& SB galaxy sample (all GOODS-N), integrated UV+IR SFRs were derived in \citet{ref:D.Liu2018}, using SED fitting of super de-blended photometry from 2$\mu$m to radio wavelengths, in addition to UV observations or an appropriate SFR$_{\rm UV}$/SFR$_{\rm IR}$ ratio. As discussed in \citet{ref:A.Cibinel2019}, the inclusion of this long-wavelength emission becomes particularly important for calculating SFRs in the highly star-forming, dust-obscured regime. For this reason, we also adopt the \citet{ref:D.Liu2018} UV+IR SFRs for galaxies from the mass-complete (HDUV area) sample whenever possible (i.e. for 247 galaxies). For the majority of the remaining 857 HDUV galaxies, we use UV+IR SFRs from \citet[based on the work of \citealt{ref:E.Bell2005}]{ref:K.Whitaker2014}, where the dust-obscured contribution to the SFR is inferred from single-band Spitzer 24$\mu$m observations. There were only a small number of HDUV galaxies ($\sim$70; $\sim$4\% of the sample) for which integrated UV+IR SFRs were not available. In these cases we use total SFRs derived by \citet{ref:K.Whitaker2014} via UV to near-IR \texttt{FAST} SED fitting. We perform a search for all galaxies with \texttt{FAST} SFRs from \citet{ref:K.Whitaker2014} that are also present in the catalogue of \citet{ref:D.Liu2018}, and find 693 matches. This allows us to calculate a median systematic offset between the two SFR derivations. Therefore, for our small fraction of HDUV galaxies where no galaxy-integrated UV+IR SFR exists, we apply this correction to the \texttt{FAST} SFRs.

In Fig.~\ref{fig:SFR_corrections}, we compare the integrated SFRs of our galaxies with the sum of their pixel-by-pixel \texttt{LePhare} SFRs that we will subsequently use to construct resolved radio maps (Section \ref{sec:sample_rad_SKA}). Only galaxies lying $<$1~dex below the galaxy MS are shown in this figure, as the SFR maps of galaxies significantly below the MS (and hence their pixel-by-pixel SFR sums) are subject to significant uncertainty. Note, however, that we do not discard these low specific SFR (SFR/$M_{\star}$, sSFR) galaxies from our sample or analysis, in order to achieve as realistic a source density as possible. The galaxy-integrated and the resolved SFR estimates for the HDUV sample (red) are in good agreement (median SFR$_{\rm int}$ = SFR$_{\rm pix}-$0.13~dex, scatter of 0.38~dex around the 1:1 line). The two SFR measurements are also consistent for the all GOODS-N sample (green; median SFR$_{\rm int}$ = SFR$_{\rm pix}-$0.06~dex, with an 0.32~dex scatter). For the remainder of this paper, we use SFR to refer to integrated SFR, unless specified otherwise. In summary, Fig.~\ref{fig:SFR_corrections} demonstrates good agreement between the resolved and integrated SFRs, implying that the map-based resolved SFRs underpinning our mock SKA images are not subject to over- or underestimation beyond the scatter observed in Fig. \ref{fig:SFR_corrections}.

In the remainder of this section we illustrate the star-formation parameter space sampled by our GOODS-N galaxy sample, and we discuss its completeness.

Fig.~\ref{fig:Cibinel_MS} shows the position of the GOODS-N sample compared to the galaxy MS, using their integrated SFRs and for three redshift bins at $0.0\leq z < 0.9$ (left; 507 galaxies), $0.9\leq z <1.6$ (center; 641 galaxies) and $1.6\leq z <2.5$ (right; 575 galaxies). For each galaxy, we calculate its offset $\Delta$MS from the MS at the galaxy's actual redshift. In each figure panel, the MS is then shown at the mean redshift of the bin, and galaxies are placed at the appropriate offset $\Delta$MS calculated above. Fig.~\ref{fig:Cibinel_MS} illustrates that our galaxy sample spans between 2-3 orders of magnitude in sSFR, ranging from starbursts to quiescent galaxies at all redshifts. We colour galaxies by their sample selection, highlighting the regions of the MS occupied by the different samples.

\begin{figure*}
\begin{minipage}{\textwidth}
\includegraphics[width=0.283\textwidth, clip, trim=0cm 0cm .9cm 0cm]{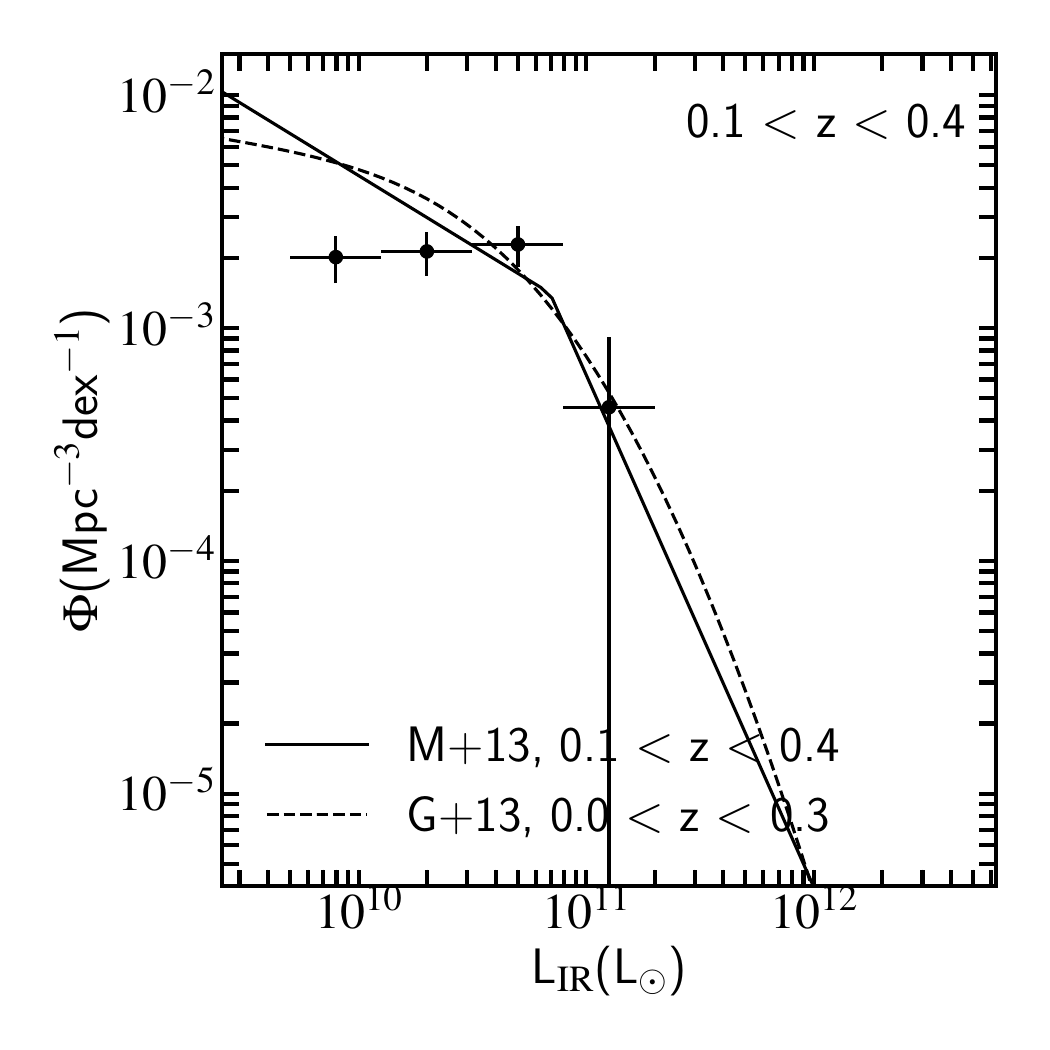}
\includegraphics[width=0.22\textwidth, clip, trim=3.75cm 0cm .9cm 0cm]{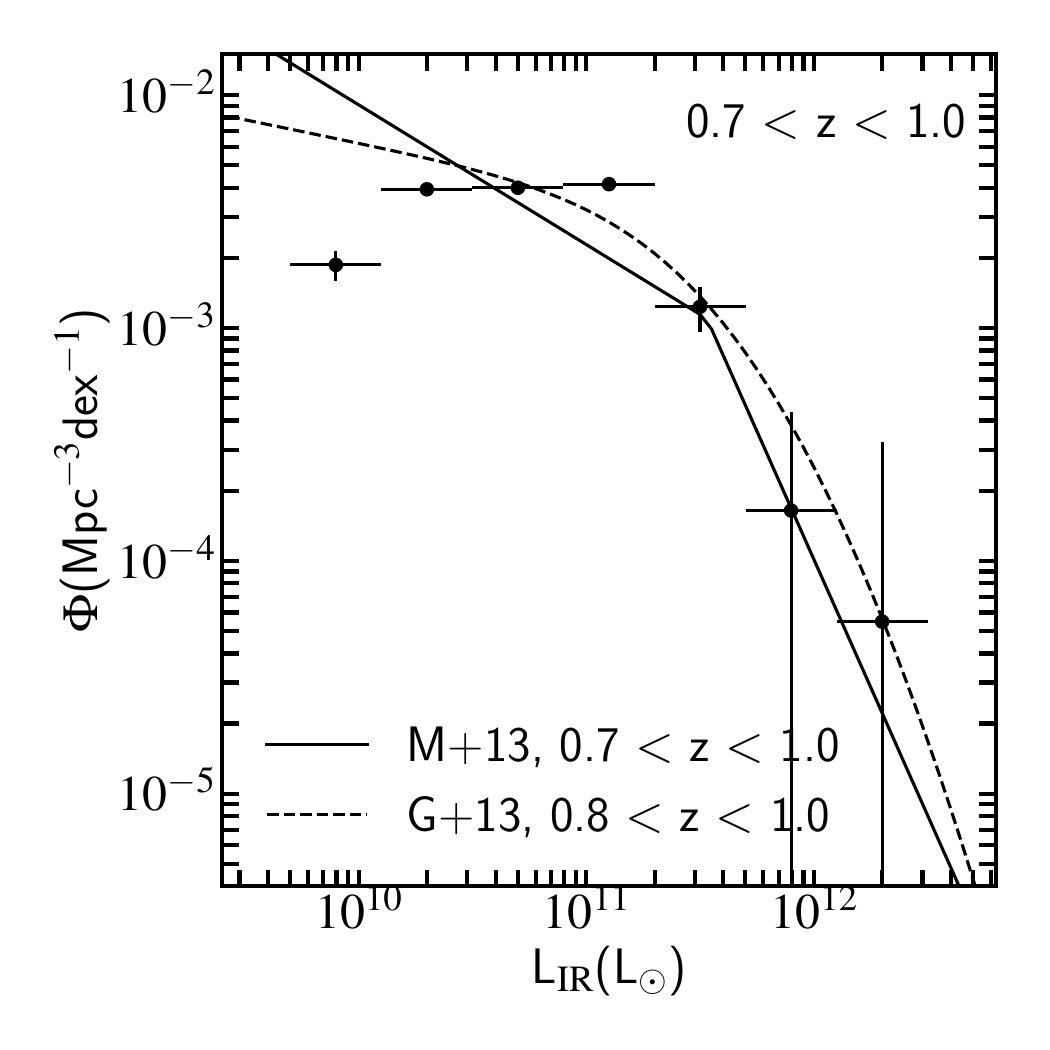}
\includegraphics[width=0.22\textwidth, clip, trim=3.75cm 0cm .9cm 0cm]{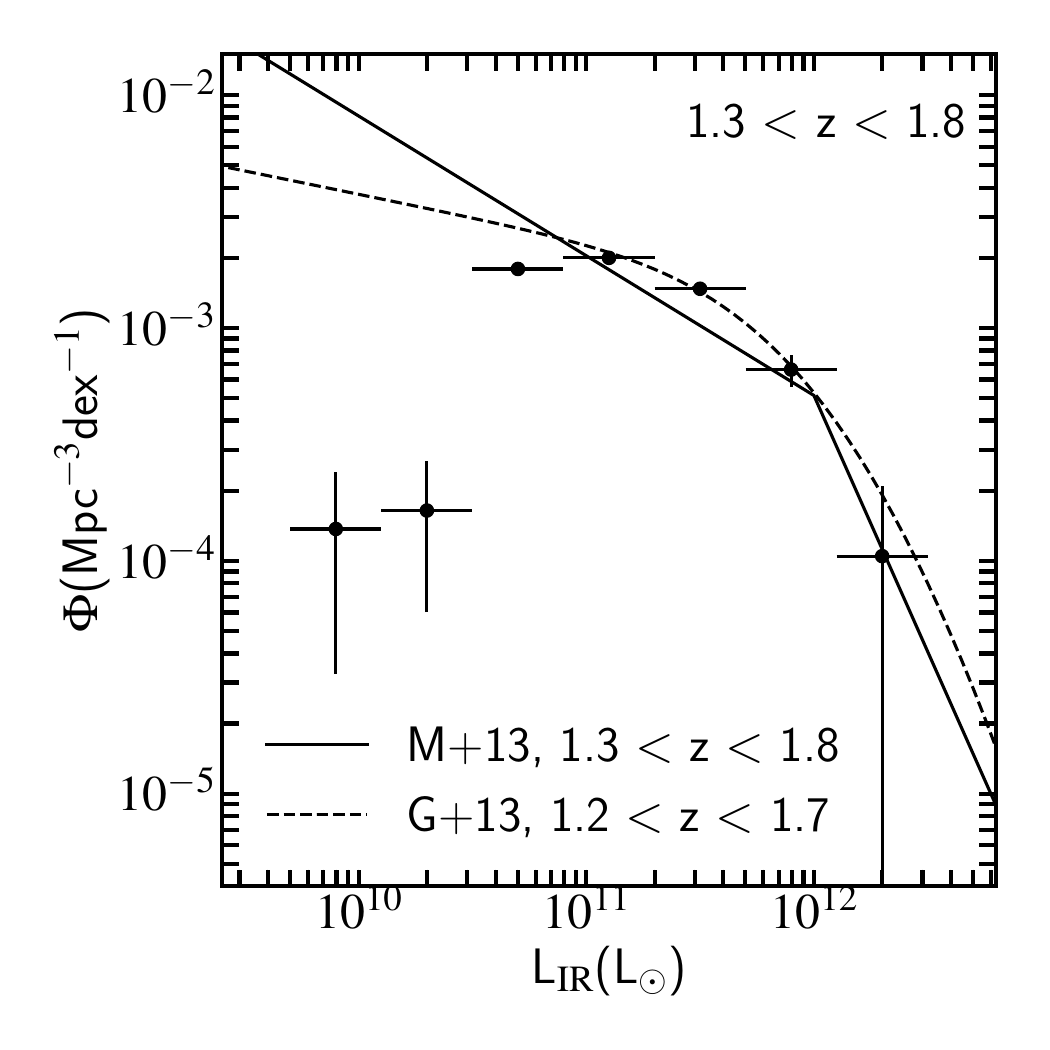}
\includegraphics[width=0.22\textwidth, clip, trim=3.75cm 0cm .9cm 0cm]{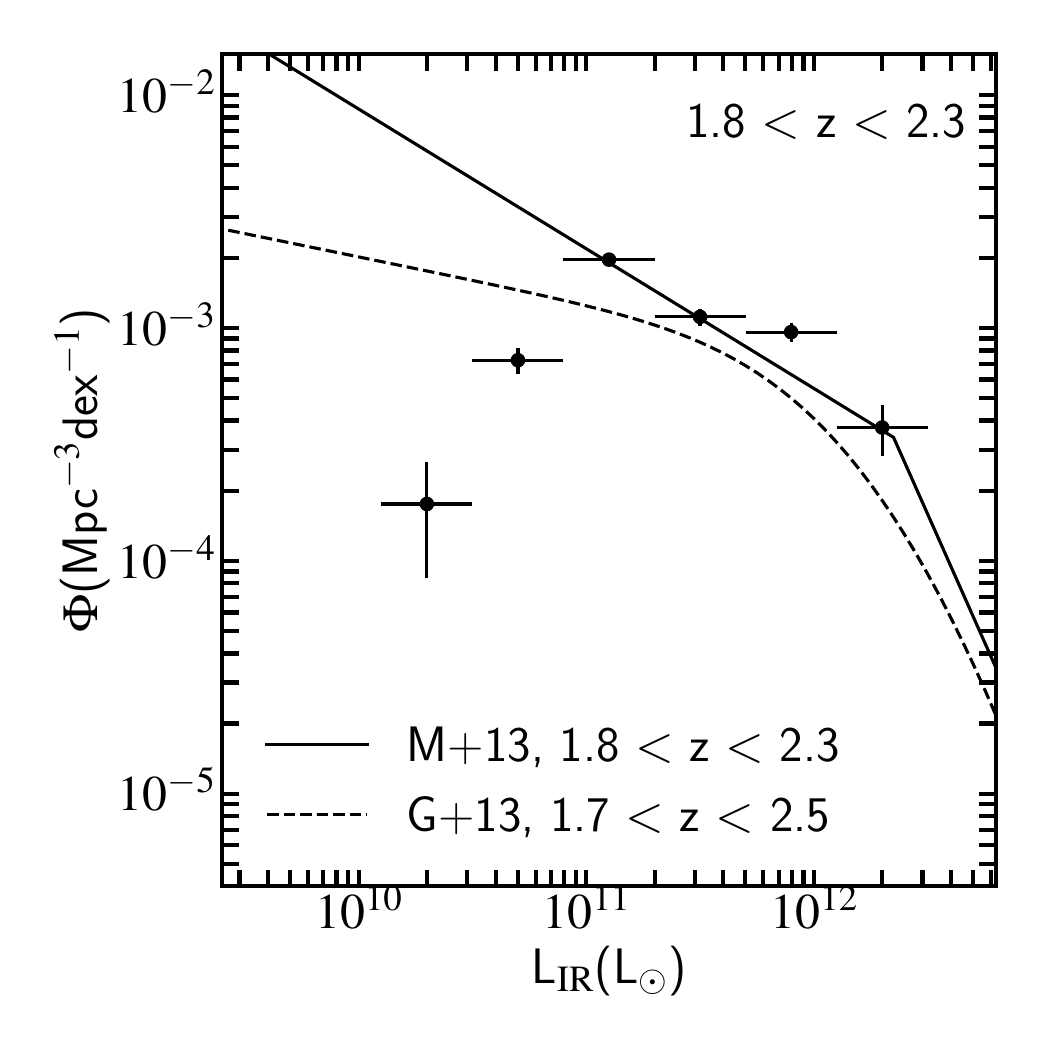}
\end{minipage}
\caption{Infrared luminosity function of our galaxy sample, for the selection of indicated redshift bins. Our galaxy sample is shown by the black circles, and the IR luminosity functions of \citet[G+13]{ref:C.Gruppioni2013} and \citet[M+13]{ref:B.Magnelli2013} are shown by the black dashed and solid lines respectively. Vertical errors are Poisson errors, horizontal errors show the $L_{\rm IR}$ range.}
\label{fig:sf_function}
\end{figure*}

Furthermore, we show in Fig.~\ref{fig:Cibinel_MS} the 5$\sigma$ SFR limits that will be probed by the SKA surveys (taken from Fig.~\ref{fig:SFRlimits}, at the median redshift of each panel in Fig.~\ref{fig:Cibinel_MS}), for the constant \qtir\ case. Between $0<z<2.5$, our HDUV sample (shown in red in Figures~\ref{fig:SFR_corrections} and \ref{fig:Cibinel_MS}) probes down to SFRs on average $\sim$5$\times$ lower than our all GOODS-N sample (shown in green). In the left panel of Fig.~\ref{fig:Cibinel_MS} at $0<z<0.9$, we see
that the minimum SFR of our all GOODS-N sample is detectable by both the UD and deep tiers in the stellar mass range of our galaxies (log$_{\rm 10}$[$M_{\star}$>9.5]), whilst there are a few all GOODS-N galaxies that have SFRs too low for the wide tier to detect. On the other hand, in order to detect all of the lowest-SFR galaxies in the HDUV sample, we would be required to use the UD tier in this redshift range.
As we move to $0.9<z<1.6$, we see a similar situation. The majority of the all GOODS-N sample (green) remains detectable by all three tiers of the survey, due to the relatively high SFRs, although the lowest-SFR galaxies in this sample are missed by the wide tier. However, the faintest star-forming galaxies in our HDUV sample (red) at these redshifts lie a few dex below the MS, meaning even the deep and UD tiers are not able to detect them.
At $1.6<z<2.5$ (right panel), the deep tier is able to detect all galaxies in our (green) all GOODS-N sample, but we see increasing numbers of galaxies in the HDUV sample at the low mass end, which can only be detected by the UD tier. Once again, there are a handful of galaxies that are undetectable even with the UD tier.
These are particularly relevant observations, as our sample allows us to explore the detectability of our galaxies as a function of survey depth, as we do not expect all of them to be present in the shallower tiers. For comparison, we note that the area of the GOODS-N field is $\sim$0.04~deg$^{2}$, which is smaller than the planned 1~deg$^{2}$ coverage of the ultradeep tier. We also note that due to the high angular resolution of SKA-MID (sub-arcsecond at 1.4~GHz), it is not expected that source confusion will limit the sensitivity of surveys such as these (e.g. \citealt{ref:J.Condon2012}). However, this high angular resolution means that a large number of sources will be resolved, introducing some sample incompleteness for resolved galaxies with integrated SFRs close to the detection threshold. To quantify the effect of resolved observations on the predictions of Fig.~\ref{fig:MS_surveys}, we revisit this parameter space for our galaxy sample in Fig.~\ref{fig:discussion_MScompl_w_depth}.

Finally, we show in Fig.~\ref{fig:sf_function} the SFR function (as probed by the inferred IR luminosity function (LF) of our galaxy sample) at different redshifts. IR luminosities for sample galaxies are calculated as $L_{\rm IR}$ ($L_{\odot}$)~=~SFR$_{\rm int}$ ($M_{\odot}$/yr)~$\times$~10$^{-10}$. We do not apply completeness corrections to our data, but calculate number densities purely based on the volume of the GOODS-N field probed, the redshift range of our sample, and the number of galaxies in each bin. For reference we show the IR luminosity functions of \citet{ref:C.Gruppioni2013} and \citet{ref:B.Magnelli2013}, with the according redshift ranges displayed in the panel legends. The completeness limit of our sample increases with redshift, from $\sim$3$\times$10$^{10}$~$L_{\odot}$ at $0.1<z<0.4$ to $\sim$10$^{11}$~$L_{\odot}$ in the higher redshift bins. The number density of high-luminosity galaxies in our sample is well-matched to the literature luminosity distributions. Our sample can therefore be considered complete down to luminosities below the LF knee, before dropping below the literature LFs at low luminosities. Fig.~\ref{fig:sf_function} therefore demonstrates that the GOODS-N galaxies used to populate our mock SKA images are representative of typically observed galaxy samples.

\subsection{Intrinsic radio flux distributions}
\label{sec:sample_rad_SKA}

For each galaxy, we create resolved observed radio flux maps at the continuum frequency of SKA band 2, $\nu_{\rm obs}\sim$1.4~GHz. SFR values were converted to total radio luminosity, on a pixel-by-pixel basis, following the constant \qtir\ case described in Section~\ref{sec:method_SKA}. This is the most conservative scenario at higher redshift ($z>0.4$), as it gives rise to the lowest predicted radio flux densities. Having created individual radio cutouts of each galaxy, these cutouts were combined into a single image (using the observed positions of the galaxies on the sky), representing the full GOODS-N field as seen at radio frequencies. Example regions of this `intrinsic' or `model' (meaning before the SKA simulation) band 2 continuum map, in addition to HST/WFC3 H-band images of the galaxies, are shown in the first two rows of the small cutouts in Fig.~\ref{fig:ska_radio_cut}. The angular resolution of the H-band images shown in the top row is FWHM=0.15", and it is from HST images of this resolution that the model SFR (and therefore radio) maps were produced\footnote{As the median $R_{\rm Kron}$ of our sample is 13.8 kpc (proper), $\sim$10 HST resolution elements would cover the Kron radius of a galaxy at $z=2$, before convolution to simulate the SKA beam}. We therefore take this original FWHM=0.15" resolution into account when convolving to a final resolution of FWHM=0.6" in our simulated SKA images. We discuss the images output from our simulations (shown in the lower three rows of small cutouts in Fig.~\ref{fig:ska_radio_cut}) in Section~\ref{sec:results}.

\begin{figure*}
\includegraphics[width=\textwidth, clip, trim=0.cm 2.cm 0.cm 0.cm]{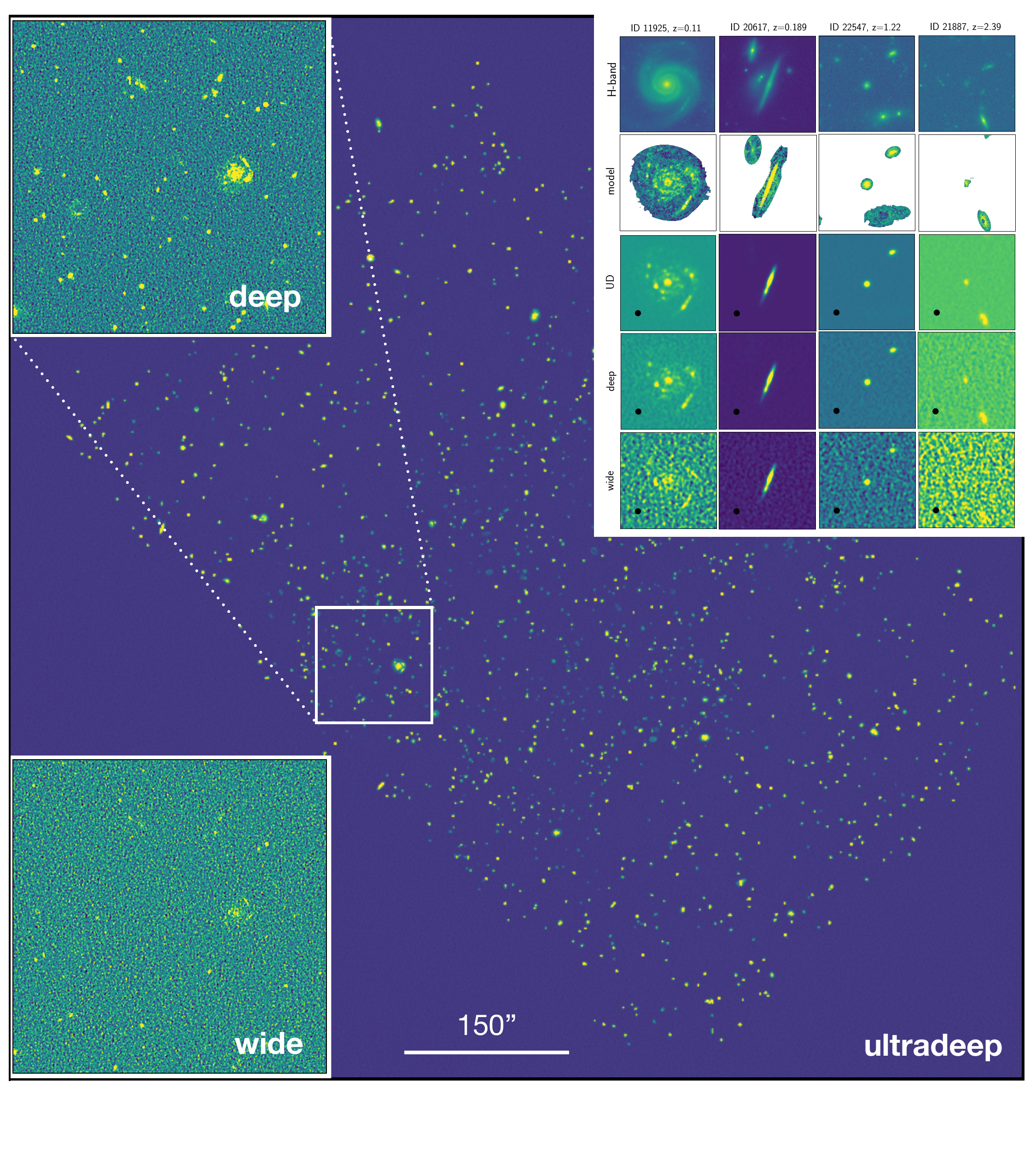}
          \caption{Background image: cutout of the final primary-beam corrected, simulated SKA image including noise, in the UD tier. 120" $\times$ 120" cutouts of the deep and wide tiers are shown in the upper and lower left respectively, for the same part of the sky. Upper right: 20"$\times$20" cutouts, taken from the full GOODS-N image, showing examples of HST/WFC3 H-band, intrinsic/model radio, and SKA-simulated radio images for the UD, deep and wide survey tiers. The rows are labelled on the left, and galaxy IDs and redshifts are shown above each column. The H-band and model images are at a scale of 0.06"/pixel, and the white regions in the model row indicate the zero-valued background. The simulated survey images are at a scale of 0.12"/pixel, and the FWHM=0.6" of the simulated synthesised beam is shown by the black circles in the bottom left of the cutouts.}
\label{fig:ska_radio_cut}
\end{figure*}

We show examples of model radio maps for four different galaxies in the top right corner (second row of cut-out stamps) of Fig.~\ref{fig:ska_radio_cut}. As the model SFR maps are truncated at $R_{\rm Kron}$, with SFR=0 imposed outside, such that the limits of the radio continuum emission can be seen. Our intrinsic model radio GOODS-N field can thus be considered noise-free. In the first column, we show a galaxy at $z=0.11$ with clearly visible H-band spiral arms, oriented face-on towards the observer and extending over a sky area of almost 20"$\times$20". The next column is a disk oriented edge-on, with a bright elongated core. The cutout in the third column is centred on a galaxy at intermediate redshift, with relatively spheroidal morphology in the H-band image. We also see several neighbouring galaxies in the H-band image, which all reappear model SFR image on the second row, as these cutouts are taken from the combined image of the full model GOODS-N field. We can see from the radio image that although the cores of the stellar emission of the Southern neighbours are separable in the H-band image, the limit between their star-formation emission regions is not visually distinct. Finally, in the fourth column of Fig.~\ref{fig:ska_radio_cut}, we see a galaxy at much higher redshift, $z=2.39$. This galaxy is much fainter in the H-band image than the lower redshift galaxies, but can still be seen through its star-formation in the intrinsic radio maps.

Finally, we show in the background of Fig.~\ref{fig:ska_radio_cut} our primary-beam corrected, simulated image for the ultradeep field. 120"$\times$120" zoom-ins for the deep and wide tiers are shown in the upper and lower left corners, respectively.

\begin{figure}
\centering
\includegraphics[clip, trim=0cm 0cm 0cm 0cm, width=0.47\textwidth]{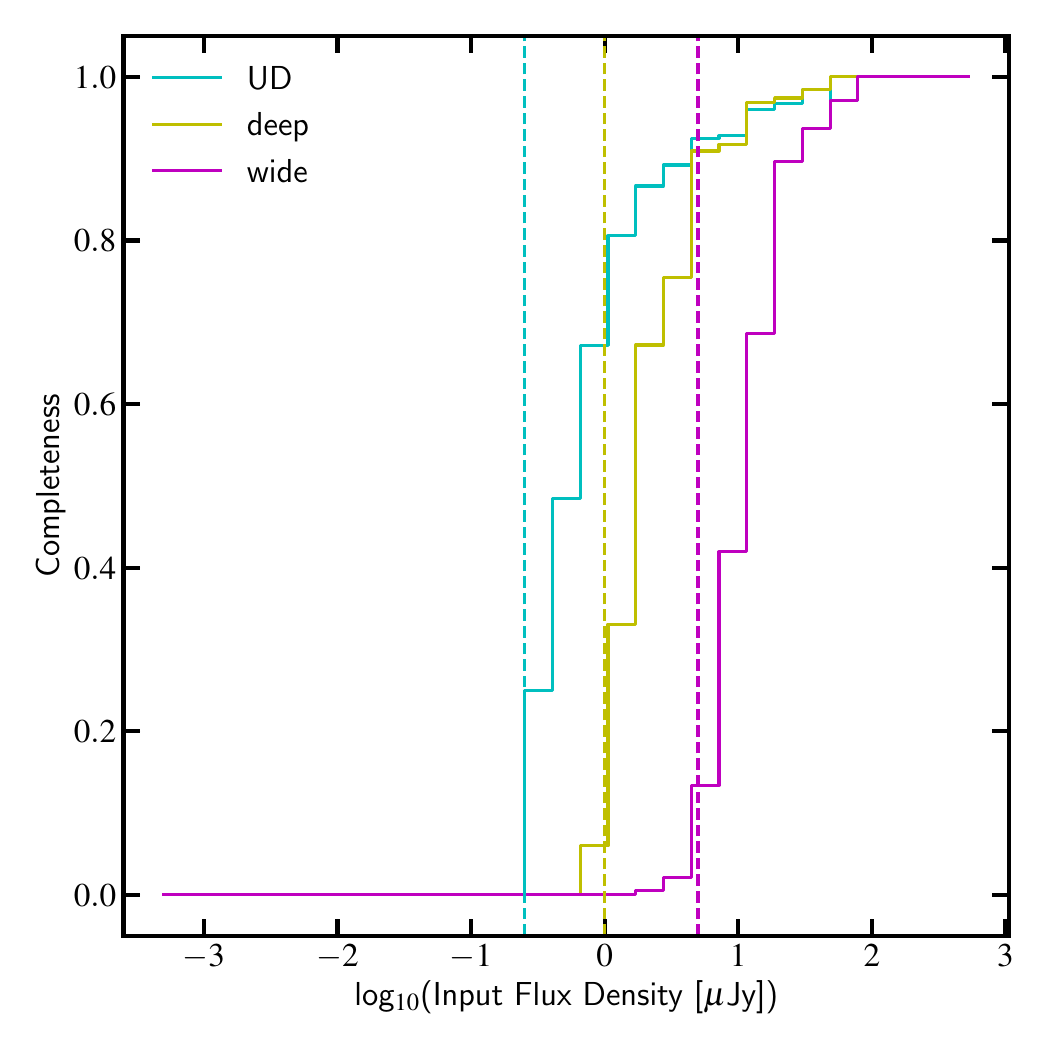}
\caption{Completeness in different input flux density bins, for all three survey tiers. The 5$\sigma$ RMS levels in the output images for the three tiers are shown by the dashed lines.}
\label{fig:completeness}
\end{figure}

\begin{figure*}
\centering
\begin{minipage}{\textwidth}
\includegraphics[clip, trim=0cm 0cm 0cm 0cm, width=0.33\textwidth]{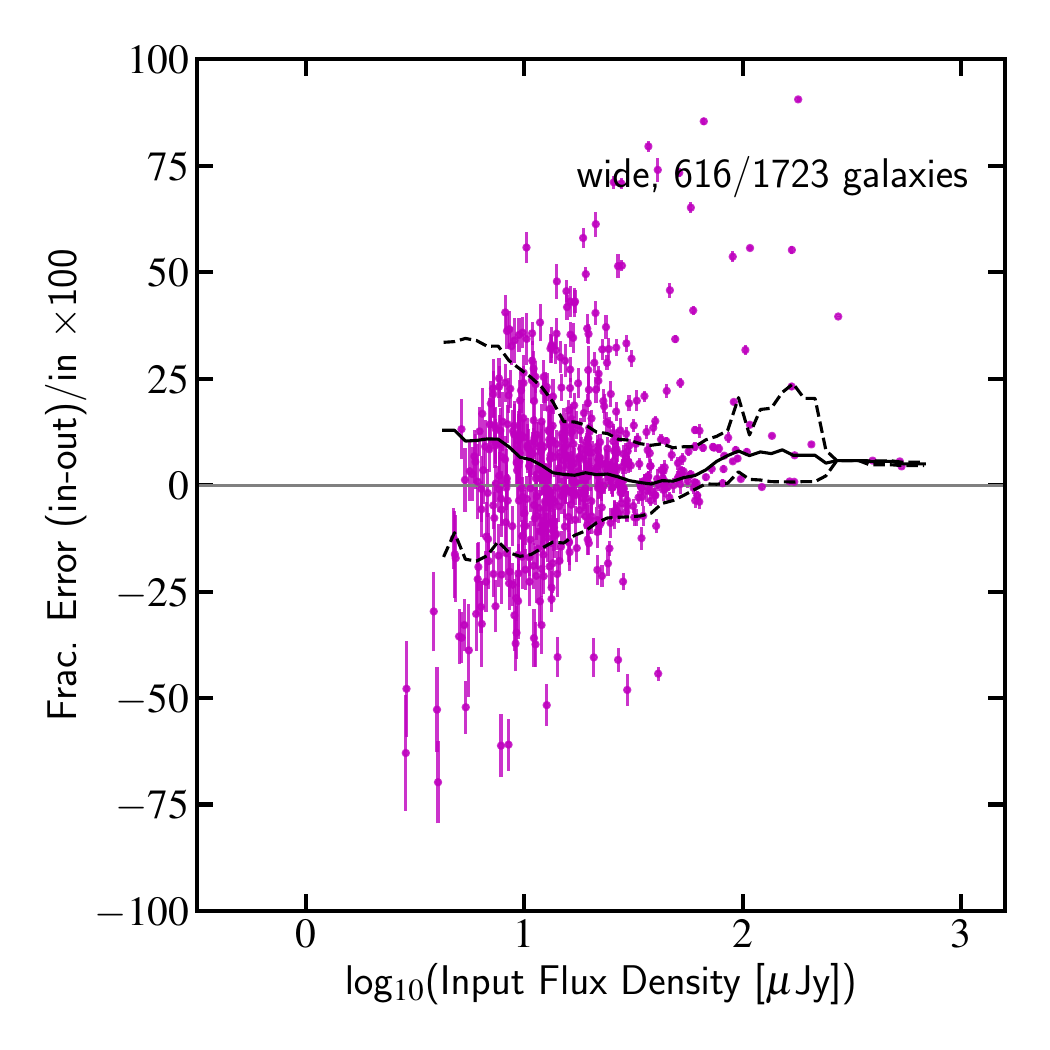}\hfill
\includegraphics[clip, trim=0cm 0cm 0cm 0cm, width=0.33\textwidth]{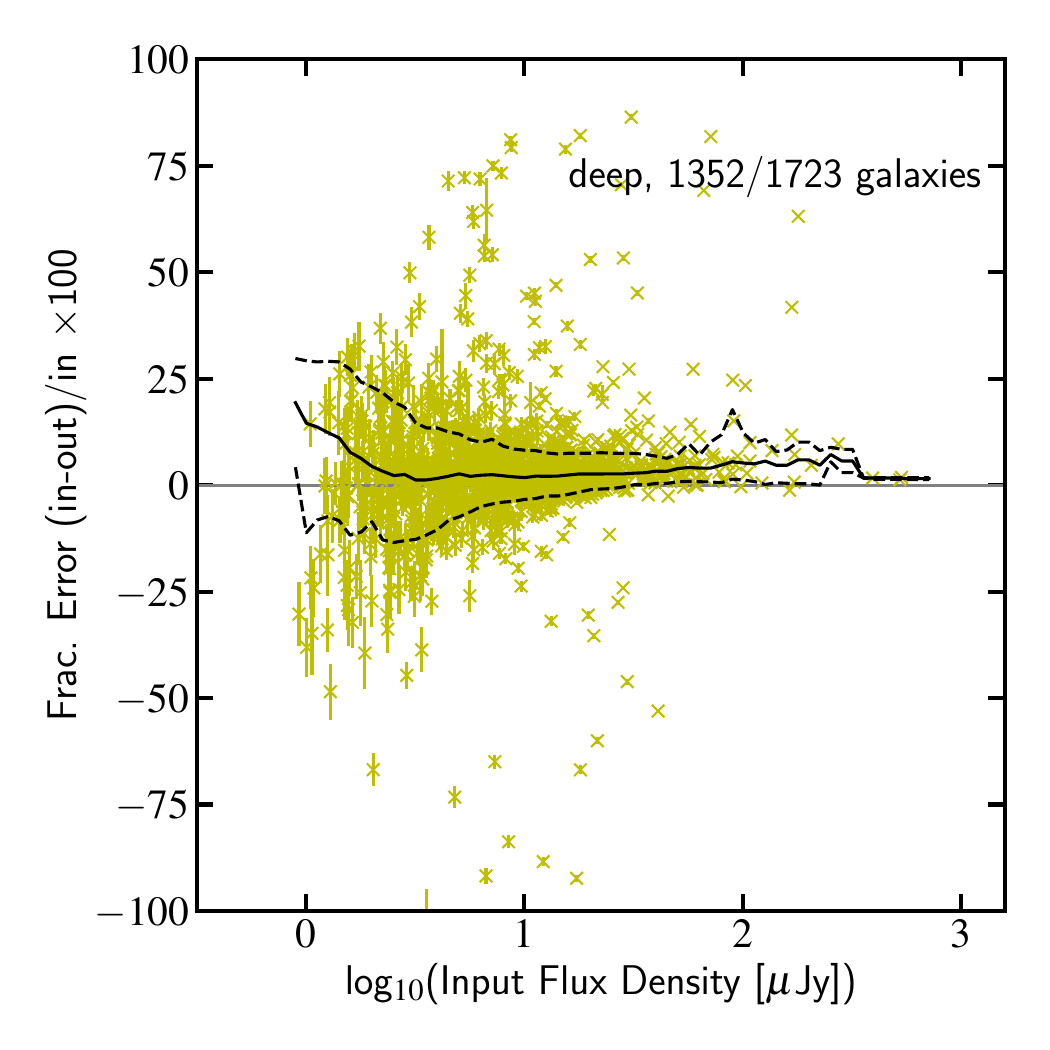}\hfill
\includegraphics[clip, trim=0cm 0cm 0cm 0cm, width=0.33\textwidth]{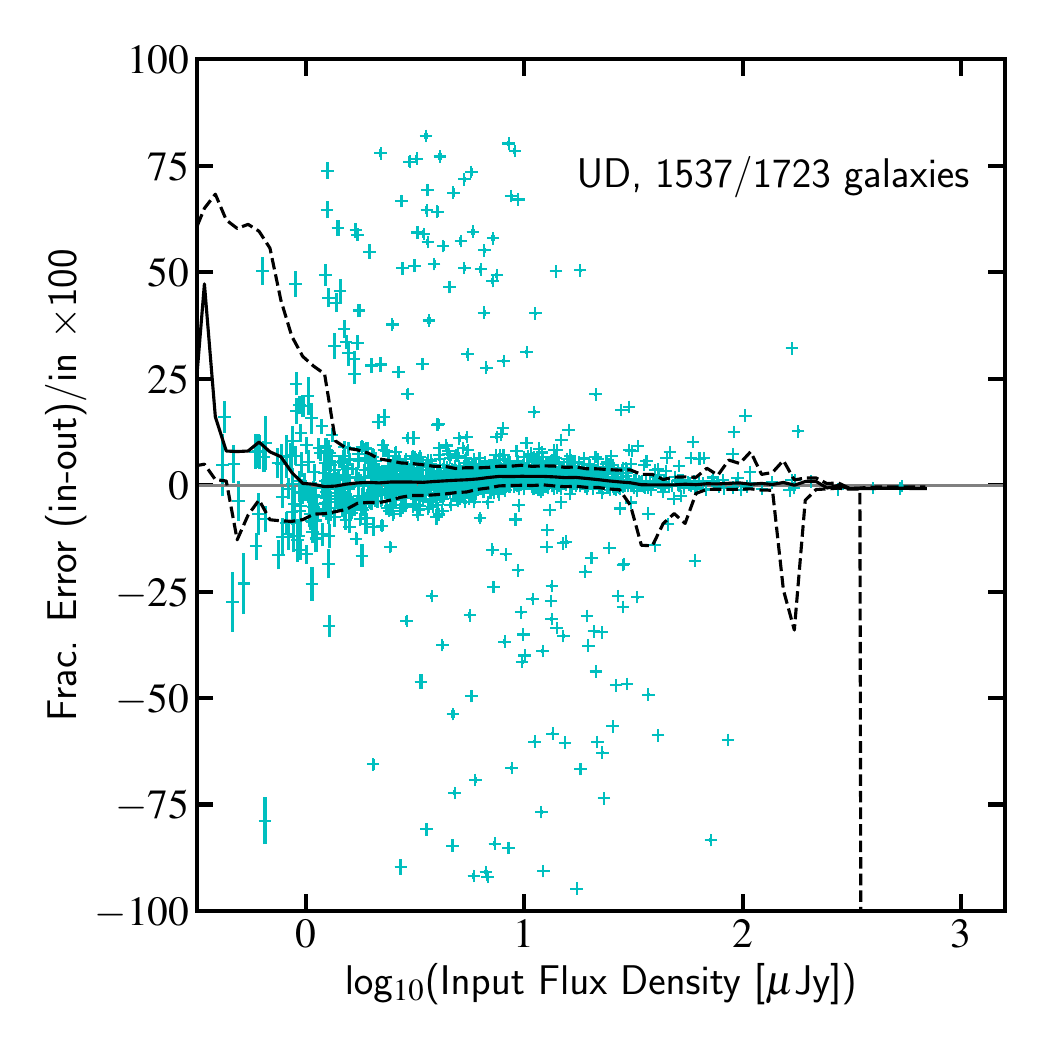}\hfill
\end{minipage}
\caption{Fractional error on the flux density as a function of input flux density. Running medians (bin width of 0.3~$\mu$Jy) and 84th/16th percentiles are shown by the solid and dashed black lines in each panel, respectively. The number of matched (detected) galaxies in each tier is shown in the upper right corner. The small ($<$1\%) output flux density corrections described in Section~\ref{sec:fluxes} have been applied. The output flux densities have also been systematically increased by 1.2\%, to account for the inherent flux density loss of the simulation, also described in the text.}
\label{fig:fluxflux}
\end{figure*}

\subsection{Simulation outline and setup}
\label{sec:SKA_sims_method} 

We perform our SKA-MID simulation in band 2 (0.95-1.76~GHz), taking a central instantaneous bandwidth (IB) of 420~MHz, corresponding to a fractional bandwidth $\delta\nu/\nu$=0.3 as also adopted for SDC1. This is a conservative expectation for the usable bandwidth in band 2 after excision of channels affected by radio frequency interference (RFI). A larger usable IB would imply a correspondingly reduced observing time, without changing the output SKA sky map as the simulation simply requires an input RMS noise level -- without reference to the on-source time required to actually achieve it (for approximate observing time estimates see Section \ref{sec:tiered_desc}).

For our image simulation we implement a circular Gaussian synthesised beam of FWHM=0.6" (as also adopted for SKA Science Data Challenge SDC1), as this is the best angular resolution that can be achieved -- given the SKA-MID baseline distribution -- whilst also maximising the sensitivity. Note that the weighting scheme implemented to achieve a well behaved Gaussian synthesised beam does come at the expense of sensitivity {\it relative} to the sensitivity achievable under natural weighting. However, this trade-off results in a synthesised beam with deviations from perfect Gaussianity that amount to $<$1$\times$10$^{-3}$ on the spatial scales relevant to our study \citep{ref:R.Braun2017}. We regrid our model image from the native pixel scale of 0.06"/pixel to 0.12"/pixel, to appropriately sample an SKA-MID synthesised beam. These resampled images are then input into the SKA simulation pipeline. The pipeline is based on the sequence of modified \texttt{Miriad} routines developed by the SKAO for SDC1 \citep{ref:A.Bonaldi2021}, and has been adapted for use with our data. The key steps of the simulation are as follows:
\begin{enumerate}
\item Take science image (observed radio flux intrinsic image) and SKA-MID band 2 primary beam response (resampled to match the pixel scale and world coordinate system (WCS) of the science image) as input.
\item Convolve the science image with a Gaussian (`clean') beam (to give a final resolution of FWHM=0.6", as well as giving image units of Jy/beam), and apply the primary beam attenuation to the science image.
\item Generate the SKA-MID mock visibility coverage. This depends upon the SKA-MID configuration of 197 dish antenna, correlator frequency setup (number of spectral windows, bandwidth, channel width), hour angle range on the sky, polarisation, source position, telescope latitude, and the Jy/K and system temperature of the telescope. We use an hour angle corresponding to 8 hours, the observation time expected for a wide tier pointing, while multiple visits of this duration will be necessary to reach the necessary on-source time for the two deeper tiers.
\item Create image of the dirty beam from these visibilities, as well as dirty images containing only noise, with the scale of the noise adjusted so that the RMS corresponds to the desired depth of observation.
\item Clip the science image (in Jy/beam) at the 3$\sigma$ flux level for an 8 hour observation. This produces a `residuals' image, containing only the data below the clip level.
\item Subtract the residuals image from the science image, leaving an image containing only peaks above this 3$\sigma$ noise.
\item Perform a linear deconvolution of the residuals with the clean beam. Then convolve them with the dirty beam.
\item Add the peak image to the dirty-beam-convolved residuals image (above), and then add the dirty noise image at the desired RMS level.
\end{enumerate}

Steps (v - viii) are somewhat equivalent to the classic `deconvolution' technique of \texttt{CLEAN}ing an interferometric image (which is always necessary when imaging an entire field containing more than one source, as we are here). When \texttt{CLEAN}ing an image, the dirty image is inspected for peaks in flux above a given threshold. The flux associated with such peaks is then subtracted from the image, and added to a `model' image. This is done iteratively, leaving a residual image below the set threshold, and a model image. The peaks in the model image are convolved with the `clean', or synthesised beam, and then added back to the residual image (which remains convolved with the dirty beam). Steps (v - viii) of our \texttt{Miriad} pipeline therefore emulate this process, by convolving the SKA dirty and clean beams with the residual and peak images derived from the input science image, respectively. As with the SDC1, a full end-to-end simulation was not performed due to the sheer volume of data that this would produce. Although our approach makes producing meaningful simulations more efficient (and therefore more accessible), it is important to recall the limitations. Our galaxies are not modelled in the uv-plane (only the noise is), and the achievable RMS sensitivity is not impacted by the dynamic range of our sources, i.e. there are no image artefacts due to very bright sources. We note that even with a full end-to-end simulation dynamic range systematics would be small, as our model GOODS-N field by construction does not contain emission from objects such as radio-loud AGN, which would typically be present in observed fields. Furthermore, our simulation does not account for calibration errors that might be introduced during data reduction and we do not consider a frequency-dependent synthesised beam, which could become a non-negligible effect for higher fractional bandwidths.

The inputs to the \texttt{Miriad} pipeline are therefore: science image to be simulated, primary beam response for the desired frequency, technical setup to generate the visibility coverage (step iii), target RMS of the output image and FWHM of the desired synthesised beam. We note that we take all 197 SKA-MID dishes to have a 15~m diameter, without modelling the 64 smaller (13.5~m) MeerKAT dishes that will be incorporated into SKA-MID. This approach was also taken for SDC1. Although the source position for our science image was given as zenith in the simulation, the actual GOODS-N field will not be observable at this elevation by the Southern hemisphere SKA. However, the galaxies in GOODS-N are representative of galaxies that the SKA will observe, and the ancillary data available for GOODS-N allowed us to create realistic resolved radio maps.

\section{Results}
\label{sec:results}

\subsection{Galaxy identification in the survey tiers}

\subsubsection{Source extraction}
\label{sec:sourcextraction}

We use the PROFOUND task of the software \texttt{ProFound} \citep{ref:A.Robotham2018} for source detection and extraction in the survey tiers. As opposed to traditional radio source extraction tools like \texttt{PyBDSF} \citep{ref:N.Mohan2015} or \texttt{AEGEAN} \citep{ref:P.Hancock2012, ref:P.Hancock2018}, \texttt{ProFound} does not assume an underlying Gaussian surface brightness distribution, but instead takes a segmentation-like approach to identifying pixels belonging to a galaxy. In this regard, \texttt{ProFound} is conceptually similar to \texttt{SExtractor}, with the difference being that \texttt{ProFound} is relatively `free-form' in the distribution of pixels that it identifies as belonging to a galaxy, rather than the more aperture-based approach of \texttt{Sextractor}. \citet{ref:C.Hale2019} investigated the performance of \texttt{ProFound} on radio images, which is of particular interest when considering galaxies that may not be well-described by smooth Gaussian light distributions. They find that \texttt{ProFound} is able to accurately recover flux densities of both simulated and observed radio sources, and performs better on extended/irregular sources than the Gaussian-based extraction tools. We therefore also adopt \texttt{ProFound} to identify sources in our simulated survey maps, as our study focuses on the observation of complex, resolved radio flux distributions. After using \texttt{ProFound} to identify source segmentation maps (and therefore galaxy positions and integrated flux densities), we will then extract morphological parameters using an aperture-based approach (see Section \ref{sec:morphologicalparams}).

Following \citet{ref:C.Hale2019} we adopt the default \texttt{ProFound} parameters, with the exception of \texttt{skycut}=3.5, \texttt{groupstats}=TRUE and \texttt{groupby}=`segim'. \texttt{skycut} is the threshold (in multiples of $\sigma$) used to identify sources in the image, and \citet{ref:C.Hale2019} found that a threshold of \texttt{skycut}=3.5 gives the best trade-off between the number of real and false detections, with a false detection rate of 2\% in the VIDEO field. We tested values of \texttt{skycut} between 3$\sigma$ and 4$\sigma$ for our data set, finding that the overall performance of \texttt{ProFound} (detection rate vs. output flux, for example), does not strongly depend on the chosen threshold for our mock images.
The other two parameters, \texttt{groupstats} and \texttt{groupby}, give rise to final segmentation maps in which individual segments that are touching one another will be considered as one segment. As segments originate from bright pixels, before `flooding' into the surrounding regions, it is possible that multiple segments could form within one galaxy, which then need to be `grouped'. The trade-off however is that separate galaxies that are close to each other in the sky may be grouped together, if their segmentation maps are adjacent. The use of these parameters therefore depends on the expected source density and sky confusion limit. With the high angular resolution of our SKA-MID simulations and finite number of input galaxies, we are in the regime where source-splitting due to multiple bright region is more likely to occur than source blending. Not grouping neighbouring segments results in over-fragmented galaxies for our sample.

\subsubsection{Cross-matching with the input catalogue}
    
For the quantitative discussion of parameter recovery in this study we require an `intrinsic' or `input' catalogue of galaxy properties, for comparison with the outputs after simulation/applying the telescope response. For galaxies selected as outlined in Section~\ref{sec:sample_SKA_sample}, the input catalogue (i.e. source positions, flux densities, ellipticities, position angles, sizes, segmentations maps) was created from the 3DHST catalogue of the GOODS-N field. The derivation of input SFRs and corresponding radio flux densities is described in Sections~\ref{sec:sample_SKA_SFRs} and ~\ref{sec:sample_rad_SKA}.

The catalogue of extracted sources from \texttt{ProFound} was cross-matched with the input galaxy catalogue, in order to assess completeness and compare input vs. output morphological parameters (see Section~\ref{sec:morphologicalparams}). Matching was performed by searching for galaxies in the output catalogue whose \texttt{ProFound} centroid lay within the input segmentation map. If more than one output galaxy centroid lay within the segmentation map of a given input galaxy, the output galaxy with the larger total (\texttt{ProFound}) flux density was taken as the match. This situation may arise when an input galaxy is fragmented into several pieces, due to fainter parts of the galaxy being buried under the noise, and individual bright peaks being classified as separate objects. This only occurs for a small fraction of our input segmentation maps, namely for $\sim$3-9\% from the UD to the wide tier respectively.
    
The resulting completeness for each survey tier is shown in Fig.~\ref{fig:completeness}. Completeness is defined as the fraction of detected sources with input flux (we do not have output flux densities for undetected sources) in a given x-axis bin. We note that we do not consider reliability within the scope of this study, as we are primarily motivated by detection and recoverability of parameters on our known sample, rather than a full optimisation of blind source detection. Depending on the science case, reliability would also be of less interest when working in a regime where galaxies are targeted at previously known positions, based on data at other wavelengths. The 5$\sigma$ noise levels of each tier are highlighted with dashed vertical lines in Fig.~\ref{fig:completeness}. A small number of sources with intrinsic flux densities below these noise levels are detected in the wide and deep tiers, as the effect of flux boosting increases these fainter galaxies' output flux densities. We note that in practice, for a blind survey, output flux densities rather than the input flux densities used Fig.~\ref{fig:completeness} would be available. If Fig.~\ref{fig:completeness} were instead plotted as a function of output flux density, the completeness curves would generally be shifted to lower values on the x-axis, as the output galaxy flux densities tend to be slightly underestimated (up to $\sim$10\%, depending on the tier, see Fig.~\ref{fig:fluxflux}). We remind the reader that this analysis is based on a sample which is not radio-selected (Section~\ref{sec:sample_SKA_sample}). Hence the completeness trends for radio-selected samples in general may not follow directly from this figure (even though one can assume radio-selected is equivalent to SFR-selected). Detailed completeness characteristics, for example, will depend on the specific extraction approach, and the normalisation of completeness curves may vary between different techniques.

\subsection{Galaxy Parameters} 

A key goal of this study is to quantify how well SKA-MID will be able to accurately recover galaxy properties such as flux and radio (SFR) morphology. As discussed in Section~\ref{sec:tiered_desc}, the band 2 SKA-MID tiered surveys will give us high-resolution observations at 1.4~GHz, for unprecedented numbers of high-redshift galaxies. In terms of our input-output comparisons, it's important to state that the segmentation maps used to quantify intrinsic galaxy properties (before simulation) do not include the emission from neighbouring galaxies. This means that the intrinsic segmentation map used for galaxies with close neighbours (see e.g. third column in upper right of Fig.~\ref{fig:ska_radio_cut}) will include only those pixels that were classified as belonging to the specified galaxy in the original resolved SFR images \citep{ref:A.Cibinel2019}. On the other hand, when we analyse the simulated SKA images in the following sections, we no longer use these intrinsic segmentation maps. The extraction therefore emulates the realistic data analysis of an SKA observation.

For our input-output recoverability tests, we consider several key galaxy parameters: the flux density (Section~\ref{sec:fluxes}), as well as morphological parameters (Section~\ref{sec:morphologicalparams}).

\subsubsection{Total Flux Densities}
\label{sec:fluxes}

The measured radio flux density from the SKA images is of particular interest, as arguably the most important physical property that will be measured using SKA continuum observations is the star-formation rates of large samples of galaxies, via their radio flux densities. In order to verify that galaxy flux density is conserved through the SKA simulation pipeline described in Section~\ref{sec:SKA_sims_method}, we measure total flux densities for the galaxies, after the SKA response has been applied but before the addition of noise. We do this by isolating each galaxy (meaning it has no neighbouring galaxies, in addition to the noise-free background), and measure the total flux contained within large apertures placed at the galaxy position. We find that the median recovered flux (in apertures up to 4 $\times$ $R_{\rm Kron}$) is 0.988$\times$F$_{\rm input, tot}$. We therefore take this 1.2\% intrinsic correction into account when comparing input-output flux densities in Fig.~\ref{fig:fluxflux}, somewhat analogous to flux calibration for a real observation. We find no strong dependence on flux density, size or redshift for these small corrections.
\citet{ref:C.Hale2019} also discuss the possibility that the flux densities of some galaxies may be underestimated using \texttt{ProFound}, if the output segmentation map covers an area smaller than the image PSF. This could arise in the case of sources with particularly faint extended wings (below the \texttt{skycut level}), for example. As demonstrated in \citet{ref:C.Hale2019}, a beam correction can be made to account for this possible effect. For each matched output segmentation map, we therefore calculate the area of the PSF covered by the segmentation map, and correct for any fractional flux losses arising from not fully sampling the PSF. We find that in our simulations, the high-resolution regime of SKA-MID, the flux correction terms are negligible ($\leq$1\%), although we take them into account nonetheless.

Output flux densities (and associated errors) for each detected galaxy are given by \texttt{ProFound}'s PROFOUND task, to which the above small correction is also applied. We show in Fig.~\ref{fig:fluxflux} the fractional error between the input and output flux densities for each tier, where a value above zero indicates that the intrinsic flux is greater than the output flux. We show a running median of the fractional errors, as well as the 86th and 16th percentiles of the distributions (solid and dashed black lines respectively). The number of recovered galaxies in each tier is shown in the upper right corner of each panel. There is a relatively flat trend of fractional error with flux density, except for the lowest flux galaxies in each tier. The output flux densities tend to be systematically underestimated, with the median fractional error rising from 1.3\% in the UD tier, 2.7\% in the deep tier, to 4.7\% in the wide tier. The 16th-84th percentile scatter on the fractional error tends to increase with decreasing survey depth. For each tier, we measure the 16th-84th percentile scatter in the flux density regime of high completeness. In the regimes where completeness $>$90\% (see Fig.~\ref{fig:completeness}), the 16th-84th\% spreads are 25\%, 21\% and 8\% for the wide, deep and UD tiers respectively, with the scatter increasing towards lower flux densities (and therefore lower completeness). The underestimation of these flux densities is predominantly due to the nature of segmentation-based source extraction based on threshold flux densities, below which the extremities of galaxies are not included in the segmentation maps.

There are a number of outlying points in these figures, for example as a result of galaxies close on the projected sky being blended, or individual galaxies with significant substructure being over-segmented. We define a catastrophic outlier as having a fractional error >50\%, as this represents approximately 1/3 of a 0.5~dex flux density bin, in which luminosity functions are often reported. In terms of absolute number, there are 69, 58 and 78 catastrophic outliers in the wide, deep and UD tiers respectively. This corresponds to 11\%, 4\% and 5\% of the detections in each tier. It can be argued that the high absolute number of catastrophic outliers in the UD tier is due to segmentation maps extending further out than in the shallower tiers, making blending of individual galaxies more likely. Over-segmentation of galaxies is also less likely in the shallower tiers if not all of the sub-peaks within a galaxy are well above the noise. As discussed in Section \ref{sec:sourcextraction}, we find that when asked not to `group', or blend, neigbouring segments, \texttt{ProFound} instead over-segments some of our galaxies in the deeper tiers, which thus doesn't present a more favourable solution.

\begin{figure*}
\centering
\begin{minipage}{\textwidth}
\includegraphics[clip, trim=0.7cm 0.7cm 0.7cm 0.7cm,width=0.33\textwidth]{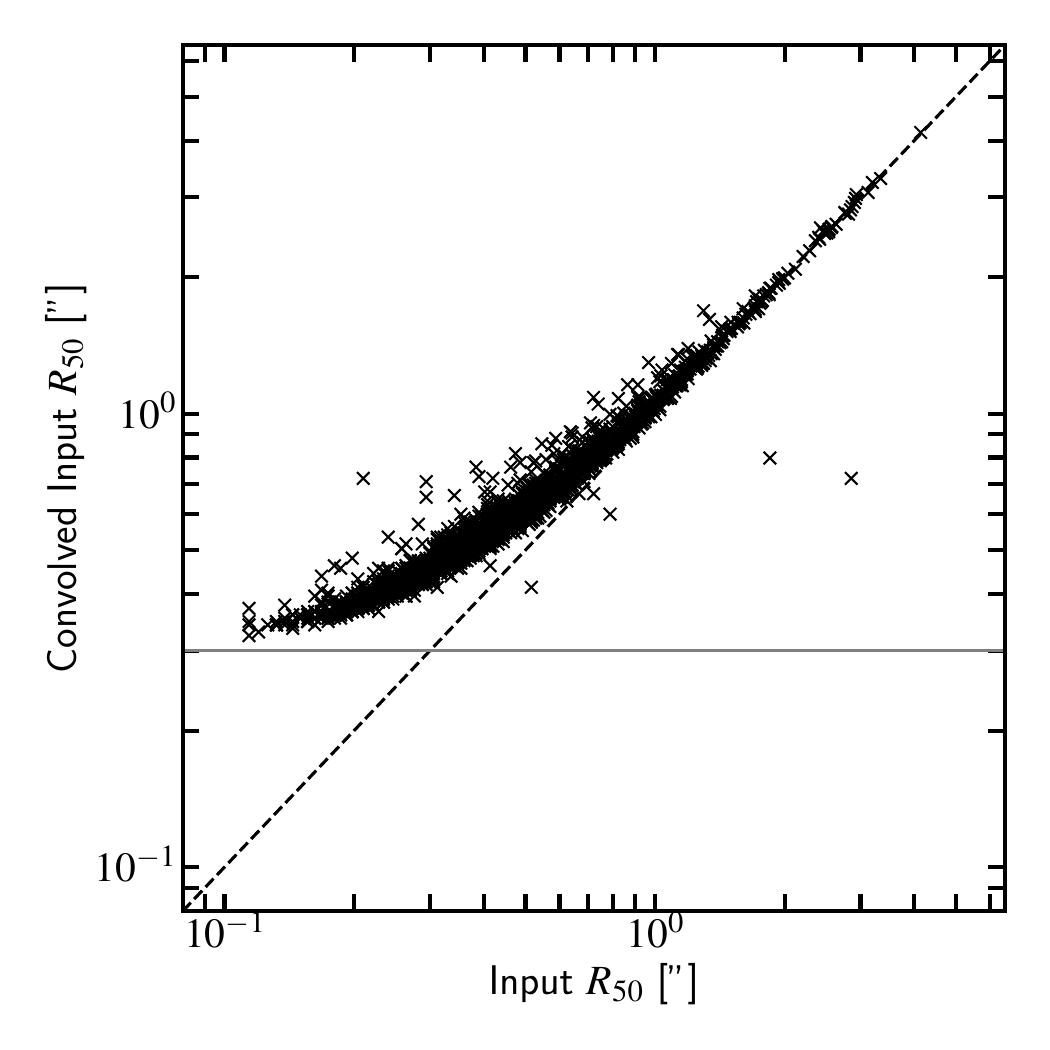}\hfill
\includegraphics[clip, trim=0.7cm 0.7cm 0.7cm 0.7cm,width=0.33\textwidth]{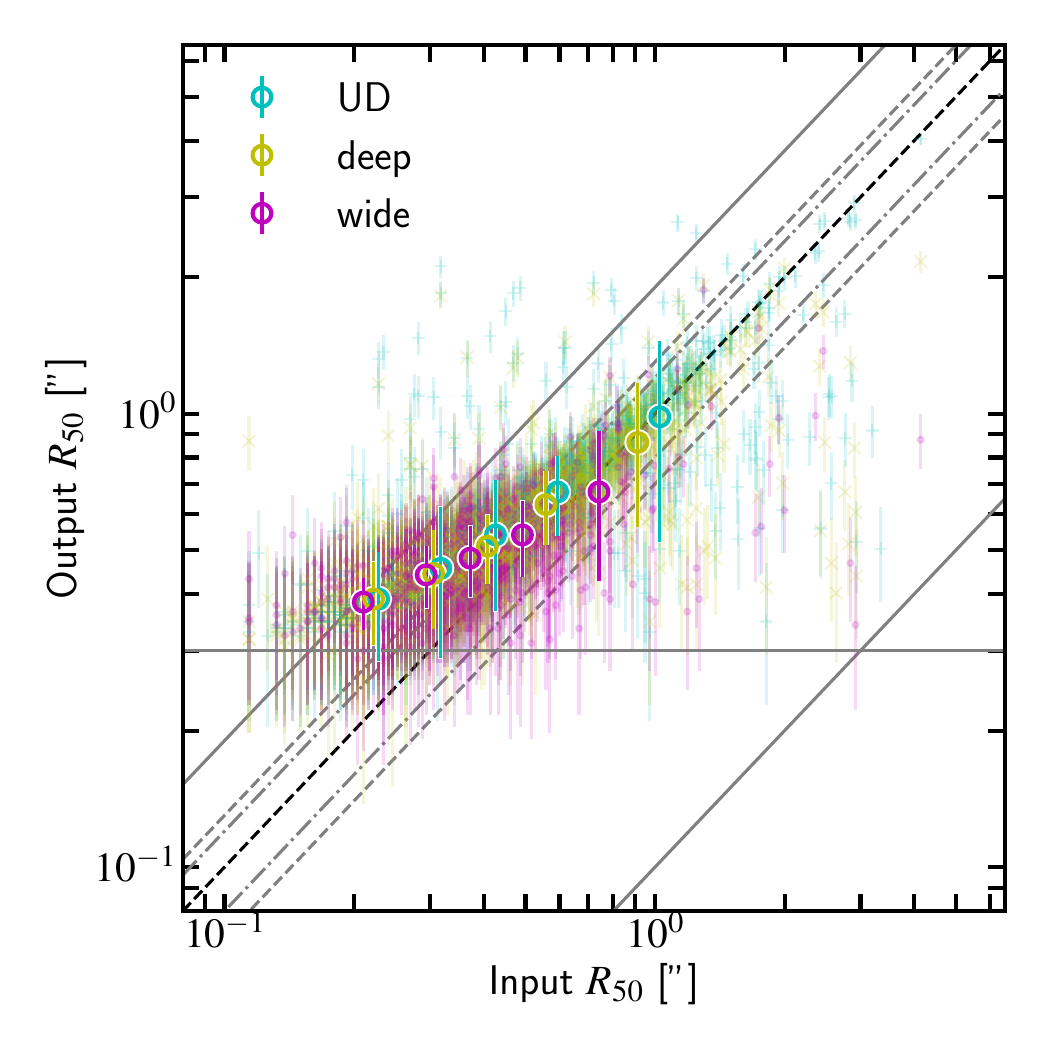}\hfill
\includegraphics[clip, trim=0.7cm 0.7cm 0.7cm 0.7cm,width=0.33\textwidth]{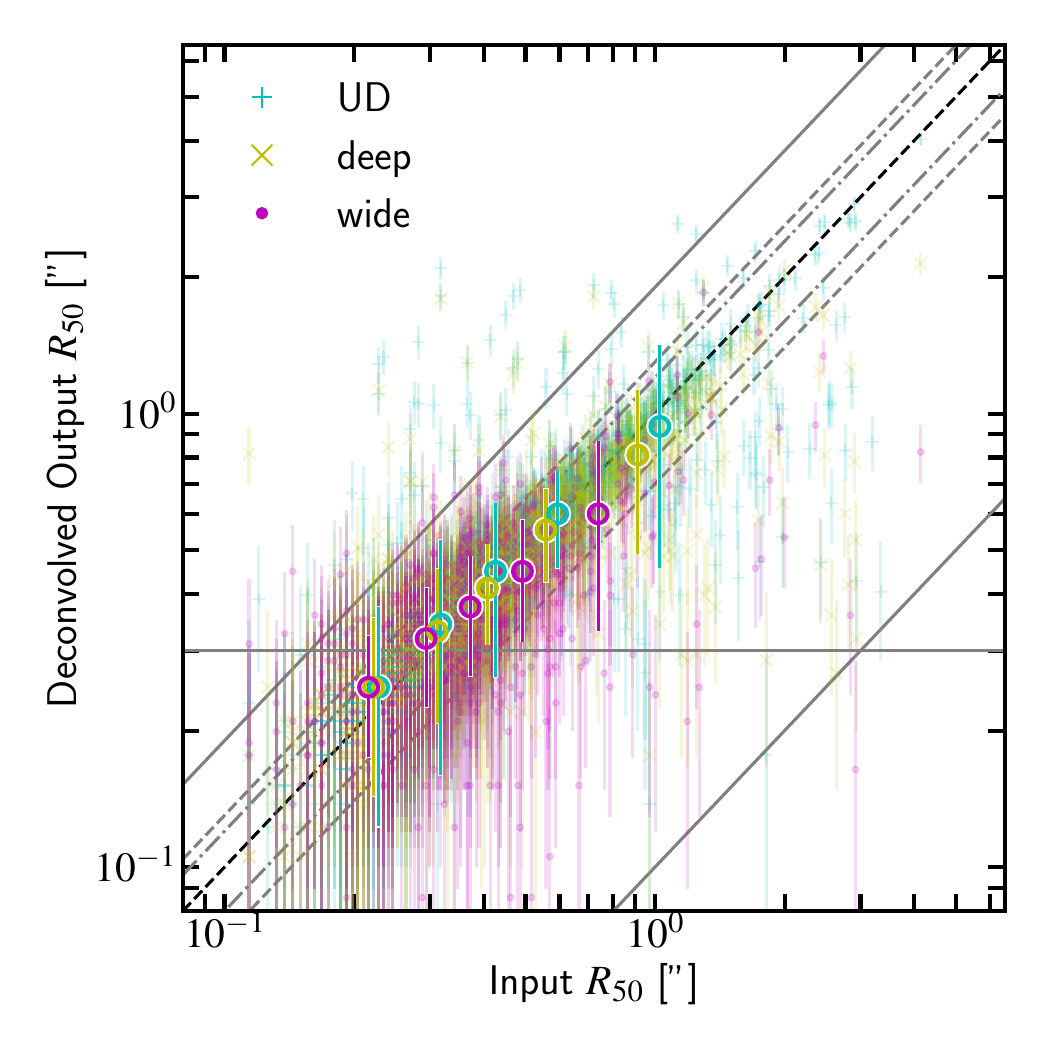}\hfill
\end{minipage}
\caption{{\it Left}: convolved vs. unconvolved input $R_{50}$ (major axis) sizes. {\it Center}: output $R_{50}$ vs. input $R_{50}$. {\it Right}: Deconvolved output $R_{50}$ vs. input $R_{50}$. Cyan: UD tier, yellow: deep tier, magenta: wide tier. One pixel ($\pm$0.12") error are shown on individual output data points. The data are also binned into equal numbers bins (large circles), where the errorbars show the 1$\sigma$ scatter in the y-direction, and median bin values are used. The input/output major axis sizes are calculated using the input/output ellipticities, respectively. The black dashed line is the 1:1 relation. The grey dash-dot, dashed and solid lines show $\pm$20\%, $\pm$30\% and $\pm$90\% offsets respectively. The horizontal solid line shows $R_{50}$ for the Gaussian PSF.}
\label{fig:R50R50}
\end{figure*}

\begin{figure*}
\centering
\begin{minipage}{\textwidth}
\includegraphics[clip, trim=0.5cm 0.5cm 0.5cm 0.5cm, width=0.31\textwidth]{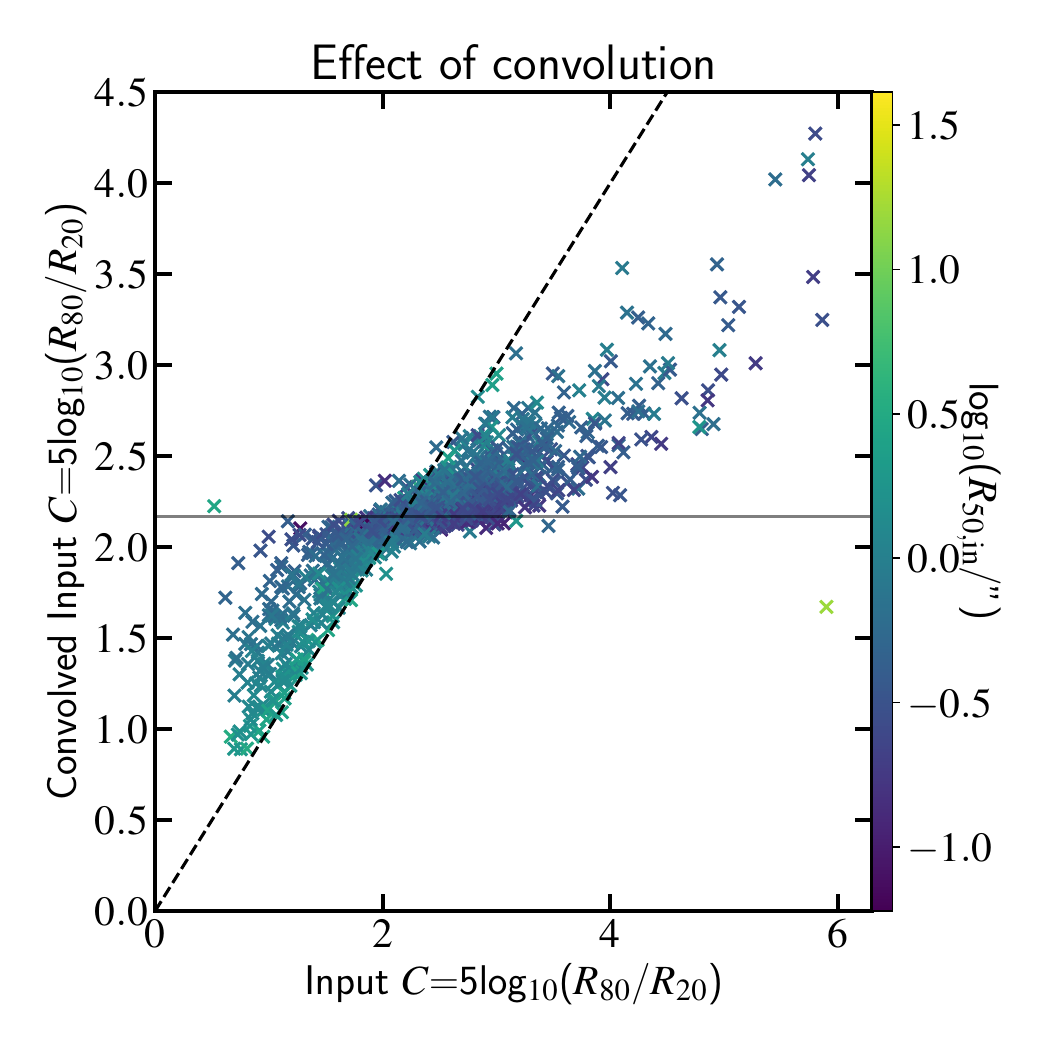}\hfill
\includegraphics[clip, trim=0.5cm 0.5cm 0.5cm 0.5cm,width=0.31\textwidth]{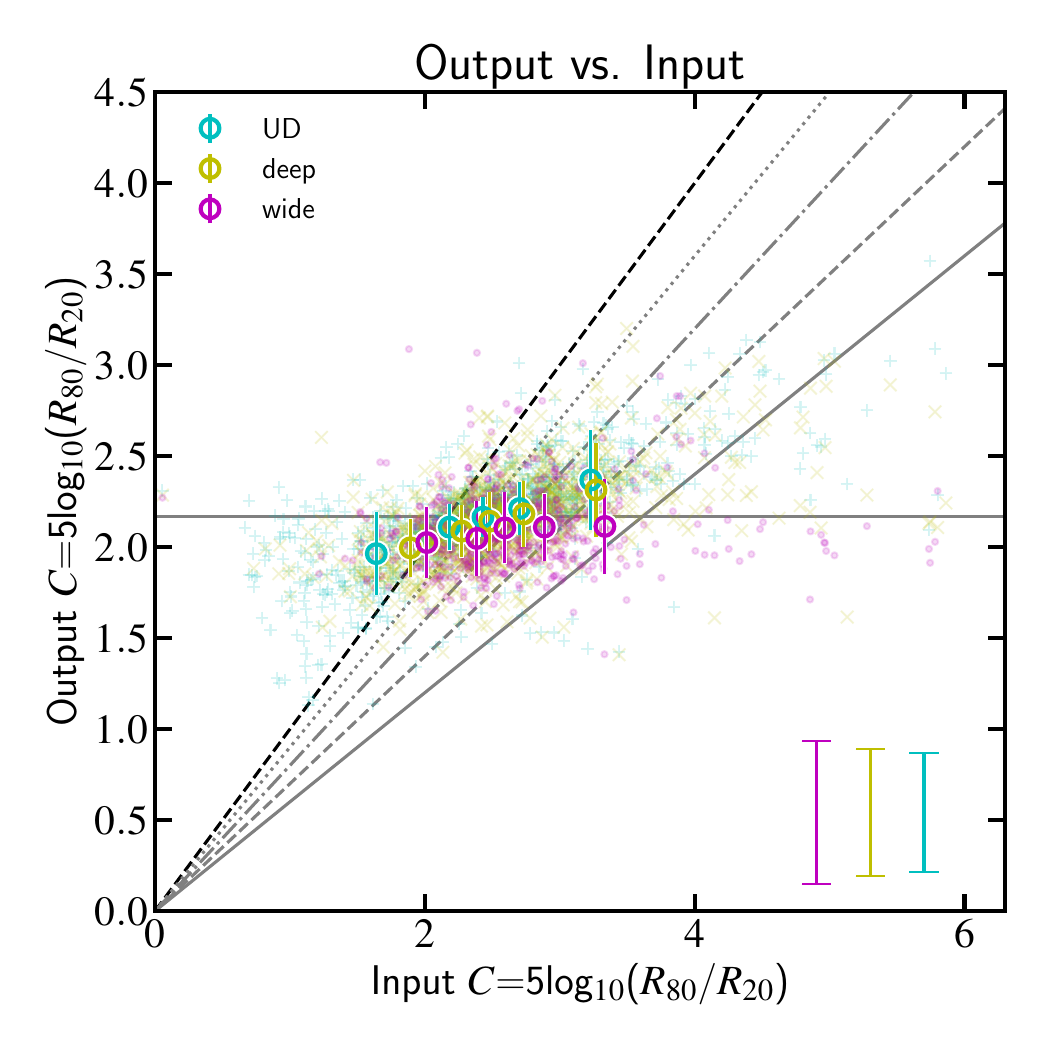}\hfill
\includegraphics[clip, trim=0.5cm 0.5cm 0.5cm 0.5cm,width=0.31\textwidth]{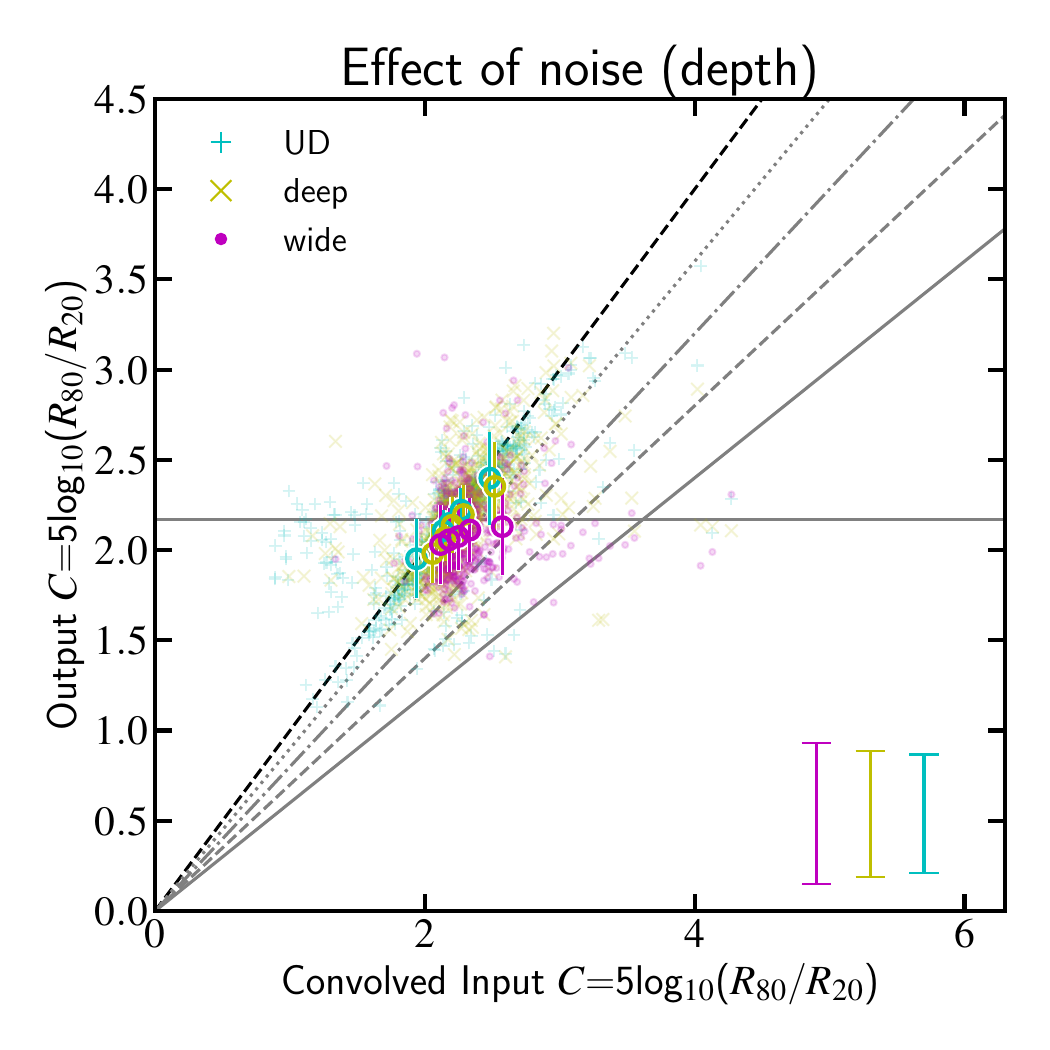}\hfill 
\end{minipage}

\centering 
\begin{minipage}{\textwidth}
\hspace{-1mm}\includegraphics[clip, trim=0.5cm 0.5cm 0.7cm 0.5cm,width=0.31\textwidth]{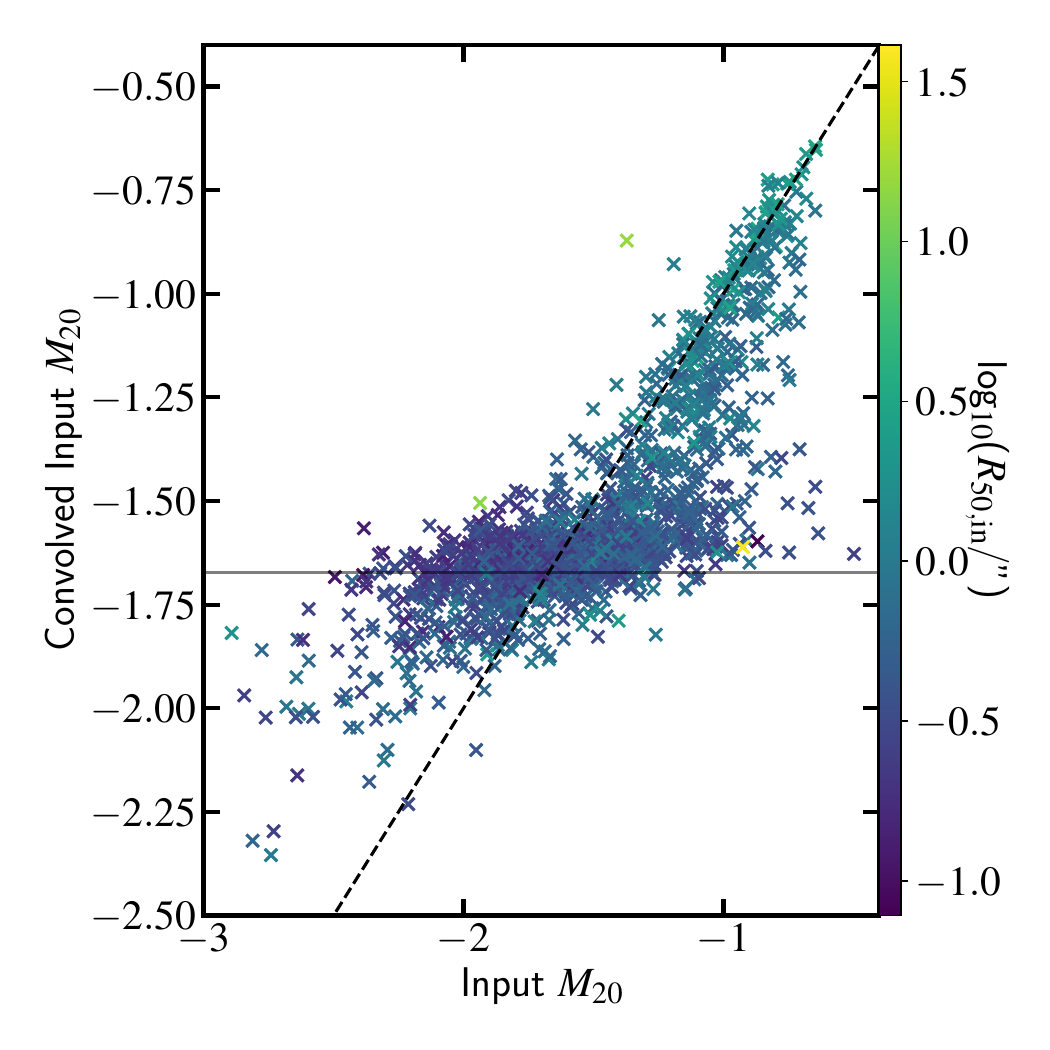}\hfill
\hspace{-8mm}\includegraphics[clip, trim=0.7cm 0.5cm 0.5cm 0.5cm,width=0.31\textwidth]{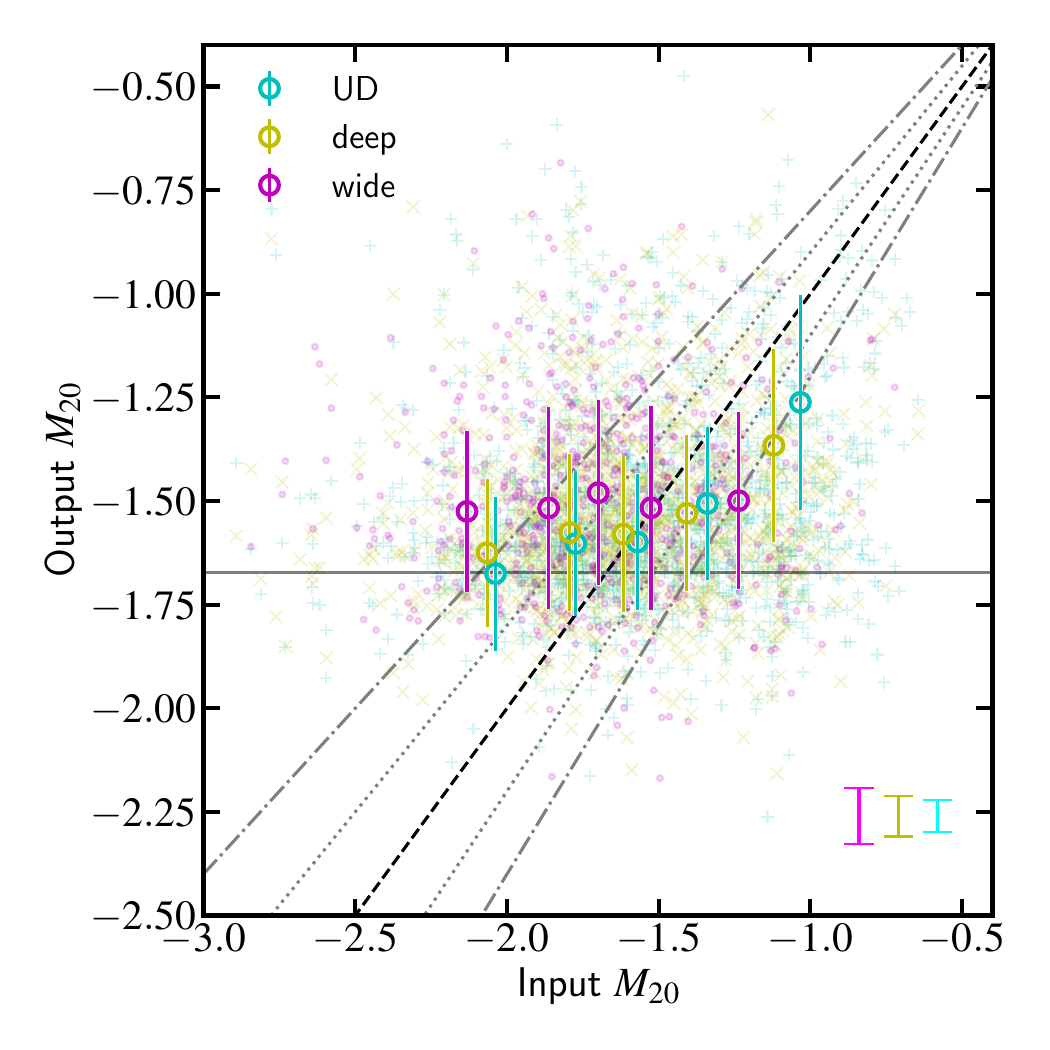}\hfill
\hspace{-9mm}\includegraphics[clip, trim=0.7cm 0.5cm 0.5cm 0.5cm,width=0.31\textwidth]{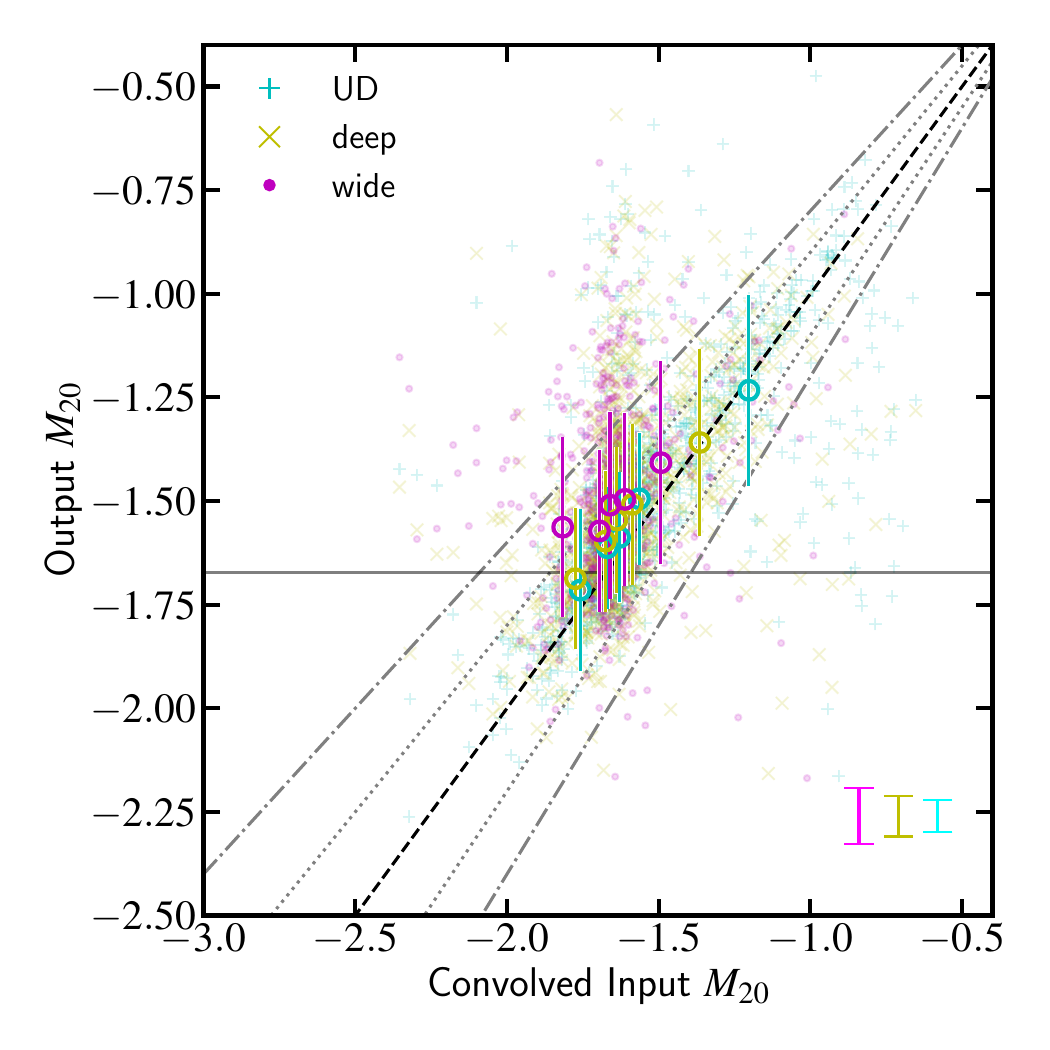}\hfill
\end{minipage}
\centering
\begin{minipage}{\textwidth}
\includegraphics[clip, trim=0.7cm 0.5cm 0.8cm 0.55cm,width=0.31\textwidth]{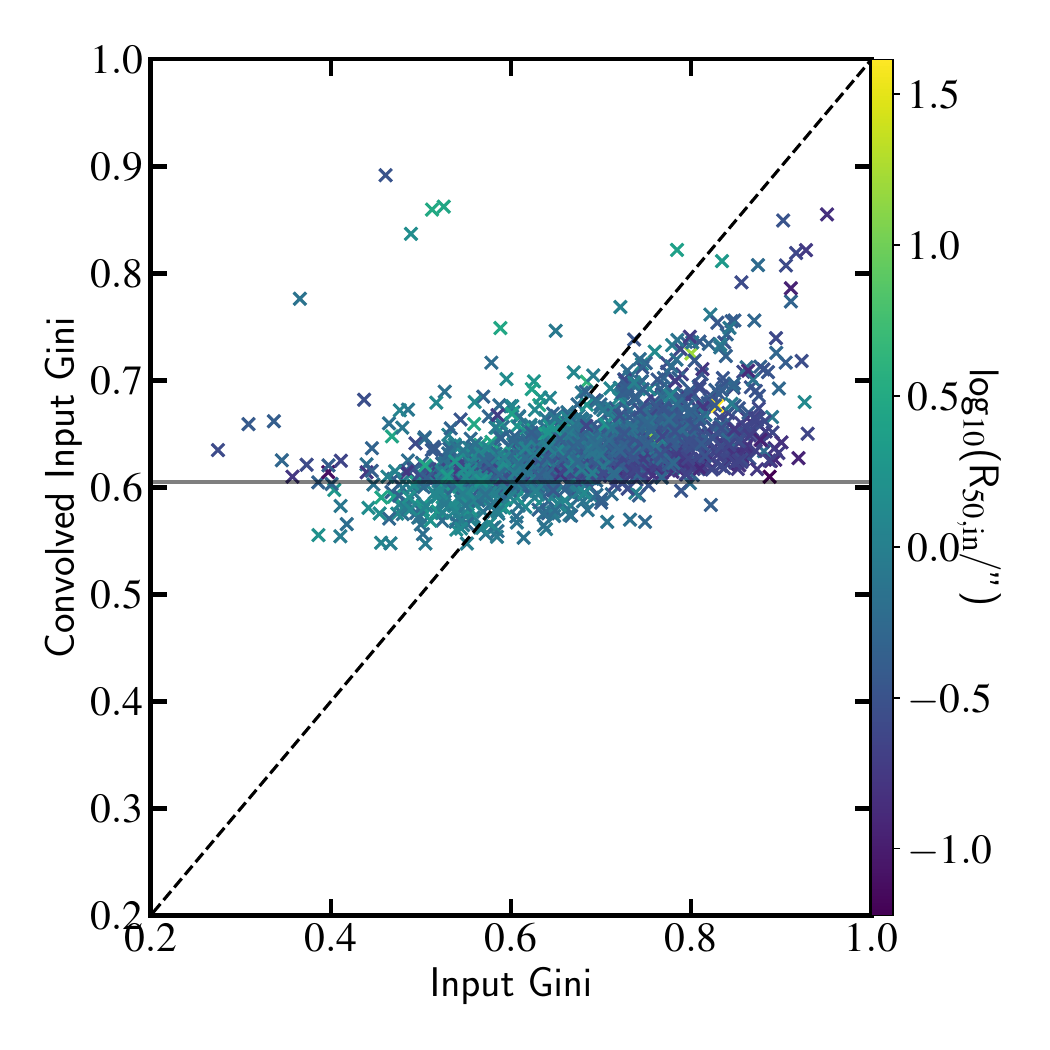}\hfill
\hspace{3mm}\includegraphics[clip, trim=0.7cm 0.5cm 0.8cm 0.42cm,width=0.31\textwidth]{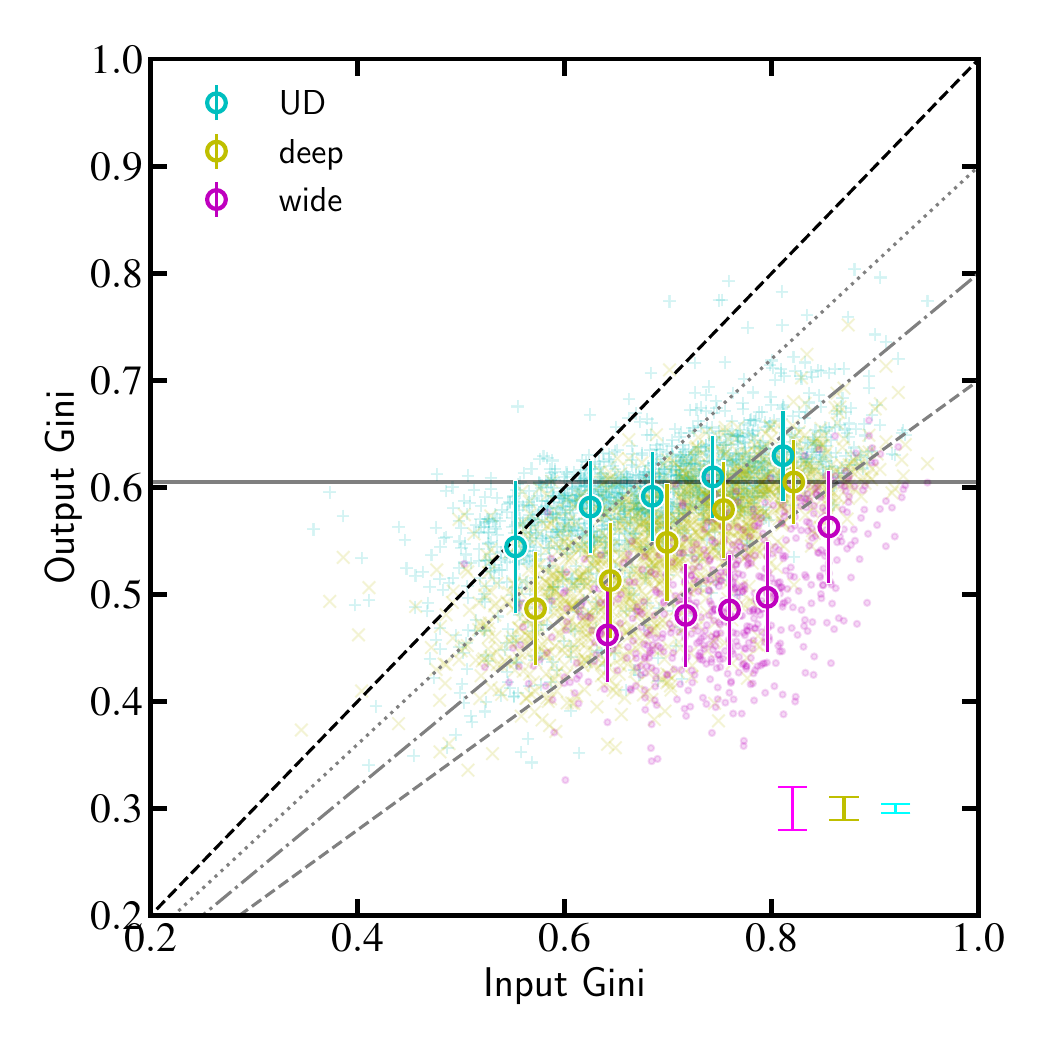}\hfill
\hspace{1mm}\includegraphics[clip, trim=0.7cm 0.5cm 0.8cm 0.42cm,width=0.31\textwidth]{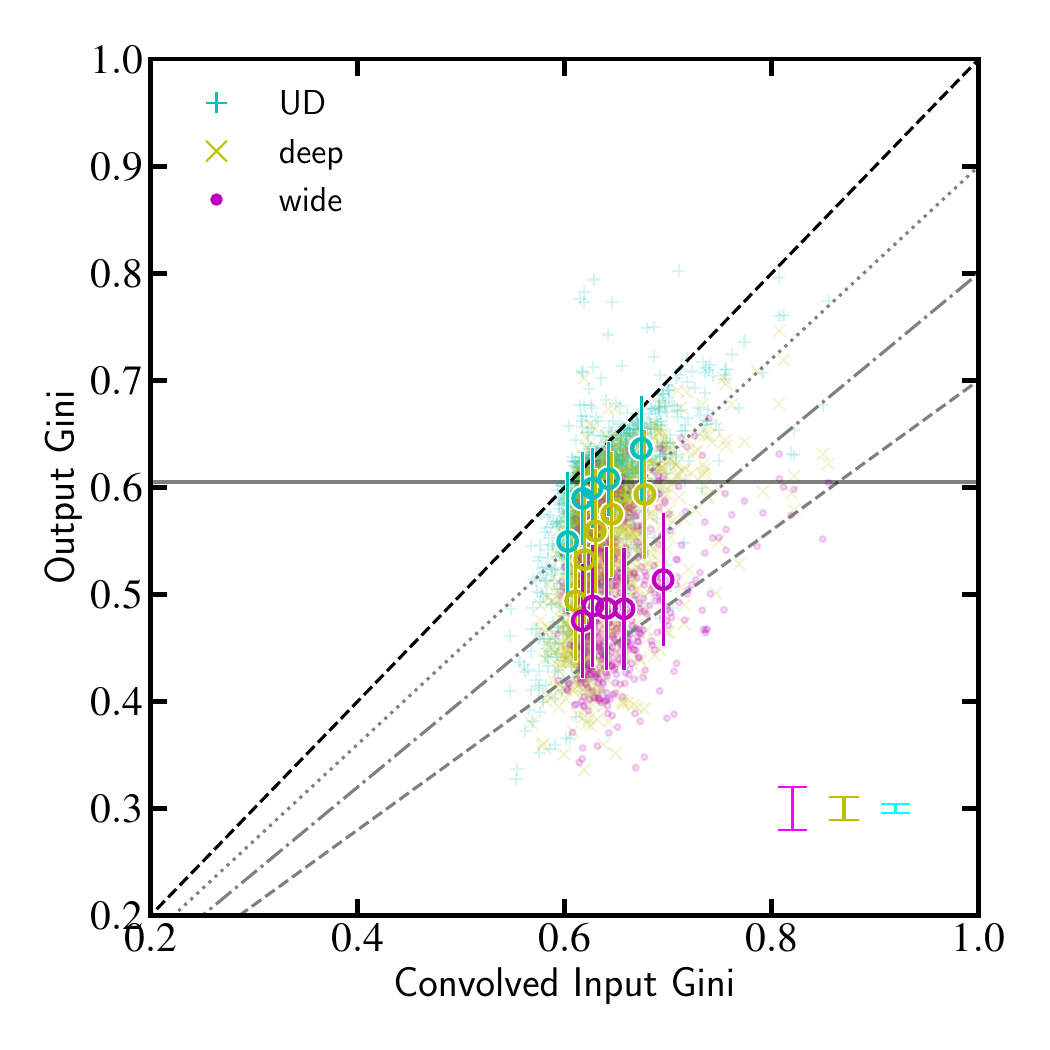}\hfill
\end{minipage}

\centering
\begin{minipage}{\textwidth}
\includegraphics[clip, trim=0.75cm 0.5cm 0.5cm 0.75cm,width=0.31\textwidth]{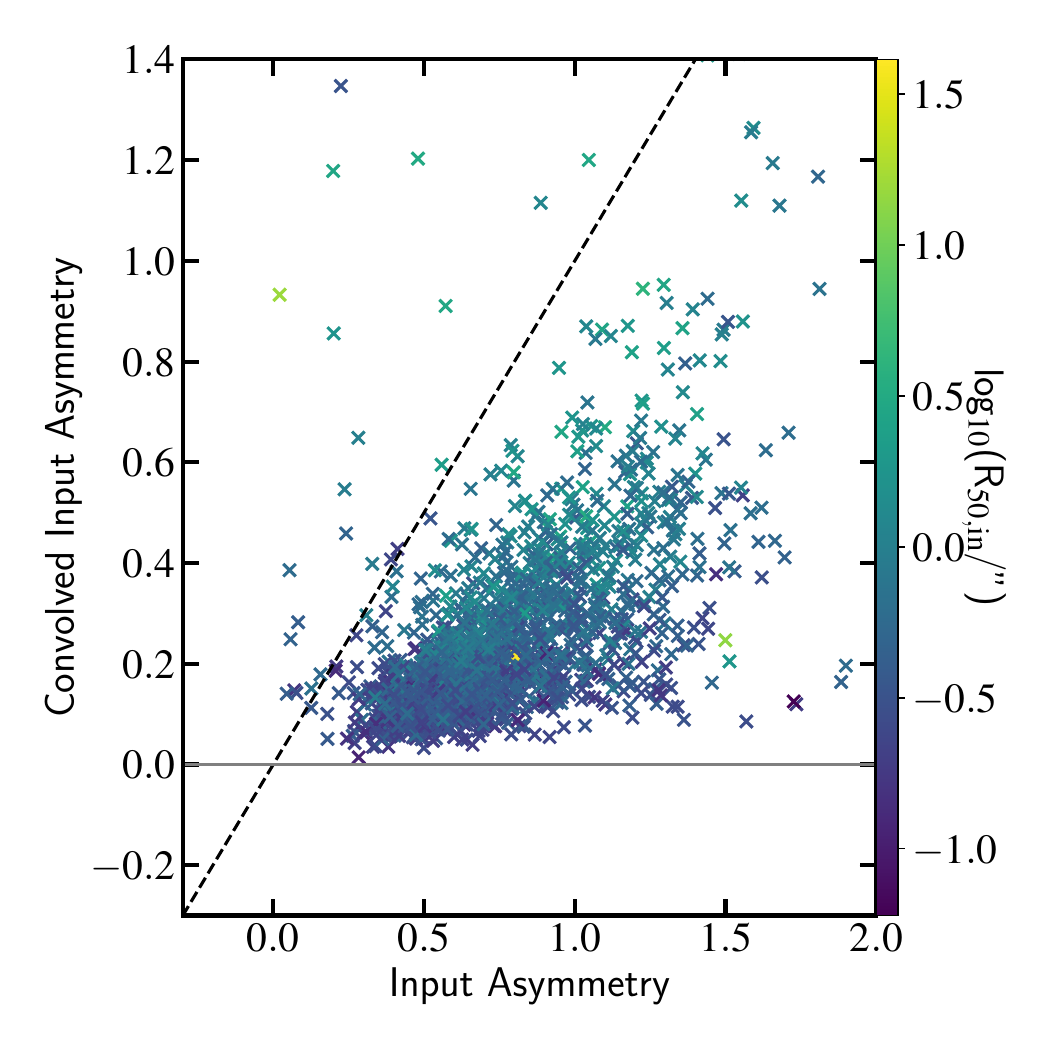}\hfill
\hspace{4mm}\includegraphics[clip, trim=0.75cm 0.5cm 0.6cm 0.75cm,width=0.31\textwidth]{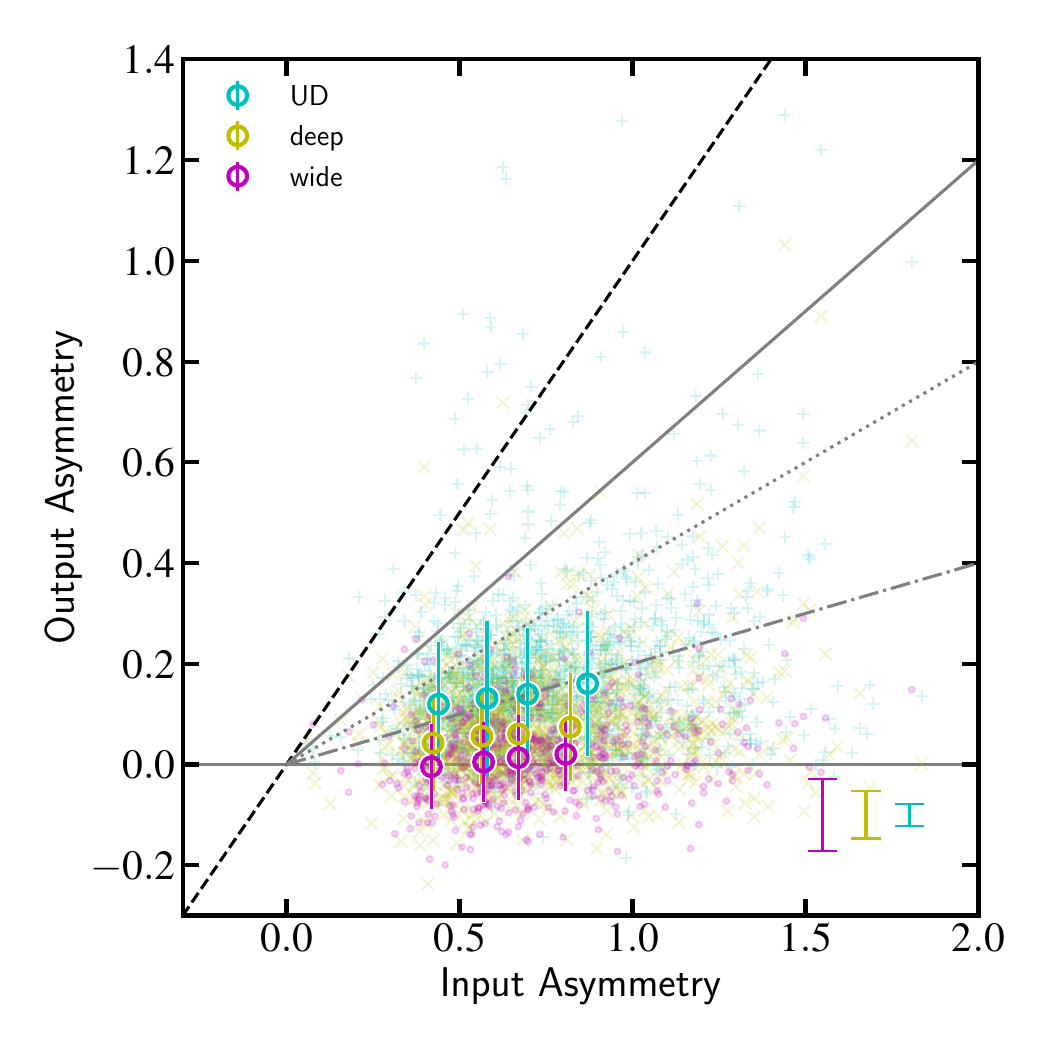}\hfill
\hspace{0mm}\includegraphics[clip, trim=0.75cm 0.5cm 0.5cm 0.75cm,width=0.31\textwidth]{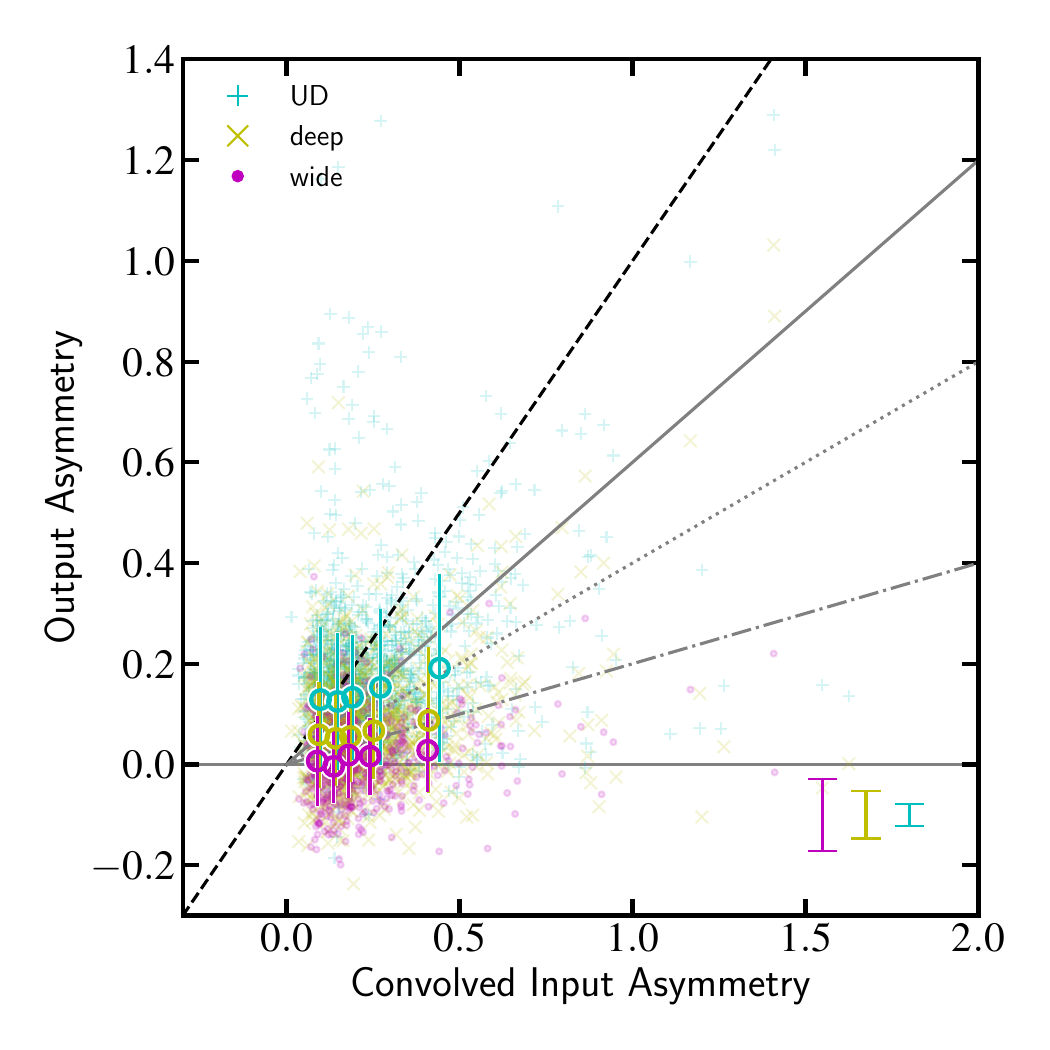}\hfill

\vspace{-3.8mm}
\caption{All panels: convolved vs. non-convolved inputs ({\it left}, colour-coded by size), outputs vs. inputs ({\it center}), outputs vs. convolved inputs ({\it right}). The black dashed line is the 1:1 relation. Top: Concentration, upper center: $M_{\rm 20}$,  lower center: Gini, bottom: Asymmetry. Individual data points are shown (dots, crosses and plus signs), along with binned data containing equal numbers of data points (empty circles), with their associated 1$\sigma$ errorbars. In the top three panels, dotted, dot-dashed, dashed, and solid grey lines indicate $\pm$10, $\pm$20 $\pm$30 and $\pm$40\% offsets around the 1:1 line respectively, where present. For asymmetry, solid, dotted and dot-dashed grey lines show 40, 60 and 80\% offsets from 1:1, respectively. The horizontal solid line in all 3 panels gives the parameter value for the Gaussian PSF. Representative (median) errorbars for each survey are shown in the bottom right corner of the center and right panels.}
\label{fig:morph}
\end{minipage}
\end{figure*}

\subsubsection{Morphological parameters}
\label{sec:morphologicalparams}

In addition to the flux density itself, an important outcome of SKA-MID will be the morphological study of galaxies, through the distribution of their star-formation. In order to quantify this, we identify several key parameters: the sizes ($R_{50}$, defined as the half-light radii of the galaxies), the second-order moment of the brightest 20\% of pixels ($M_{\rm 20}$, indicating how far the brightest flux is located from the center of the galaxy), the Gini parameter ($G$, a measure of whether the total flux of a galaxy is evenly distributed over all of its pixels), the concentration ($C$, to what degree a galaxy's flux is centrally concentrated) and the asymmetry ($A$, a measure of how symmetrically the flux of a galaxy is distributed around the centre point). Quantitative definitions of these parameters are given in the following subsections. We first measure the above parameters for our galaxies using their intrinsic segmentation maps (at 0.06"/pixel, before resampling). Having been measured for the intrinsic maps, within the input segmentation maps corresponding to a radius $R_{\rm Kron}$, these morphological parameters are then compared to those measured in the output SKA-simulated images.

\subsubsection*{Sizes}

Sizes of the galaxies were measured using aperture photometry for both input and output, as segmentation map-based tools for measuring size cannot be run on noise-free images such as the radio input image (see small cutouts in upper right of Fig.~\ref{fig:ska_radio_cut}). Small apertures were placed at the flux density centroid positions of the galaxies in the simulated survey maps, where the flux centroid is calculated according to the segmentation maps (input or \texttt{ProFound}). The apertures were increased in size in increments of 0.006" until the aperture contained more than or equal to a given fraction of the flux density, e.g. 50\%, 90\%. A sub-pixel size increment was chosen, as we consider fractional pixel contributions to the total flux density when only part of a pixel is contained within the aperture. The total input flux density is defined as the pixel-by-pixel sum of the input segmentation map, and the input ellipticity was taken from the SExtractor 3DHST catalogues. For the output measurements, the output ellipticity and the total flux density are given by \texttt{ProFound}.

In order to distinguish between the effect of survey depth and the resolution of the observations, i.e. the convolution of the image by the synthesised beam, we additionally measure morphological parameters in input maps that have been convolved with the PSF, a Gaussian beam of FWHM=0.6". Each input galaxy is isolated from its neighbours (as with the other input parameter measurements), before being convolved with the pure Gaussian. We then measure new `convolved input' sizes for the galaxies. We follow these steps to define apertures on the convolved input galaxies:

(i) Measure approximate 1$\sigma$ Gaussian sizes of the non-convolved input galaxies, by measuring the sizes containing 68\% of the total input flux. The flux is measured in apertures with ellipticities and position angles defined by the input SExtractor catalogue.

(ii) Analytically estimate a new 1$\sigma$ Gaussian size for the convolved input galaxies, by convolving the old Gaussian $\sigma$ with the 1$\sigma$ Gaussian size of the convolving PSF. Do this for both the major and minor axes.

(iii) Measure the new ellipticity for the convolved input galaxies using the ratio of the convolved major and minor 1$\sigma$ Gaussian sizes.

(iv) Define the aperture for the convolved inputs using the original (non-convolved) position angle (as the convolving Gaussian beam was symmetric with position angle = 0, it leaves the position angles of our galaxies unchanged), and new (convolved) ellipticity.

(v) Calculate the sizes containing certain flux levels (e.g. $R_{50}$, $R_{90}$), for the convolved input galaxies, using these apertures.

(vi) Define $R_{\rm Kron}$ apertures for each convolved input galaxy, using the sizes found in step (v, also see below). These will be used to measure morphological parameters on the convolved input maps.

We also calculate sizes for the new convolved input maps using the original, non-convolved ellipticities, and verify that the effect of using the old vs. new ellipticities is negligible, in terms of the morphological parameters calculated within apertures of $R_{\rm Kron}$. We therefore continue using the sizes derived using the new ellipticities, which are consistently smaller than those calculated using the old ellipticities, with a median difference of 0.3~pixels (0.02"). We show the results for non-convolved inputs, convolved inputs, and outputs, for all morphological parameters. Fig.~\ref{fig:R50R50} shows a comparison of input and output $R_{50}$ sizes between the survey tiers. A comparison of $R_{50}$ for the convolved and non-convolved inputs can be seen in the left panel. At large $R_{50}$, the convolved and non-convolved inputs agree well, with the non-convolved $R_{50}$s becoming increasingly smaller than the convolved inputs towards smaller sizes. The plateau at a minimum of $\sim$0.3-0.4" for the convolved input sizes is caused by flux smearing with the PSF, where no such lower limit exists in the input maps (down to 1 pixel).

In the center panel, we show the input vs. output sizes. The output sizes tend to lie above the 1:1 line, and there is a clear lower limit on the output galaxy sizes, at around $\sim$0.33", just above the $R_{50}$ of the PSF (grey horizontal line). For the smallest input galaxies, the output sizes are 1.8-1.9 times larger than the input galaxies, with this offset from the 1:1 line decreasing steadily with size. Finally, in the right panel, we show a comparison of input sizes with deconvolved output sizes (deconvolved from the FWHM=0.6" of the Gaussian beam).} The curvature seen in the centre panel is no longer present, reiterating that this effect was caused by the PSF. The galaxies with the most overestimated sizes are found in the deep and ultradeep tiers, which is likely an effect of the galaxy blending in these deeper tiers, discussed in the previous section. This panel indicates that the UD tier is deep enough to provide a reliable characterisation of the sizes of galaxies used in our sample - keeping in mind the caveat that this result applies to galaxies satisfying the selection criteria given in Section~\ref{sec:sample_SKA_sample}. The UD tier gives sizes closest to the 1:1 relation, with the wide tier tending to give the smallest output sizes, as the fainter outskirts of galaxies may become buried under noise. The binned output size measurements in all tiers are less than 20\% offset from the 1:1 relation.

In the following, all input parameters are measured within $R_{\rm Kron}$, as defined from the input segmentation map sizes. We find from size analysis of the input segmentation maps that $R_{\rm Kron}$ = 1.46$R_{90}$. All output morphological parameters (and input convolved parameters) are therefore also measured within 1.46$R_{90}$, where $R_{90}$ is found from the size derivations described above. Additionally, for the concentration, Gini and $M_{\rm 20}$ parameters, we also calculate the value for the FWHM=0.6" PSF beam (also sampled at 0.12"/pixel), measured within an aperture of 1.46$R_{90}$.

\subsubsection*{Concentration}

Concentration is defined as in \citet{ref:J.Lotz2004}:
    
    \begin{equation} 
    C=5 \log \left(\frac{R_{80}}{R_{20}}\right)
    \label{eqn:conc}
     \end{equation} 

\noindent where we take $R_{80}$ and $R_{20}$ (major axes sizes for the apertures containing 80\% and 20\% of the flux density, respectively) as calculated by the output size measurements. The results are shown in Fig.~\ref{fig:morph} (first row), where the panel layout is as for Fig.~\ref{fig:R50R50}: convolved input vs. non-convolved input (left), non-convolved input vs. output (center), convolved input vs. output (right). We have assumed one pixel errors on both $R_{20}$ and $R_{80}$. In the left panel, colour-coded by galaxy size, the result of the convolution is to reduce the concentration for galaxies with input concentrations $C$$\gtrsim$2, as might be expected if a concentrated central flux distribution is becoming spread over a larger portion of the galaxy. Interestingly the concentration is increased for galaxies that have intrinsic $C$ below this value, which may arise from low surface brightness galaxies having their flux distributed more smoothly/Gaussian-like as a result of the convolution. Unlike $M_{20}$ and Gini (see below), there does not appear to be a minimum concentration for the convolved galaxies, but rather a turning point at $C\sim$2 below which the concentration is increased by convolution, with the turning point centered around the concentration value of the PSF. The effect of convolution is dependent on the intrinsic flux distribution, with the concentration often increasing for those galaxies with a lower concentration than the PSF (and vice-versa). Another factor may be the size of a galaxy in relation to the PSF, with galaxies much larger than the PSF having their overall concentration modified less, and therefore lying closer to the 1:1 line, than much smaller galaxies. In the right panel, the output concentrations increase towards the deeper tiers, moving closer to the 1:1 line, with a flattening of the slope towards the shallower tiers. A possible explanation for this flattening is that the higher the image RMS, the more likely the noise is to overshadow the flux distribution of a galaxy, decreasing the contrast between central and outskirt regions and giving rise to more uniform (noise-dominated) flux distributions. Whilst the binned concentrations in the wide tier reach 10-20\% offset from the 1:1 relation, those in the deep and UD tiers are $<$10\% offset. When we examine the total combined effect of survey depth and telescope response in the center panel, there is little difference between the three tiers, with the binned data reaching 30-40\% offset from 1:1 at large input concentrations. We note a small shift closer to the 1:1 line at high input concentrations for deeper tiers. The output concentrations are generally lower than the intrinsic values for inputs above $C$=2, with the output concentrations scattering around $C_{\rm output}$=2, as in the left panel.

\subsubsection*{M20}

The second-order moment of the brightest 20\% of pixels, $M_{\rm 20}$, is calculated according to \citet{ref:J.Lotz2004}. This is done by first calculating the total second order moment:

\begin{equation}
M_{\mathrm{tot}}=\sum_{i}^{n} M_{i}=\sum_{i}^{n} f_{i}\left[\left(x_{i}-x_{c}\right)^{2}+\left(y_{i}-y_{c}\right)^{2}\right].
\label{eqn:M20_1}
\end{equation}

This total moment is the sum of the flux density in each pixel ($f_{i}$) multiplied by the squared distance to the center of the galaxy. The center of the galaxy is given by coordinates ($x_{c}, y_{c}$), and the positions of each pixel in the galaxy segmentation map are the ($x_{i}, y_{i}$). We take the centre coordinate of the galaxy to be the radio flux density-weighted centroid position, both before and after simulation. $M_{\rm 20}$ is then calculated by rank-ordering the galaxy pixels by flux density, and taking the sum of the \textit{i}th brightest pixels until the sum reaches 20\% of the total galaxy flux density. $M_{\rm 20}$ is the sum of the moments of these brightest pixels, normalised by the total moment:

\begin{equation}
M_{20} \equiv \log_{10}\left(\frac{\sum_{i} M_{i}}{M_{\mathrm{tot}}}\right), \text { while } \sum_{i} f_{i}<0.2 f_{\mathrm{tot}}.
\label{eqn:M20_2}
\end{equation}

The normalisation by $M_{\rm tot}$ removes the dependence on total galaxy flux or size. In equation~\ref{eqn:M20_2}, $f_{\mathrm{tot}}$ is the total flux density contained within the segmentation map, and the $f_{i}$ are the individual pixel flux densities, sorted from brightest to faintest such that $i$=1 is the brightest pixel. Combining equations~\ref{eqn:M20_1} and \ref{eqn:M20_2}, we see that $M_{\rm 20}$ returns negative values, where a less negative $M_{\rm 20}$ indicates a larger second-order moment, and therefore a distribution of the brightest flux further from the centre of the galaxy (and vice-versa).

The results for input-output $M_{20}$ are shown in Fig.~\ref{fig:morph} (second row), where errors are calculated assuming the image RMS value as the flux density error on one pixel, and one pixel as the error on the distance from the center of the galaxy\textcolor{black}{, where we consider the center position to be well-defined}. In the left panel, we see once again on the y-axis that the parameter values cover a considerably narrower range after convolution with the PSF, and \textcolor{black}{there appears to be a dense locus of $M_{20}$ convolved input values at $\sim$-1.7, scattering around the $M_{20}$ of the PSF}. Interestingly, several galaxies with high $M_{20}$ values ($\gtrsim$-1.25) branch upwards towards the 1:1 line, away from the rest of the distribution, indicating that the convolution does not degrade $M_{20}$ for some galaxies with the largest $M_{20}$ values. Inspection of our data suggests that those galaxies which do not retain their high $M_{20}$ values here are on average smaller than the PSF, in the regime where the size-independence of $M_{20}$ may break down. More generally, whether or not $M_{20}$ increases or decreases after convolution will again depend on the intrinsic flux distribution. For example, galaxies with a bright ring-like structure may often have their $M_{20}$s decrease, whereas galaxies whose flux declines with radius may have bright flux moved further from the center during convolution, thus increasing $M_{20}$.

Interestingly, when we look at the right panel of Fig.~\ref{fig:morph} (second row) to investigate the effect of noise on the measurements, we find that adding the noise distribution corresponding to the SKA-MID telescope response increases the output values of $M_{20}$, bringing them above the 1:1 line. We also see in this panel that the slope of the binned data is dependent on the depth of the survey tier, with steeper gradients (closer to 1) with increasing observation depth. We observe somewhat vertical structures in the data, particularly prominent in the wide tier (shown also by the narrower x-range of the wide tier binned data), highlighting that the output values of $M_{20}$ span a large range for a smaller input range. The binned data in all tiers lie less than 10\% offset from the 1:1 line, apart from the wide tier with the smallest input $M_{\rm 20}$ values. In the center panel, we find that for the final output measurements, the relationship between input and output $M_{20}$ is essentially flat and indistinguishable between tiers, spanning at least $\pm$20\% either side of the 1:1 line. This makes the recovery of intrinsic $M_{20}$ values challenging.

\subsubsection*{Gini parameter}

The Gini parameter is defined as \citep{ref:G.Glasser1962}:

\begin{equation}
G=\frac{1}{|\overline{X}| n(n-1)} \sum_{i}^{n}(2 i-n-1)|X_{i}|
\label{eqn:Gini}
\end{equation}

\noindent where n is the total number of pixels, and $X_{i}$ is the flux density value in the \textit{i}th pixel, having first sorted $X_{i}$ into increasing order. This definition returns Gini values between 0 and 1, where a value of 0 would indicate that the flux is distributed equally over all pixels in the galaxy, and conversely $G$=1 implies that all of the galaxy flux is located in one pixel. Although the previously-discussed concentration parameter $C$ (Eqn.~\ref{eqn:conc}) is often used in morphological studies, $C$ is measured at the core of a galaxy, and is therefore insensitive to potential off-center concentrations of flux. $C$ has been shown to correlate with both $M_{\rm 20}$ and the Gini parameter (Eqn.~\ref{eqn:Gini}), in opposite senses (see \citet{ref:C.Scarlata2007} for morphological studies in the COSMOS field, for example). Concentration is positively correlated with Gini, as $G$ gives a measure of how equally distributed (i.e. concentrated vs. diffuse) the total flux of a galaxy is, taking into account all of the galaxy pixels. On the other hand, concentration is negatively correlated with $M_{\rm 20}$, as the second-order moment of the brightest 20\% of the galaxy increases with distance from the core. When we combine $M_{\rm 20}$ and $G$, we therefore gain an indication of whether the flux is relatively concentrated or extended throughout the galaxy, in addition to the spatial distribution of the most star-forming regions. Where both AGN activity and nuclear starbursts would have high $C$ and $G$ parameters (with low $M_{\rm 20}$), bright star-forming regions can also exist in the outskirts of a galaxy, for example in tidal tails of galaxies undergoing an interaction with their large-scale environment (low $C$, high $G$ and $M_{\rm 20}$).

We show the input-output results for the Gini parameter in Fig.~\ref{fig:morph} (third row). Looking at the left panel, we can see that there is a lower boundary on the minimum Gini parameter measured in the input convolved maps, at $G\sim$0.6. The input convolved Gini values scatter both sides of the 1:1 line, with the majority of galaxies having lower Gini parameters in the convolved inputs maps than in the non-convolved inputs. This can be understood by the convolution with the beam distributing the flux density of each pixel over multiple pixels. In the case of galaxies that have high Gini concentrations in the intrinsic maps, the beam will more `fairly' distribute the flux densities over a larger number of pixels, lowering the Gini parameter. In the case of galaxies with lower intrinsic Gini parameters (i.e. $<$0.6 in this case, corresponding approximately to the Gini of the PSF), the smearing of small regions of similarly-bright pixels by the PSF may lead to regions of flux with more-Gaussian like flux distributions, thus increasing the Gini parameter to that of the PSF - as we also saw for galaxy concentrations.

In the center and right panels of Fig.~\ref{fig:morph} (third row), output Gini parameters tend to be systematically underestimated in the survey tiers. By looking at the right panel, we see that the effect of adding noise to our galaxies, the underlying distribution of which is dictated by the uv-coverage of our observations and therefore the dirty beam, is to systematically reduce Gini. Both the normalisation and slope of the binned data are dependent on the survey tiers. The data move further from the 1:1 line with decreasing survey depth, and the slopes also become increasingly shallow due to the higher noise. We mark grey lines at 90\%, 80\%, 70\% and 60\% of input values in the centre and right tiers, as indicated in the caption. The effect of the noise (seen in the right panel by comparing quantities on the x- and y-axis that are both convolved with a FWHM=0.6" Gaussian beam) is to reduce output Gini values to $\sim$95\% of their input values in the UD tier, $\sim$88\% in the deep tier, falling to $\sim$78\% by the wide tier. When both convolution and noise are taken into account, i.e. the SKA response and survey depth (center panel), the output Gini values fall further, by another $\sim$10\% on average. The relatively flat slope and narrow output (y-axis) range seen in the center panel is the effect of the convolution shown in the left panel.

\subsubsection*{Asymmetry}

Asymmetry is defined as in \citet{ref:C.Conselice2000}:

\begin{equation}
A=\min \left(\frac{\Sigma\left|I_{o}-I_{\phi}\right|}{\Sigma\left|I_{o}\right|}\right)-\min \left(\frac{\Sigma\left|B_{o}-B_{\phi}\right|}{\Sigma\left|I_{o}\right|}\right)
\end{equation}

\noindent where $I_{o}$ is the intensity distribution in the original (non-rotated image), and $I_{\phi}$ is the intensity distribution in the rotated image. We measure asymmetry after a rotation of $\phi$=180~deg around the flux centroid, including an average background correction (which is non-zero for the output maps, as they contain noise), where $B_{o}$ and $B_{\phi}$ are the original (rotated) background intensity distributions. The background correction is derived from the average of 100 background asymmetry measurements. In order to minimise the effect of correlated noise in these images, apertures used to measure background have the same ellipticity and position angle as those used to measure the galaxy asymmetry, but with an area of 25 times greater. The sum of pixel intensities was then scaled down correspondingly to calculate 100 background asymmetries, before they were averaged. The background aperture positions were checked against the input segmentation map, to ensure that there were no galaxies at that position. The possible asymmetry values can be between 0 and 2. In the input and the output, the center position is defined as the flux-weighted centroid position, based on the corresponding segmentation map of the galaxy. The pixels in the background apertures were also scaled to have the same primary-beam correction as the galaxy (although this effect is usually minor, given the large SKA-MID FoV). 

The asymmetry results are shown in Fig.~\ref{fig:morph} (lower). \textcolor{black}{The uncertainty comes from the error on flux differences, and assumes that the center position of the galaxy is well defined}. We find that asymmetries are systematically underestimated, and span a narrower range of values than the intrinsic values, most likely once again due to the convolving effect of the synthesised beam, which smooths out features occurring on scales smaller than the size of the beam. In all three tiers, several galaxies have negative (and therefore unphysical) asymmetry values. This arises when the background asymmetry value is larger than the galaxy asymmetry value. In an attempt to understand this further, we measure the Spearman's correlation coefficient between output flux density and output asymmetry, for the wide tier. Indeed we find a clear correlation between these quantities, giving correlation coefficient R = 0.5 and p-value (probability of the null hypothesis that flux density and asymmetry are uncorrelated) of p = 2$\times$10$^{-41}$. The lowest (total) flux density sources may contain pixels at the level of the image RMS noise itself or lower, thus giving rise to galaxy asymmetries that are comparable to background asymmetries and thus negative overall asymmetries. The non-zero background asymmetries may also be partially an effect of the correlated noise found in interferometric images, which not usually relevant in optical images. To quantify this further, we calculate the number of galaxies in which negative asymmetries occur in each tier. We find 74 (5\% of detected sources in this tier) occurrences in the UD tier, 321 (24\%) in the deep tier, and 294 (48\%) in the wide tier, indicating a clear trend with the depth of the observations. Taking this altogether, we find that signal-to-noise is an important factor when measuring galaxy asymmetries in the SKA images. Measurements approach the 1:1 line for low input asymmetry values in the deep and UD tiers, with the locus of output points moving closer to 1:1 with increasing depth. It's interesting that the best parameter recovery doesn't happen for the most asymmetric galaxies, demonstrating either that low-moderate asymmetry is more robust to the effects of telescope convolution and noise, and/or implying that these low-moderate output asymmetry values should not be trusted, and may simply arise due to the noise. Fig.~\ref{fig:morph} clearly demonstrates that high noise is very effective at lowering asymmetry values. However, even in the UD tier, the output asymmetry values form a `cloud' well below the 1:1 line, making accurate recovery of intrinsic asymmetry difficult.

\begin{figure*}
\centering
\begin{minipage}{\textwidth}
\includegraphics[clip, trim=0.6cm 0.7cm 0.6cm 0.5cm, width=0.5\textwidth]{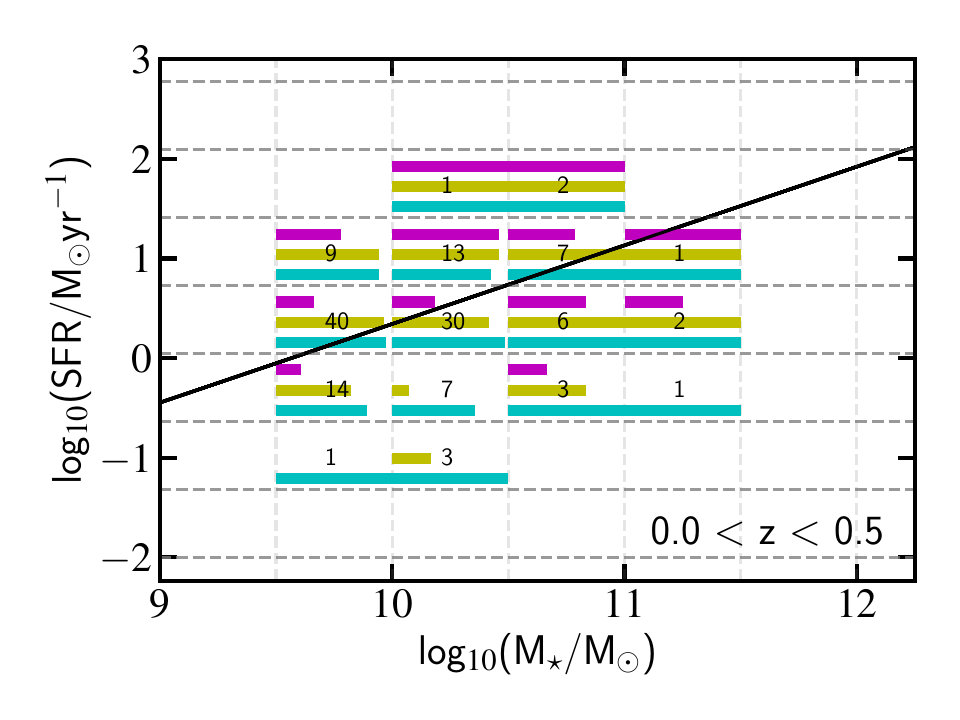}\hfill
\includegraphics[clip, trim=0.6cm 0.7cm 0.6cm 0.5cm, width=0.5\textwidth]{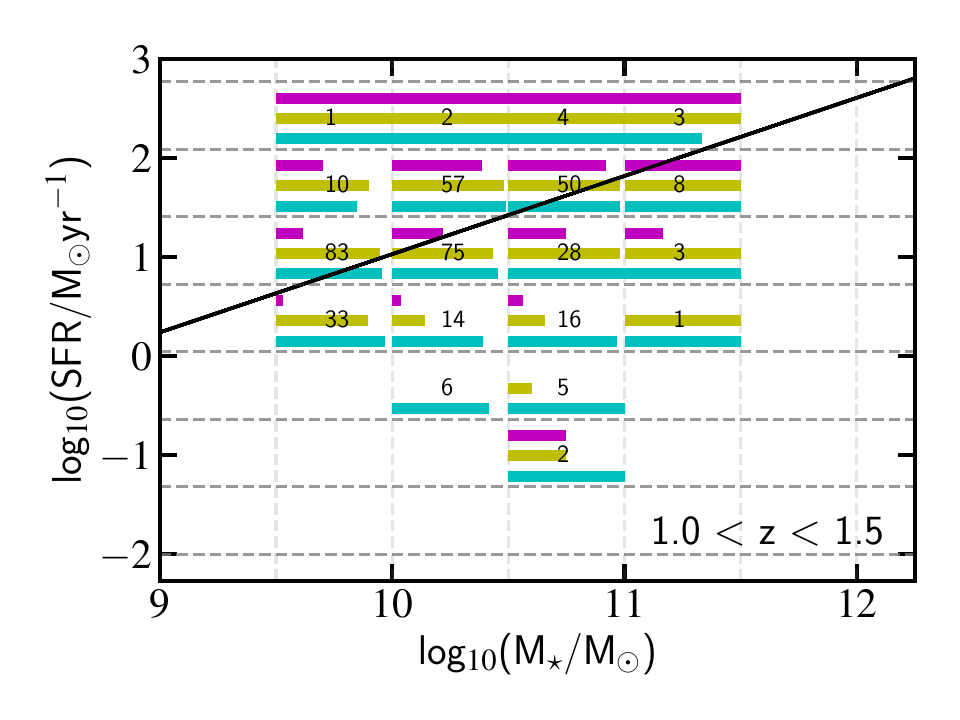}\hfill
\end{minipage}
\begin{minipage}{\textwidth}
\includegraphics[clip, trim=0.6cm 0.7cm 0.6cm 0.5cm, width=0.5\textwidth]{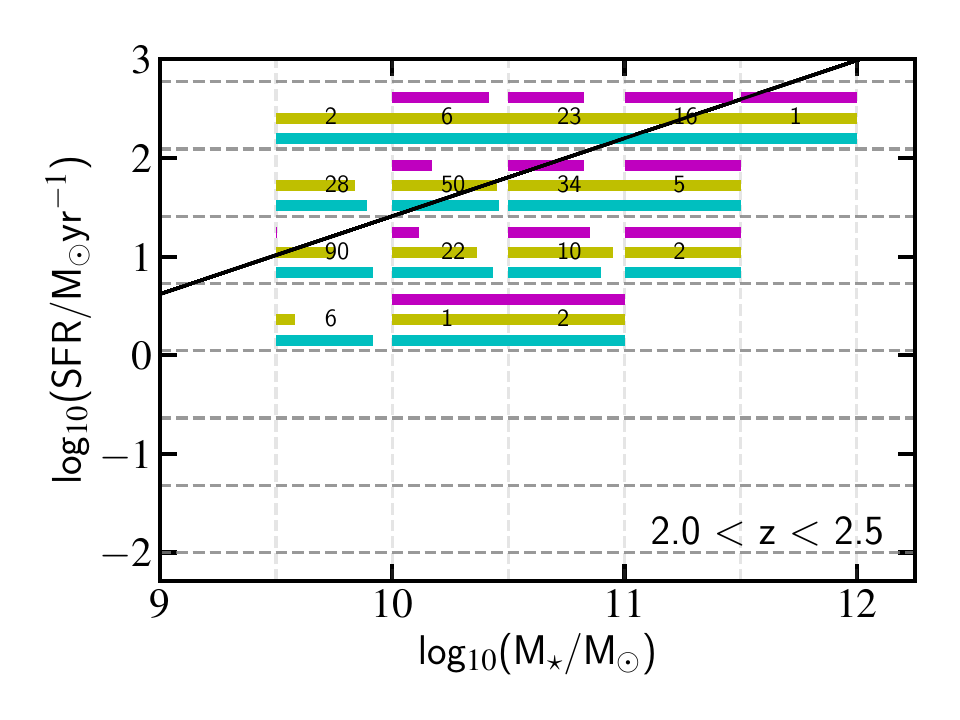}\hfill
\includegraphics[clip, trim=0.6cm 0.75cm 0.8cm 2cm, width=0.5\textwidth]{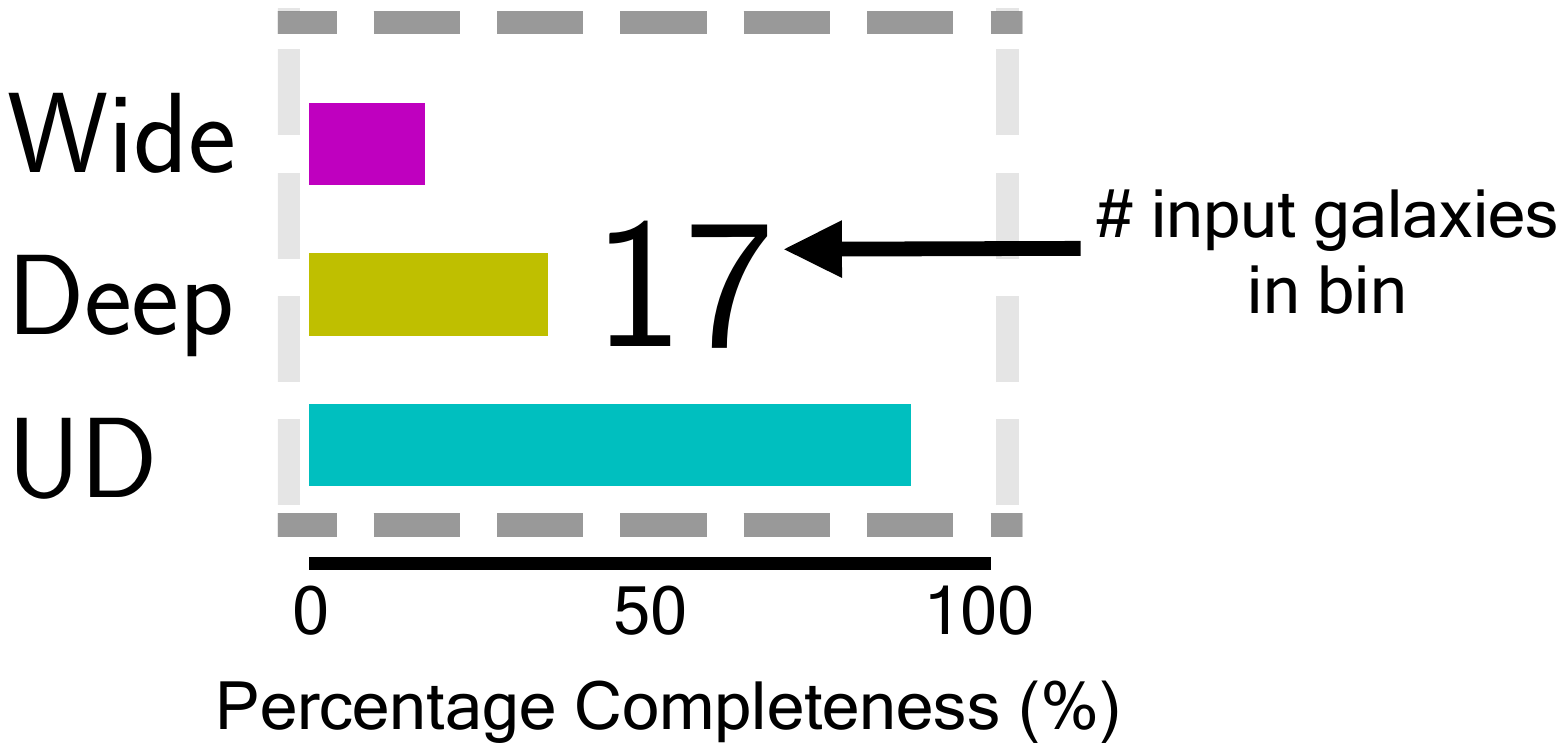}\hfill
\end{minipage}
\caption{Percentage completeness of each survey, shown in bins of stellar mass and integrated SFR with respect to the Main Sequence (black solid line, \citealt{ref:M.Sargent2014}). The panels are also divided into bins of redshift, as indicated in the bottom right corner. Three bars are shown in each SFR-stellar mass bin, one for each survey (yellow: wide, magenta: deep, cyan: UD), where the length of the bar represents the percentage completeness (as shown in the legend, lower right panel). The number written in each bin is the number of input galaxies.}
\label{fig:discussion_MScompl_w_depth}
\end{figure*}

\section{Discussion}
\label{sec:discussion}

\subsection{Galaxy detection across the Main Sequence}
\label{sec:dec_MS}

There are several considerations when interpreting the above results, based on the science one wishes to achieve with SKA-MID. Figs~\ref{fig:completeness} and \ref{fig:fluxflux} demonstrate the achievable survey completeness levels and the accuracy of flux density measurements for the planned wide, deep and ultra-deep SKA-MID surveys, based on our sample of $0<z<2.5$ galaxies in the GOODS-N field. We show the recovered completeness of each survey tier in Fig.~\ref{fig:discussion_MScompl_w_depth}, as a function of a galaxy's star-formation rate and stellar mass (i.e. its position with respect to the MS of star formation). We remind the reader that the completeness is more likely to be affected by extraction technique than the input-output comparative properties discussed in this study, but we expect overall trends to be robust to different approaches. A full comparison of different extraction softwares is out of the scope of this paper, but once a full mock SKA pipeline exists, studies can be carried out to finely characterise the resulting differences. Three example redshift bins are shown in Fig.~\ref{fig:discussion_MScompl_w_depth}: $0<z<0.5$, $1.0<z<1.5$ and $2.0<z<2.5$, and the remaining redshifts are shown in Appendix~\ref{sec:appendixA}. Completeness here is defined as the percentage of detected galaxies in each joint SFR-$M_{\star}$ bin, and completeness tends to increase as a function of SFR. As the SFR is directly related to the radio continuum flux density (see Section~\ref{sec:method_SKA}), we would indeed expect higher completeness levels with increasing SFR. We note however that because the flux density distributions of our galaxies contain significant substructure, the survey completeness is not only affected by the total SFR (flux density) of a galaxy, but also by the distribution of the flux. If a galaxy contains a bright localised region of star-formation, for example, it may be more likely to be detected than a galaxy of the same total flux that may have its flux distributed over a much wider area (meaning each pixel is at a relatively low S/N). On the other hand, we do see a hint that survey completeness may increase with stellar mass (and therefore galaxy size) for our sample in the wide tier at $2<z<2.5$, for example. This may be due to relatively high-noise pixels in the wide tier being able to `mask' or `hide' large fractions of smaller galaxies, as this effect doesn't appear as prominently in deeper tiers. We note that, for a mass-dependent \qtir\ as in \citet{ref:I.Delvecchio2021}, we would likely reach higher levels of completeness at log$_{\rm 10}$($M_{\star}$/$M_{\odot}$)$>$9.2, as their \qtir\ at these stellar masses implies higher radio luminosities than we have used here. The use of a mass-dependent \qtir\ would also give rise to mass-dependent completeness levels, with completeness increasing with stellar mass.

Particularly important from Fig.~\ref{fig:discussion_MScompl_w_depth}, we find that the survey tiers are not always complete along the Main Sequence of star-formation, depending on the survey depth and the SFR-$M_{\star}$ bin. This is a relevant consideration when planning survey-like observations with SKA-MID. In addition to the overall completeness levels for different survey depths shown in Fig.~\ref{fig:completeness}, the completeness varies significantly between different galaxy populations (i.e. `starburst' galaxies, MS galaxies, sub-MS galaxies), which needs to be taken into account depending on the specific science goal. For example, Fig.~\ref{fig:discussion_MScompl_w_depth} demonstrates that if a completeness of at least 50\% is required along the Main Sequence of star-formation at $M_{\star}>$10$^{10}$ and z$\sim$2-2.5, we must go to either the deep or ultradeep survey tiers, as the completeness in this regime in the wide tier falls below 50\% at $M_{\star}\sim$10$^{10.5}$. We see evidence for MS-completeness dropping off most significantly below this stellar mass in all three redshift bins in the wide tier. As might be expected, Fig.~\ref{fig:discussion_MScompl_w_depth} also highlights that significantly sub-MS galaxies are only accessible to the deep and UD tiers.

Finally, we can explore the differences between the completeness shown in Figs~\ref{fig:discussion_MScompl_w_depth} and \ref{fig:MS_surveys}, which will arise due to the effect of resolving the galaxies as discussed in Sec.~\ref{sec:int_SFRs}. Taking the first redshift bin in Fig.~\ref{fig:discussion_MScompl_w_depth}, we find a median redshift of 0.44 for our sample. Taking the same, point-source approach as for Fig.~\ref{fig:MS_surveys}, we would expect the UD tier to detect (at 5$\sigma$) all MS galaxies down to log$_{\rm 10}$($M_{\star}$/$M_{\odot}$)$=$8.4, the deep tier to detect MS galaxies down to log$_{\rm 10}$($M_{\star}$/$M_{\odot}$)$=$9, and the wide tier to detect MS galaxies down to log$_{\rm 10}$($M_{\star}$/$M_{\odot}$)$=$9.9. We are unable to make a comparison in the UD and deep tiers as our sample does not reach log$_{\rm 10}$($M_{\star}$/$M_{\odot}$)$<$9.5, but for the wide tier in Fig.~\ref{fig:discussion_MScompl_w_depth} (top left), MS completeness levels start at $\sim$30\% at log$_{\rm 10}$($M_{\star}$/$M_{\odot}$)$=$9.9 and increase with SFR (and therefore $M_{\star}$). In the top right panel, at median z = 1.22, we would expect unresolved MS galaxies to be detected down to log$_{\rm 10}$($M_{\star}$/$M_{\odot}$)$=$9.1 in the UD tier, log$_{\rm 10}$($M_{\star}$/$M_{\odot}$)$=$9.7 in the deep tier and log$_{\rm 10}$($M_{\star}$/$M_{\odot}$)$=$10.5 in the wide tier. As before, we are limited to comparisons at log$_{\rm 10}$($M_{\star}$/$M_{\odot}$)$>$9.5, but the deep tier has relatively high level of completeness down to the log$_{\rm 10}$($M_{\star}$/$M_{\odot}$)$=$9.5-10 MS mass bin, and the wide tier also performs relatively well above log$_{\rm 10}$($M_{\star}$/$M_{\odot}$)$=$10.5. We find similar results (completeness comparable with point-source expectations for deep and UD tiers) in the third panel ($2.0<z<2.5$), when comparing with predictions from Fig.~\ref{fig:MS_surveys}. Comparing expectations from unresolved approaches with the results of high-resolution studies such as these gives us a first indication of the drop in completeness we can expect due to resolving galaxies and their flux densities with SKA-MID. Our results suggest that completeness of resolved sources will be more affected in shallower data, and also that the drop in completeness becomes less significant at higher redshifts, where galaxies may be resolved into fewer beams, or comparable to the size of the PSF.

\begin{figure*}
\centering
\begin{minipage}{\textwidth}
\includegraphics[clip, trim=0.cm 0.cm 0.cm 0.cm, width=0.5\textwidth]{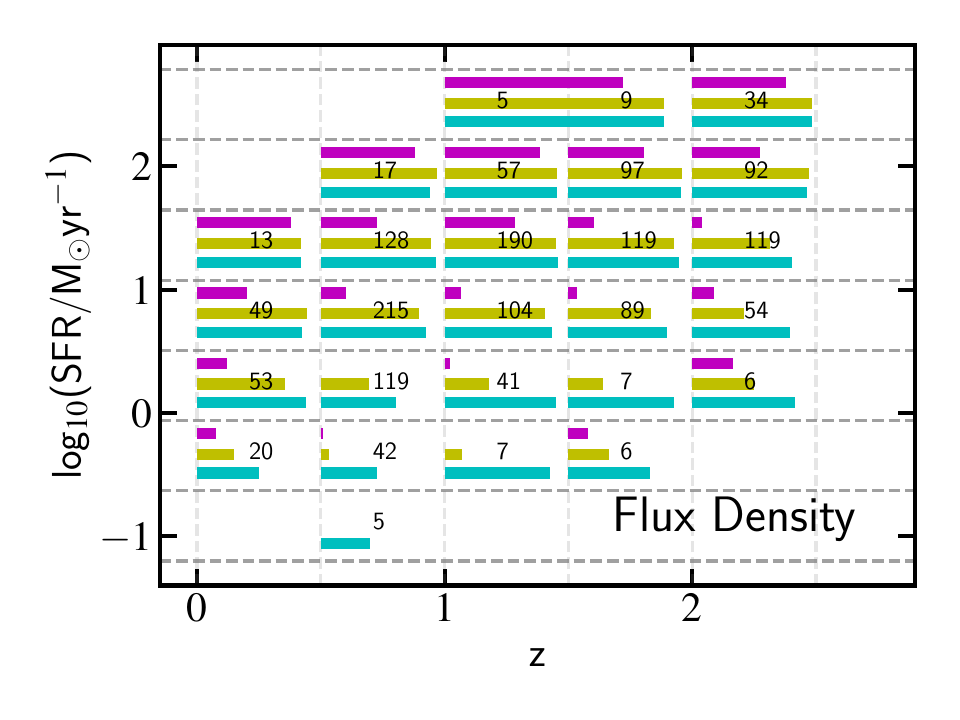}\hfill
\includegraphics[clip, trim=0.cm 0.cm 0.cm 0.cm, width=0.5\textwidth]{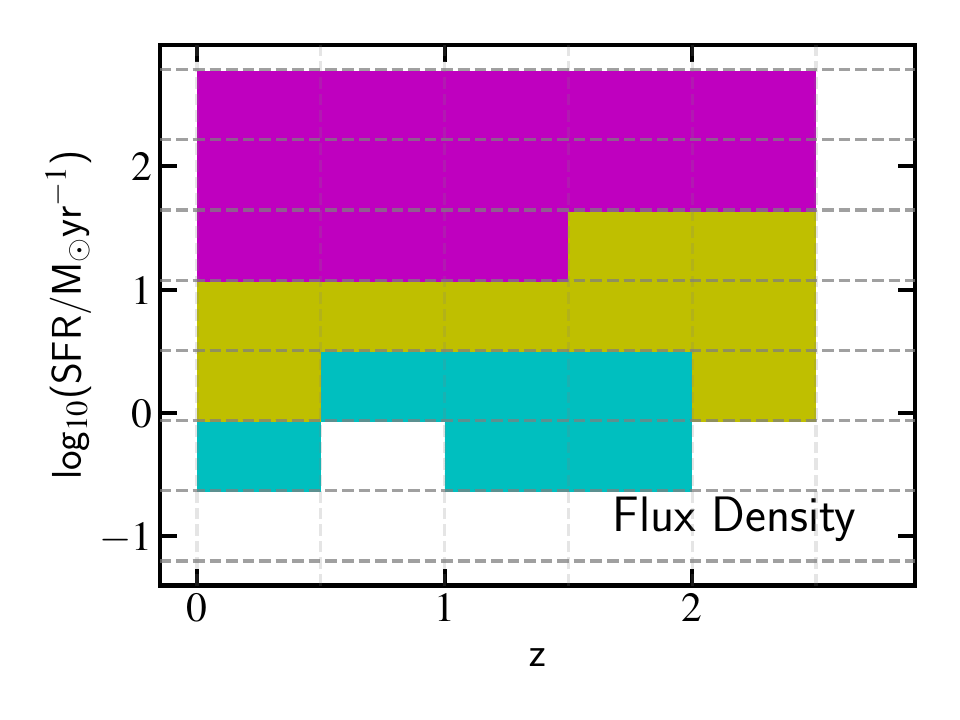}\hfill
\end{minipage}
\begin{minipage}{\textwidth}
\includegraphics[clip, trim=0.cm 0.cm 0.cm 0.cm, width=0.5\textwidth]{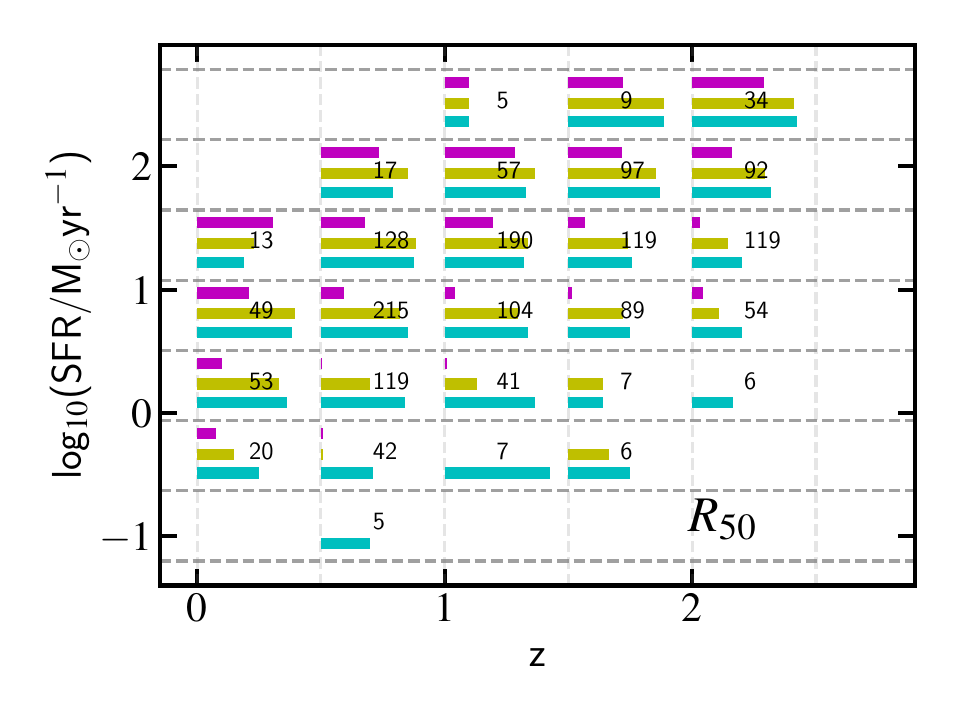}\hfill
\includegraphics[clip, trim=0.cm 0.cm 0.cm 0.cm, width=0.5\textwidth]{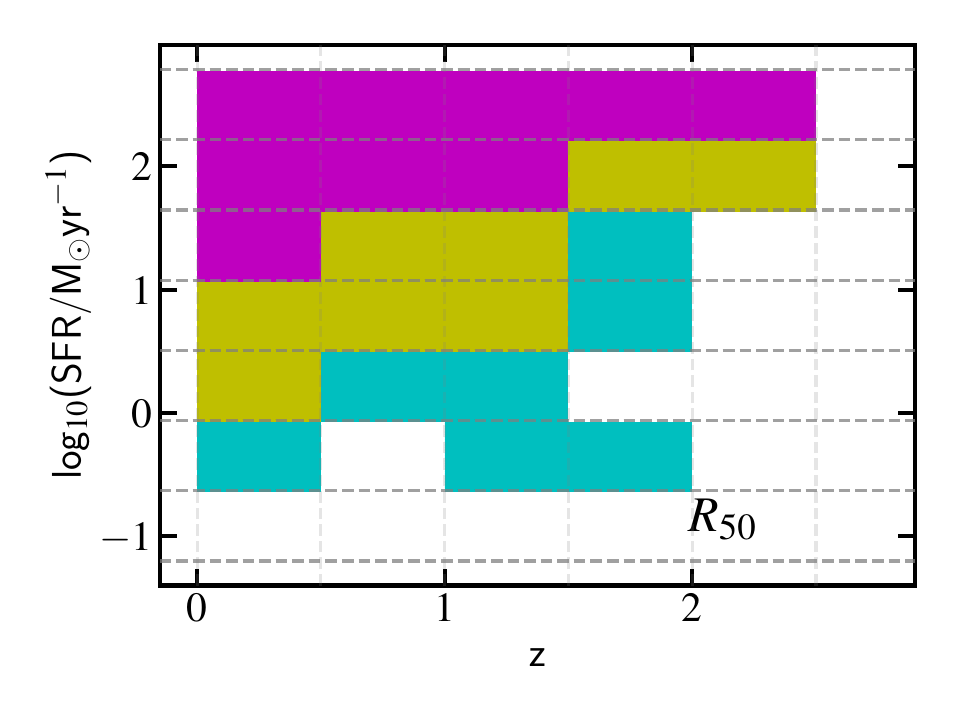}\hfill
\end{minipage}
\begin{minipage}{\textwidth}
\includegraphics[clip, trim=0.cm 0cm 0.cm 0cm, width=0.5\textwidth]{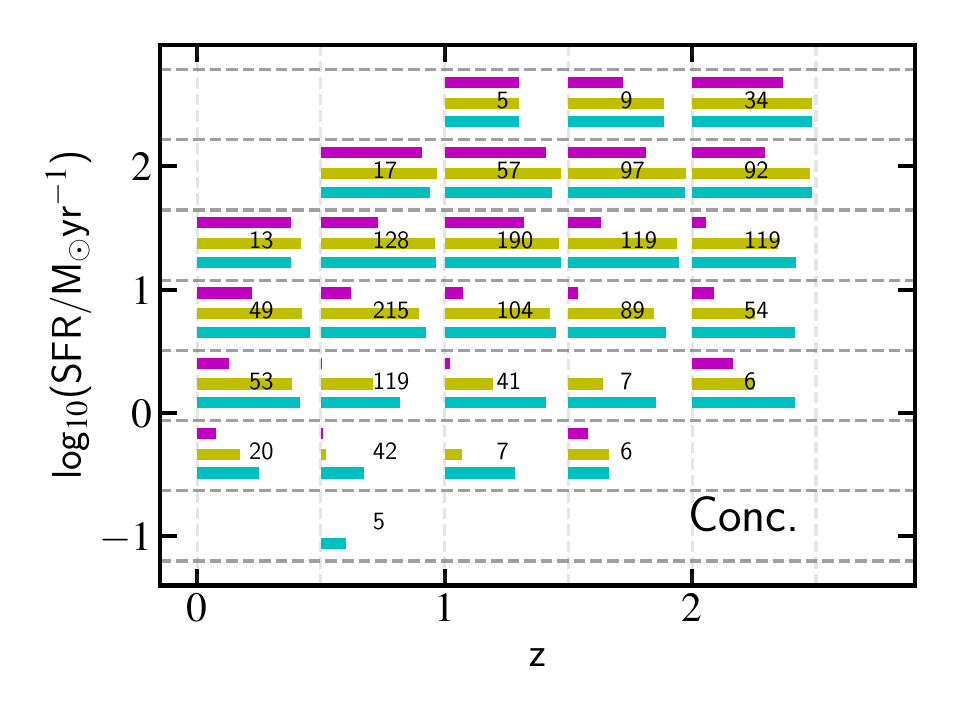}\hfill
\includegraphics[clip, trim=0.cm 0.cm 0.cm 0cm, width=0.5\textwidth]{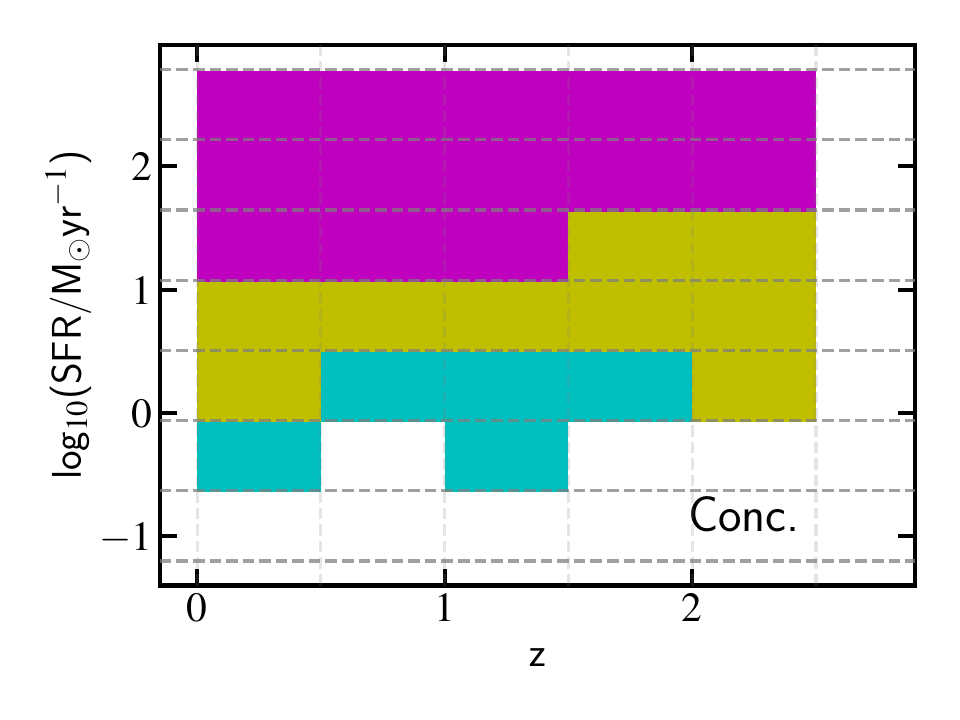}\hfill
\end{minipage}
\caption{Parameter recovery as a function of SFR and redshift. Each row corresponds to one parameter (indicated in the bottom right), and our galaxy sample is divided into bins of SFR (y-axis) and redshift (x-axis). {\it Left}: In each bin, the length of the bars indicates the fraction of all input galaxies in that bin that had the given parameter recovered to within 50\% of the input value. This is shown separately for each tier (magenta: wide, yellow: deep, cyan: UD). The number of total input galaxies in each bin is indicated in black, as for Fig.~\ref{fig:discussion_MScompl_w_depth}, and we only show bins that contain at least 5 galaxies. {\it Right}: The same SFR - z bins, colour-coded by the shallowest tier in which $\geq$50\% of the detected galaxies have the given parameter recovered to within 50\% of the input value.}
\label{fig:param_recovery_fct_sfr}
\end{figure*}

\subsection{Parameter recovery as a function of galaxy SFR}
\label{sec:param_recovery_fct_size}

As demonstrated by Figs~\ref{fig:R50R50} and \ref{fig:morph}, the values of morphological parameters such as Gini coefficient, $M_{20}$ etc. are altered when a galaxy is observed by a telescope, in this case SKA-MID, due to physical limitations on the spatial resolution, telescope response and noise. By looking at the left and right panels of Figs~\ref{fig:R50R50} to \ref{fig:morph}, we find that a large amount of the flattening of the output parameter distribution is due to the convolution with the 0.6" PSF alone, which has the effect of increasing galaxy sizes, and making their flux distributions (and therefore parameters such as asymmetry, Gini, $M_{20}$) appear more homogenous across the population than they are intrinsically. In the cases of $R_{50}$, $M_{20}$ and concentration, there is little appreciable difference between the input-output trends in the different tiers. In the case of Gini on the other hand (and arguably asymmetry), we do see improvement with increasing observation depth - i.e. the input-output relation moves closer to the 1:1 relation.

Thinking now about flux density and morphological analysis as a function of physical parameters, in Fig.~\ref{fig:param_recovery_fct_sfr} we explore parameter recovery as a function of galaxy SFR and redshift. Here, we focus on galaxy flux density, size, and concentration, with the remaining parameters shown in Appendix~\ref{sec:appendixB}. The coloured bars in the left column show the fraction of galaxies per SFR-z bin that are both detected and have the given parameter recovered to within 50\% of the input value. For a simplified view, in the right column we show the same axes, but instead colour each SFR-z bin by the shallowest tier in which $\geq$50\% of detected galaxies have the given parameter measured to within 50\% of the input value. We emphasise that the choice of 50\%, both for the parameter accuracy and proportion of the population, is somewhat arbitrary, as the accuracy required will depend entirely on the individual science case being pursued as well as the sample size. We therefore choose these values for Fig.~\ref{fig:param_recovery_fct_sfr}, as they are conservative enough to be useful for a number of science cases (such as those discussed in Section~\ref{sec:discussion}, e.g. star-formation concentration in galaxies across different environments, star-formation drop-off with redshift, and possible links to morphology). We see a clear dependence on SFR, with hints at diagonal divides between the tiers, suggesting also a dependence on redshift, particularly for $R_{50}$. In order to minimise colour discontinuities, which are likely caused by small-number statistics or sample variance, we have manually changed the colour of a handful of cells\footnote{2 cells were adjusted from deep to wide and 1 from UD to deep, following criterion (i), and the 3 cells in the left corners that were not covered by our sample were filled following criterion (ii)} to that of the next shallower tier (i) if these cells come to within a few percent of the 50\% threshold, or (ii) for cells not covered by our sample at lower z and higher SFR, where the ability to recover the physical properties can be expected to improve. For all parameters shown, Fig.~\ref{fig:param_recovery_fct_sfr} indicates the boundary SFRs for each tier, assuming we want $\geq$50\% of detected galaxies to have the parameter measured to within 50\% of the input value. Below log$_{10}$(SFR)$\sim$1-1.5, we need to start probing to deeper depths than the wide tier provides, and an order of magnitude lower than this, below log$_{10}$(SFR)$\sim$0-0.5, the sensitivity of the UD tier is required. Above $z=2$, for log$_{10}$(SFR)$\lesssim$1.5, none of the reference surveys are able to measure sizes within an accuracy of 50\% for the majority of the population.

\begin{table*}
\centering
\begin{tabular}{lccccccc}
\hline
& & R & & & p-value &\\
\hline
 & Wide & Deep & UD & Wide & Deep & UD\\
\hline
$R_{50}$ & 0.658 & 0.848 & 0.887 & 0.0 & 0.0 & 0.0 \\
Conc. & 0.182 & 0.582 & 0.714 & 5.5$\times$10$^{-6}$ & 0.0 & 0.0 \\
Gini & 0.500 & 0.695 & 0.681 & 0.0 & 0.0 & 0.0 \\
$M_{20}$ & 0.234 & 0.389 & 0.540 & 4.3$\times$10$^{-9}$ & 0.0 & 0.0 \\
Asymm. & 0.138 & 0.146 & 0.176 & 5.8$\times$10$^{-4}$ & 6.9$\times$10$^{-8}$ & 0.0 \\
\hline
\end{tabular}
\caption{Matrix of Spearman's rank correlation coefficients (R) and associated p-values for various morphological parameters, separated by tier, for the non-convolved, intrinsic inputs. Entries are ordered by descending R (for the UD tier).}
\label{tab:Spearman}
\end{table*}

\begin{figure*}
\centering
\begin{minipage}{\textwidth}
\includegraphics[clip, trim=0.6cm 0.6cm 0.6cm 0.6cm, width=0.2\textwidth]{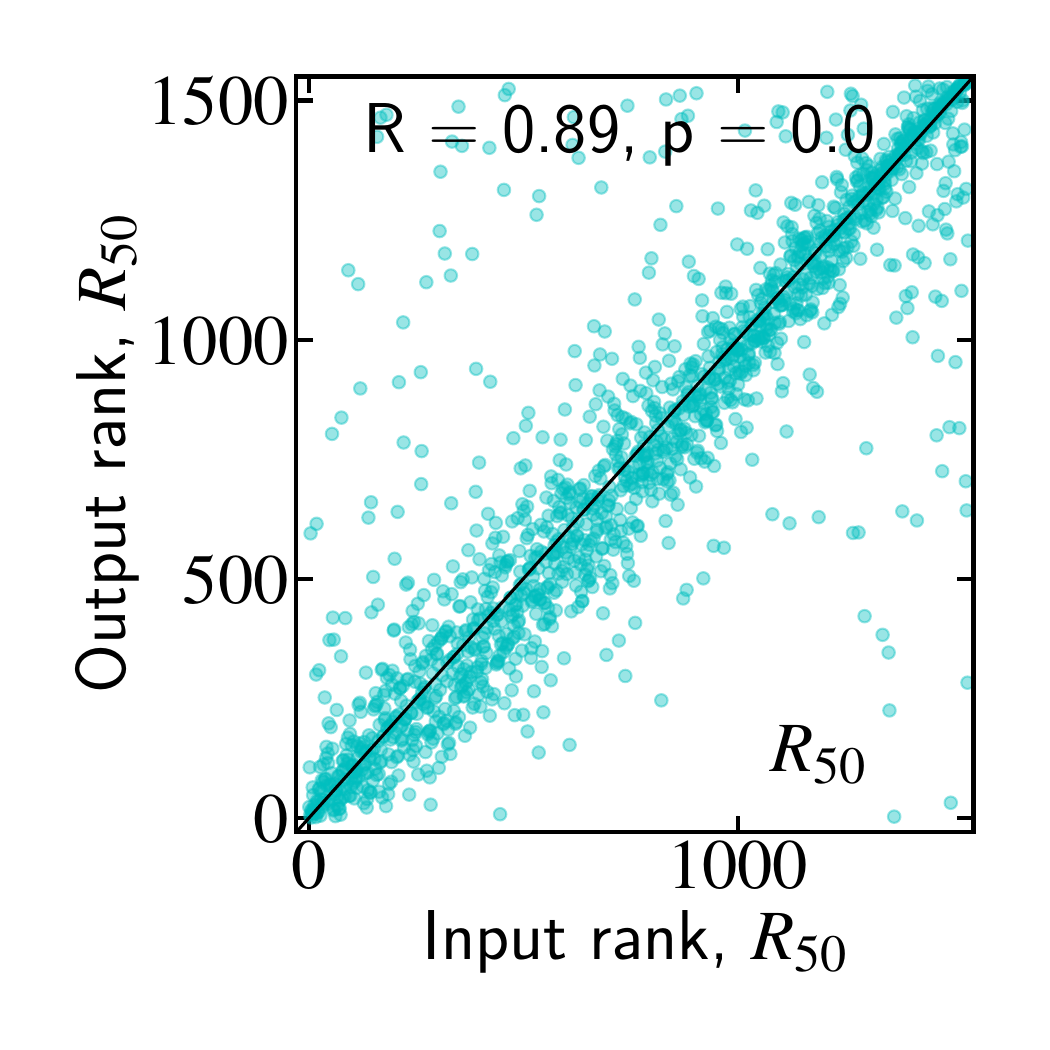}\hfill
\includegraphics[clip, trim=0.6cm 0.6cm 0.6cm 0.6cm, width=0.2\textwidth]{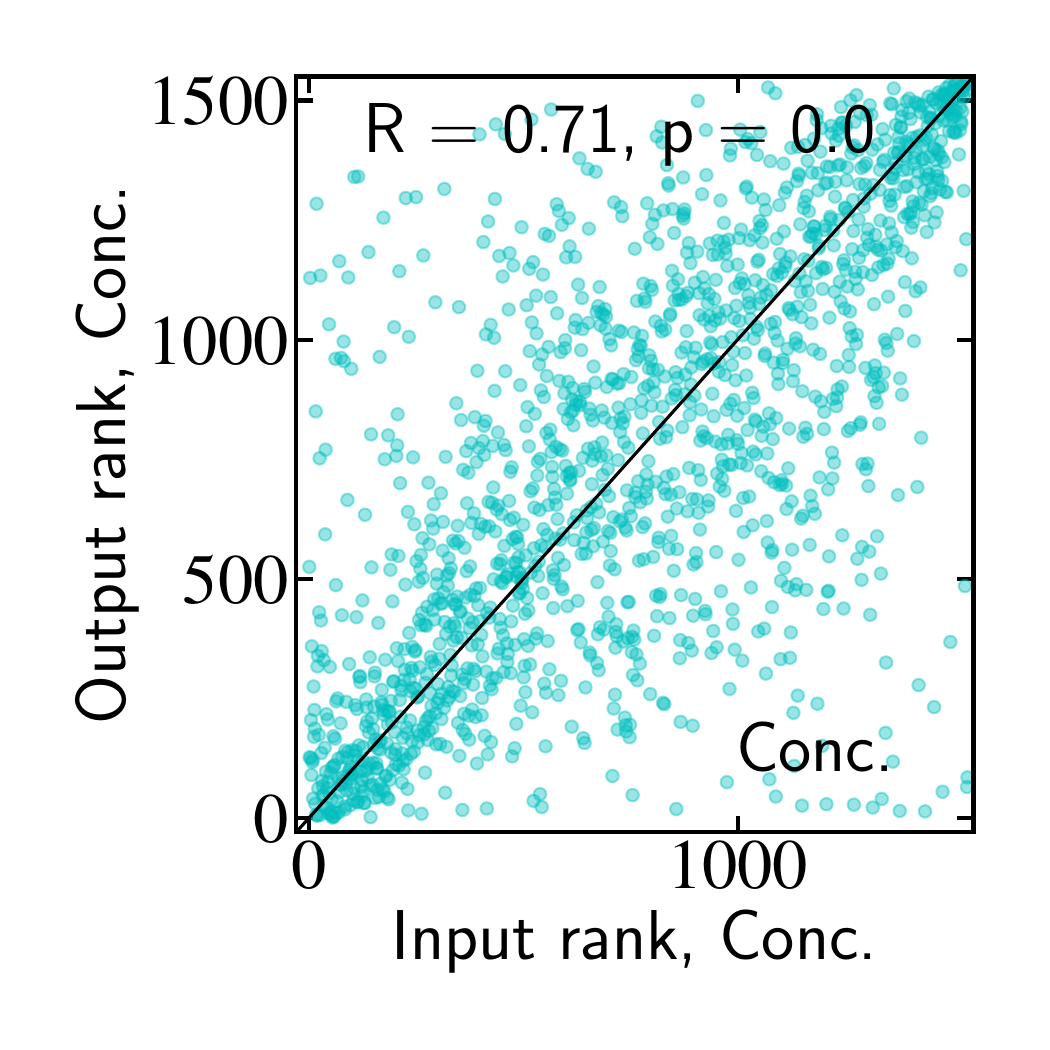}\hfill
\includegraphics[clip, trim=0.6cm 0.6cm 0.6cm 0.6cm, width=0.2\textwidth]{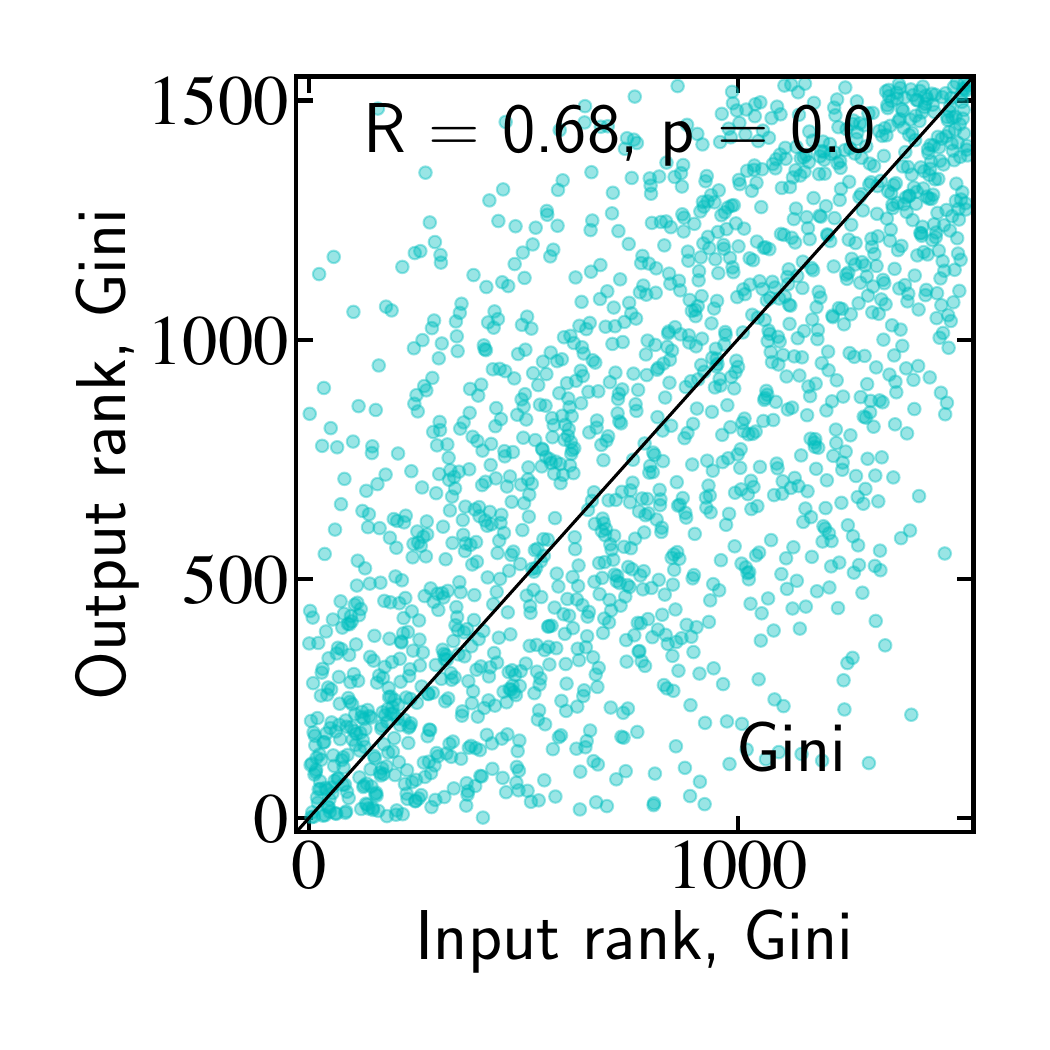}\hfill
\includegraphics[clip, trim=0.6cm 0.6cm 0.6cm 0.6cm, width=0.2\textwidth]{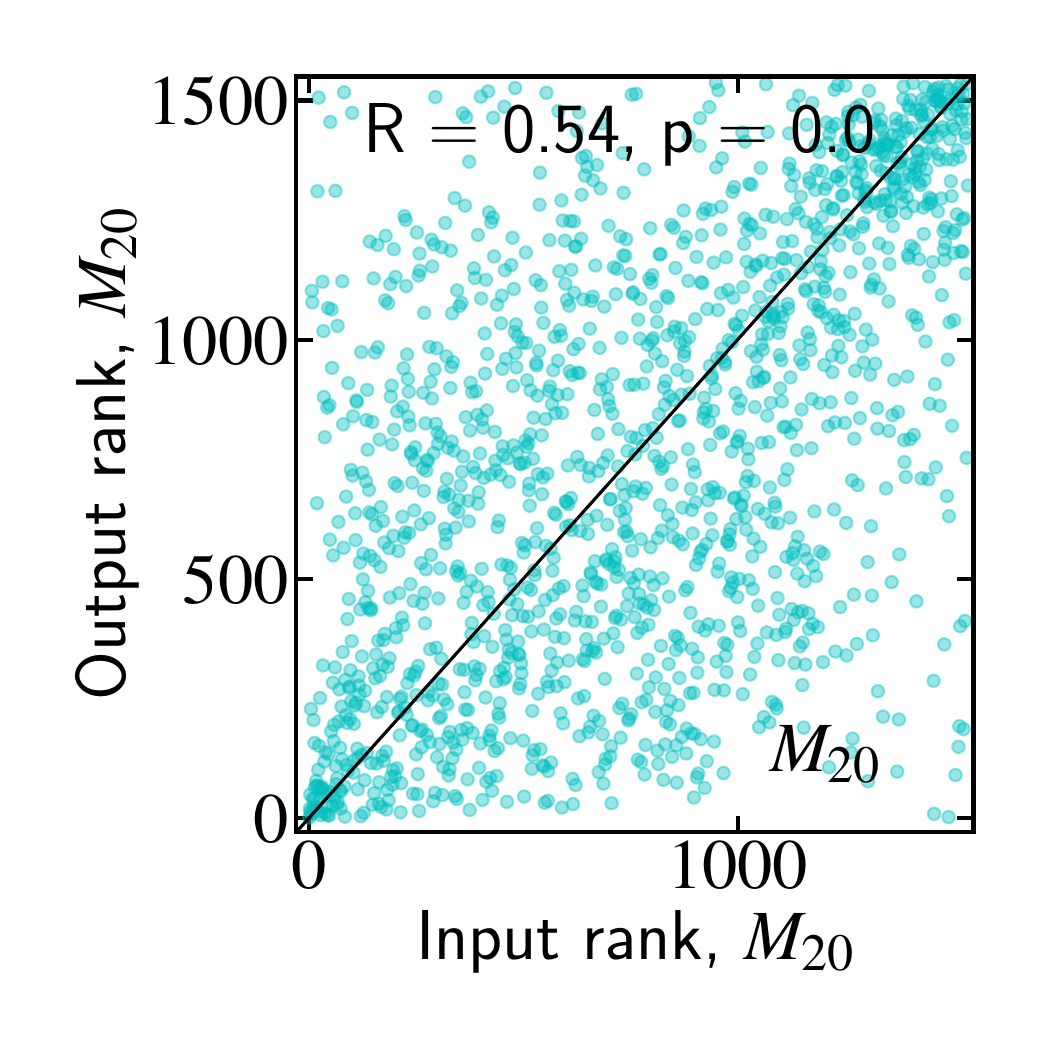}\hfill
\includegraphics[clip, trim=0.6cm 0.6cm 0.6cm 0.6cm, width=0.2\textwidth]{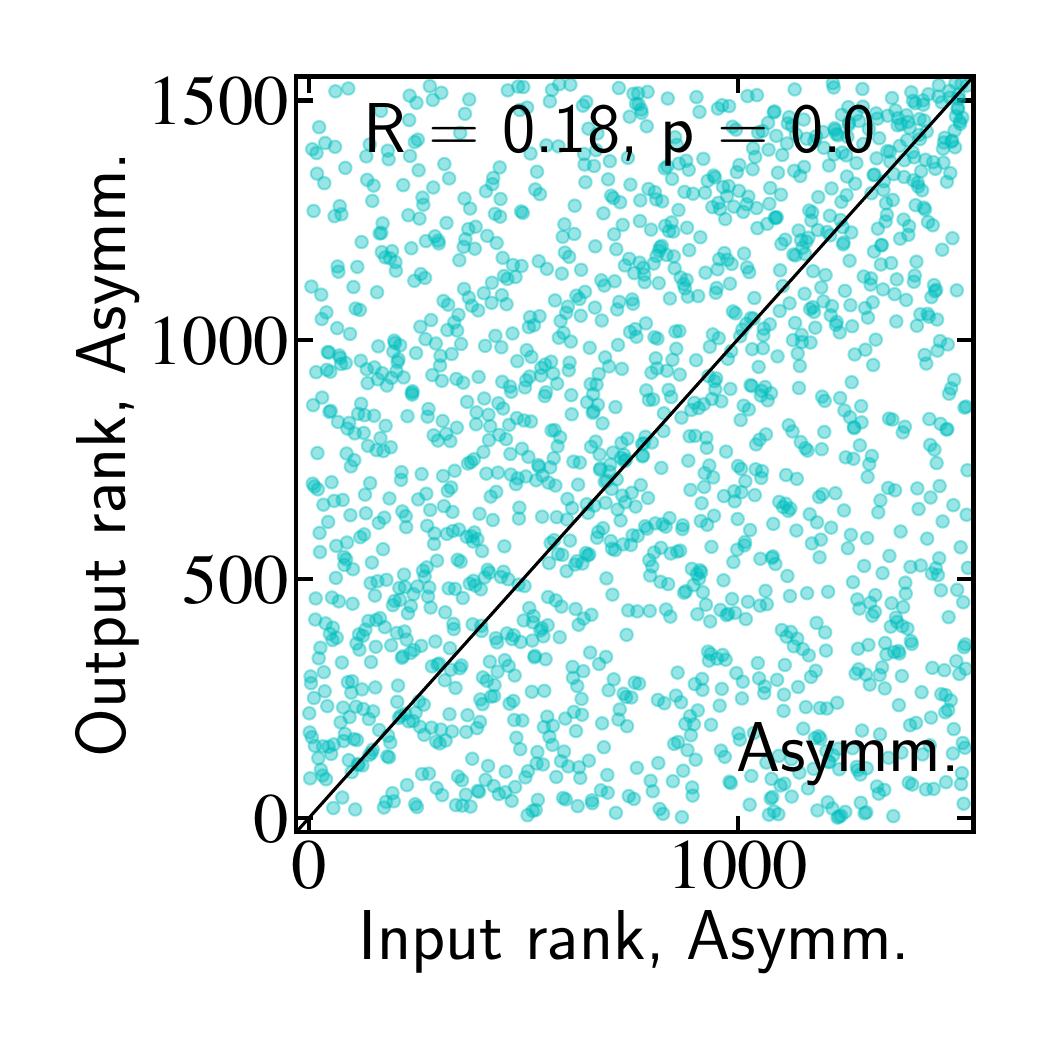}\hfill
\end{minipage}
\caption{Plots of output rank vs. input rank, for the $R_{50}$, concentration, Gini, $M_{20}$ and asymmetry parameters (left to right, as labelled). Only the UD tier is shown. Spearman's correlation coefficients (R) and p-values are given in the top left of each panel.}
\label{fig:Spearman}
\end{figure*}

\subsection{Rank ordering of galaxy parameters}
\label{sec:ranking}

What is often of interest when investigating the mechanisms of galaxy evolution are the morphologies of galaxies \textit{relative} to one another, as a function of environment or SFR, for example. Using our input and output morphology measurements, we can investigate to what degree the rank ordering of different morphological parameters is preserved by SKA observations. To do this, we compare the intrinsic (non-convolved) input parameters with the output parameters in each tier. To assess the rank ordering, we use Spearman's rank coefficient, R, a non-parametric measure of the correlation between two datasets. R can vary between -1 and 1, corresponding to perfect negative and positive monotonic correlations, respectively, and where 0 implies no correlation. The Spearman's rank test also returns a `p-value', which is the probability of an uncorrelated system producing datasets that have a Spearman correlation coefficient at least as extreme as that calculated.

Table~\ref{tab:Spearman} shows that the Spearman's rank correlation coefficient increases from the wide to the UD tier in all cases (with the exception of Gini in deep/UD), meaning that the rank ordering is retained better as the observations get deeper. Interestingly, there is only a very small difference in the results between the deep and UD tiers for both Gini and $R_{50}$, although we note that this may be affected by the input source selection. For all parameters, R is positive, indicating a positive correlation between the input and output rank values. We show the correlation between input and output ranks in Fig.~\ref{fig:Spearman}, for the UD tier. In addition to the R values given in Table~\ref{tab:Spearman}, Fig.~\ref{fig:Spearman} helps to visually identify the parameters that are most successful in maintaining their rank ordering during observation with the SKA. We see in both Table~\ref{tab:Spearman} and Fig.~\ref{fig:Spearman} that asymmetry has the lowest correlation coefficients in all tiers, ranging from $\sim$0.14 in the wide tier to R$\sim$0.18 in the UD tier, indicating a weak correlation between the input and output rankings. It is also interesting to see that although concentration performs relatively well in the UD tier, with R=0.71, the wide tier is significantly less useful for assessing this rank ordering, with R~=~0.18 - the largest difference between wide and UD tiers seen in all of our morphological parameters. The size parameter $R_{50}$ gives the highest correlation coefficient in all tiers, meaning that relative sizes between galaxies are retained better than any other parameter, for a given tier. Small p-values are found in all cases, meaning that although not all correlations necessarily approach monotonic, the probability of these datasets (input vs. output morphology values) being unrelated is small. These very small p-values are most likely influenced by our relatively large sample size, even if the correlations themselves are not always strong.

Interestingly, although the top row of Fig.~\ref{fig:morph} suggests that going to the UD tier for the measurement of absolute concentration values is only marginally beneficial, the Spearman's rank analysis in Table~\ref{tab:Spearman} suggests that the difference between the wide and UD tiers is fairly marked (R=0.18 and R=0.71 respectively), in terms of ordering galaxies by the concentration. Similarly, although the third row of Fig.~\ref{fig:morph} implies that survey depth noticeably impacts measured Gini parameters, we do not see any improvement between deep and UD tiers in terms of the rank ordering. On the other hand, Fig.~\ref{fig:R50R50} indicates that the recovery of $R_{50}$ improves only a little towards deeper tiers, and Table~\ref{tab:Spearman} confirms that the measurement of relative sizes in the wide tier is already much better than for other morphological parameters, and improves less with increasing survey depth than e.g. concentration. It's therefore important to plan observations or studies with the SKA-MID according to whether the relative parameters, or absolute values or parameters, will provide the most useful information. Band 2 observations with SKA-MID will have a theoretical maximal resolution of $\sim$0.35" (based on the maximum baseline of the array), but the sensitivity on those finer scales will be somewhat less than it is at the 0.6" scale, which formed the motivation for adopting a FWHM=0.6" beam in this study. However, our results highlight the importance of high-resolution follow-up observations with SKA-MID in band 5, particularly at high redshift, and ultimately extending the SKA to longer baselines in future deployment stages.

\subsection{Some science cases enabled by resolved galaxy morphologies with SKA-MID}
\label{sec:science}

We finish by discussing implications for the scientific questions to be addressed by the SKA. One of the particularly interesting outcomes of observations such as these will be the comparison of radio/star-formation morphologies in comparison to optical morphologies.

Several recent studies have already demonstrated that the resolved picture of molecular gas and obscured star-formation may differ strongly from what is seen in the optical (e.g. \citealt{ref:A.Cibinel2017, ref:M.Franco2020a, ref:W.Rujopakarn2019}). \citet{ref:A.Cibinel2017} find physically distinct UV-bright star-forming clumps offset from the obscured star-formation in a MS galaxy at $z=1.5$ seen in the submillimeter, calling into question the star-formation efficiency across different regions of the galaxy disk and core. Moving to more generalised trends, several efforts to characterise the star-formation and cold dust distributions of galaxies have already been carried out using existing submillimetre-radio instruments (e.g. \citealt{ref:S.Ikarashi2015, ref:D.Guidetti2017, ref:W.Cotton2018, ref:T.Muxlow2020}). These have successfully begun to characterise how galaxy sizes change within the radio bands, and also how radio sizes compare to dust spatial distributions, stellar emission sizes and dust-corrected star-formation sizes at shorter wavelengths (e.g. \citealt{ref:E.Murphy2017, ref:M.Bondi2018}). Increasing numbers of studies are finding a systematic offset between optical galaxy sizes and IR-submillimeter sizes, with galaxy dusty star-forming regions being systematically more compact than their optical light distribution, giving information also on the physical conditions of the ISM. This suggests that some star-forming galaxies are hosting centrally enhanced star-formation activity, relative to their more extended outer disks \citep{ref:J.Simpson2015, ref:M.Bondi2018, ref:E.JimenezAndrade2019, ref:E.JimenezAndrade2021, ref:A.Puglisi2021}. However, many of these studies have been fairly time consuming efforts for relatively small number statistics, or somewhat marginally resolved galaxies. The SKA is therefore poised to significantly improve upon the efficiency of what can currently be achieved.

Such comparisons between morphologies and star-formation distributions in optical vs. submillimeter-radio data are also crucial to unveil the physical processes experienced by galaxies (e.g. major vs. minor mergers, gas rich-instabilities, \citealt{ref:C.Gomez-Guijarro2018, ref:W.Rujopakarn2019}), and sub-arcsecond resolution SKA-MID surveys will provide a wealth of data on the morphology of obscured star-formation in galaxies to complement existing UV-optical observations.

We have seen in Fig.~\ref{fig:R50R50} and Table~\ref{tab:Spearman} that both absolute and relative sizes of galaxies are already accessible to the wide tier at a comparable level to that of the deeper tiers. A key question in galaxy evolution today is the effect of galaxy environment on the evolution of galaxies, which can be investigated through comparisons of properties such as star-formation rate and morphology, for galaxies across a wide range of environments (e.g. groups, (proto)clusters, field). In order to observe a range of environmental densities, a large area of the sky must be observed, making the wide tier an ideal place to gather the necessary statistics. A study of this kind in the wide tier would be reliant on relative properties between galaxies in dense and less dense environments, so it's important to focus on parameters such as $R_{50}$ that preserve their rank order well. The characterisation of relative sizes of galaxies evolving in different environments, at fixed stellar mass and/or SFR, is of significant interest (e.g. \citealt{ref:B.Moore1998, ref:M.Cappellari2013, ref:R.Allen2015, ref:J.Matharu2019}). Although the absolute values of Gini are strongly dependent on the survey tier, the rank ordering of Gini parameters is relatively well preserved in the wide tier, and could be used to make additional comparisons. A comparative study of the Gini parameters across different environments would give an indication of whether star-formation is less homogeneously distributed in galaxies in dense environments, perhaps due to disruptions experienced due to their surroundings (e.g. interactions with other galaxies or with the hot intra-group/cluster medium). Additionally, a wide survey area such as this is well-suited to searching for rare populations of massive, dusty star-forming galaxies, so-called `optically-dark' galaxies that are faint in UV-optical observations, but can be detected at long wavelengths through their star-formation activity (e.g. \citealt{ref:M.Franco2018, ref:C.Williams2019, ref:T.Wang2019}).

On the other hand, if we are interested in measuring more accurate galaxy flux densities and morphological parameters, the deep tier provides a good balance between sample size and reliability of measurements. The absolute values of concentration are not strongly dependent on survey tier, and the relative rankings of Gini in the deep tier are already as well-ordered as in the UD tier. Concentration is a key parameter for assessing the prevalence of nuclear star-formation or AGN activity, although we note that at 0.6" spatial resolution, the SKA-MID alone would not be able to distinguish between star-formation and AGN activity, without the addition of mid-far infrared observations, for example. Assessing and resolving which galaxy populations are most likely to display nuclear activity as a function of (specific) SFR and redshift would give important insight into the changes in star-formation mode (extended vs. nuclear) over cosmic time. The addition of Gini would further give an indication of whether star-formation is localised to certain regions of a galaxy, even if the central concentration is low.

Finally, although the smallest tier, the ultradeep survey will give the most accurate galaxy flux densities and morphological parameters, for both absolute and comparative studies. In addition, the advantage of the UD tier is its higher levels of completeness than the shallower tiers, such that meaningful conclusions on galaxy physical properties can be drawn for complete, albeit smaller samples of galaxies. As demonstrated in Figs~\ref{fig:MS_surveys} and \ref{fig:discussion_MScompl_w_depth}, sub-MS, MS and `starburst' galaxies will be accessible in the UD tier at high completeness up to distant redshifts, meaning that the numbers of galaxies in each of these populations can be quantified, and their star-formation rates accurately measured. This will naturally give an insight into the timeline and progression of the drop-off of star-formation as a function of redshift, which, combined with detailed measurements of galaxy morphologies, could be used to investigate prevalent quenching mechanisms as a function of cosmic time.

\section{Conclusions}
\label{sec:concs}

We have created simulated SKA-MID images of a representative $\sim$0.04~deg$^{2}$ region of the GOODS-N field, using resolved, multi-band HST images of 1723 galaxies at $0<z<2.5$. As these galaxies display significant substructure in their flux distributions, this allows us to assess the future ability of SKA-MID at 1.4~GHz and 0.6" resolution to recover galaxy integrated flux densities and morphological properties. The simulated image depths correspond to the wide, deep and ultradeep (UD) tier of the band 2 reference surveys proposed in \citet{ref:PrandoniSeymour2015}. We find the following key results:

\begin{itemize}
\item Using the source extraction tool \texttt{ProFound} \citep{ref:A.Robotham2018}, we find a median underestimation of output flux densities rising from $\sim$1\% in the UD tier to $\sim$5\% in the wide tier (Fig.~\ref{fig:fluxflux}). Source blending and oversegmentation can lead to substantial over- or underestimation of galaxy-integrated flux densities especially in the UD tier.
\item 616, 1352 and 1537 of our total 1723 galaxies are detected in the wide, deep and UD survey tiers, respectively, with the completeness in each survey tier rising rapidly as a function of flux density above the $5\sigma$ detection limit (Fig.~\ref{fig:completeness}). Survey completeness for resolved galaxies along, above, and below the galaxy Main Sequence (MS) is presented in Fig.~\ref{fig:discussion_MScompl_w_depth}.
\item We explore flux density, size and CAS parameter recovery as a function of galaxy SFR and redshift (Fig.~\ref{fig:param_recovery_fct_sfr}). We find that to recover parameters to within 50\% of their input values for $>$50\% of the population, the wide tier is sufficient above log$_{10}$(SFR)$>$1-1.5, with the deep tier needed in the regime log$_{10}$(SFR)$\sim$0-0.5, and the UD tier below these SFRs (the ranges quoted reflect the fact that higher SFRs are needed to fulfil the criterion at higher redshifts). Sizes are not well-recovered for large fractions of detected galaxies at $z>2$ and log$_{10}$(SFR)$\lesssim$1.5.
\item We find that it is more beneficial to go to deeper observations for recovering some morphological parameters (Gini, asymmetry) than for others ($R_{50}$, concentration, $M_{20}$). We find that $R_{50}$ sizes are slightly over-estimated from the SKA survey images, but that the trend with survey depth is fairly weak (Fig.~\ref{fig:R50R50}, center).
\end{itemize}
From our detailed analysis of the recoverability of galaxy structural indicators ($R_{50}$, plus Gini, $M_{20}$, concentration and asymmetry parameters) we conclude that:
\begin{itemize}
\item Convolution with the SKA-MID PSF leads to clear lower boundaries on output Gini parameters, at $G$$\sim$0.6 (Fig.~\ref{fig:morph} third row, left panel). Output Gini parameters tend to be systematically underestimated, with the highest input Gini values being the most underestimated (Fig.~\ref{fig:morph} top row, centre). Gini parameter recovery is more affected by survey depth than $R_{50}$, $M_{20}$ and concentration, making it more favourable to go to deeper surveys if targeting absolute values of Gini.

\item Output $M_{20}$ values lie fairly flat at $\sim$-1.7, for all galaxies except those with very high input $M_{20}$ ($\gtrsim$-1.25, Fig.~\ref{fig:morph}, second row, center panel). There is very little difference seen between survey tiers in the final images.

\item Galaxy concentrations are reduced in the SKA images compared to the intrinsic values, except for those galaxies that started with the lowest intrinsic concentrations ($C<2$), which increase after convolution. The difference between survey tiers does not appear to be significant when we take errors into account, although there is some evidence for better recovery of high concentration parameters ($C>3$) in deeper surveys.

\item Galaxy asymmetries are underestimated in all survey tiers, particularly for high intrinsic asymmetry (Fig.~\ref{fig:morph}, bottom row).

\item The best morphological parameter for assessing relative ordering of galaxies is the size, $R_{50}$, demonstrated by the moderate-strong correlations between rankings of the input and output parameters (Fig.~\ref{fig:Spearman}). The relative ranking of Gini values is maintained relatively well in shallower tiers. Most likely due to the poor recovery of asymmetry values in general, the asymmetry also gives the weakest correlation between input-output rank orderings.
\end{itemize}

Our analysis confirms that searches for massive, rare, star-forming galaxies will be well-suited to the wide tier (1000~deg$^{2}$), as well will the investigation of the environmental effect on galaxy evolution, where measurements of the physical properties of galaxies over a wide range of environmental densities are necessary. We would recommend a comparative study of size and Gini parameters, for example. In the deep tier, more reliable galaxy concentrations become accessible, meaning that this tier could be used to perform a census of extended vs. nuclear star-formation, over a moderate area of the sky (10-30~deg$^{2}$). Finally, we verify that the most accurate flux and morphological measurements can be made in the ultradeep tier (1~deg$^{2}$), where populations of MS, starburst, and sub-MS galaxies will all be detectable by SKA-MID over a wide range of redshifts $z\lesssim3$), as one normally requires of an UD survey. Characterisation of the star-formation rates of these galaxy populations, combined with morphological measurements in the UD tier, could be a powerful tracer of quenching mechanisms and timescales across cosmic time.

There are a myriad of science questions that the SKA will be primed to investigate. Depending on the science goal of a particular observation (e.g. galaxy completeness, flux density measurements, morphological comparisons etc.), simulation studies such as these inform us about the usefulness of increasing the observation time, the choice of galaxy parameters to study, as well as the levels of accuracy that can be achieved. In this way, we will be ready to take full advantage of the SKA, when it becomes the world's largest radio telescope.

\section*{Acknowledgements}

We thank the anonymous referee, whose comments improved the accuracy and clarity of the paper. We thank Robert Braun for both sharing the SKAO pipeline and for the helpful tutorage on its implementation. We also thank Catherine Hale for useful discussions on the use of \texttt{ProFound} and overall feedback on the study. MTS acknowledges support from a Scientific Exchanges visitor fellowship (IZSEZO\_202357) from the Swiss National Science Foundation. MF acknowledges support from the UK Science and Technology Facilities Council (STFC, grant ST/R000905/1), NSF grant AST-2009577 and NASA JWST GO Program 1727. This work builds on material from the doctoral thesis ``The impact of environment on galaxy evolution: Starburst and AGN activity'' by Rosemary Theresa Coogan, submitted for the degree of Doctor of Philosophy, University of Sussex, June 2019.

\section*{Data Availability}

The data underlying this article will be shared on request to the corresponding author.




\bibliographystyle{mnras}
\bibliography{EnvironmentNew}

\appendix

\section{Completeness as a function of SFR and stellar mass}
\label{sec:appendixA}

\begin{figure*}
\centering
\begin{minipage}{\textwidth}
\includegraphics[width=0.5\textwidth]{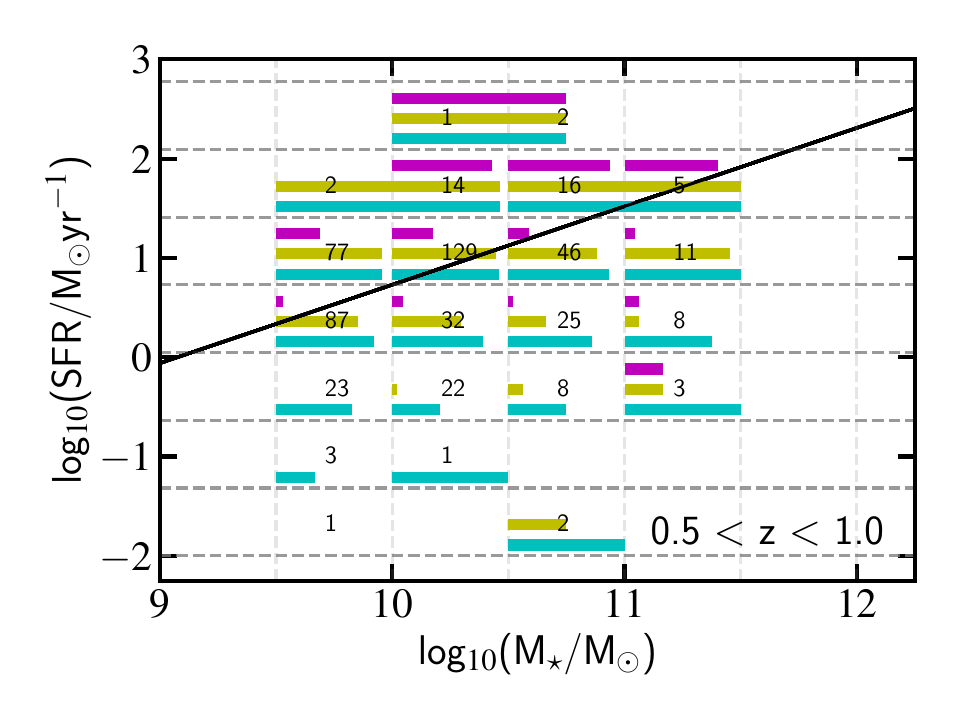}\hfill
\includegraphics[width=0.5\textwidth]{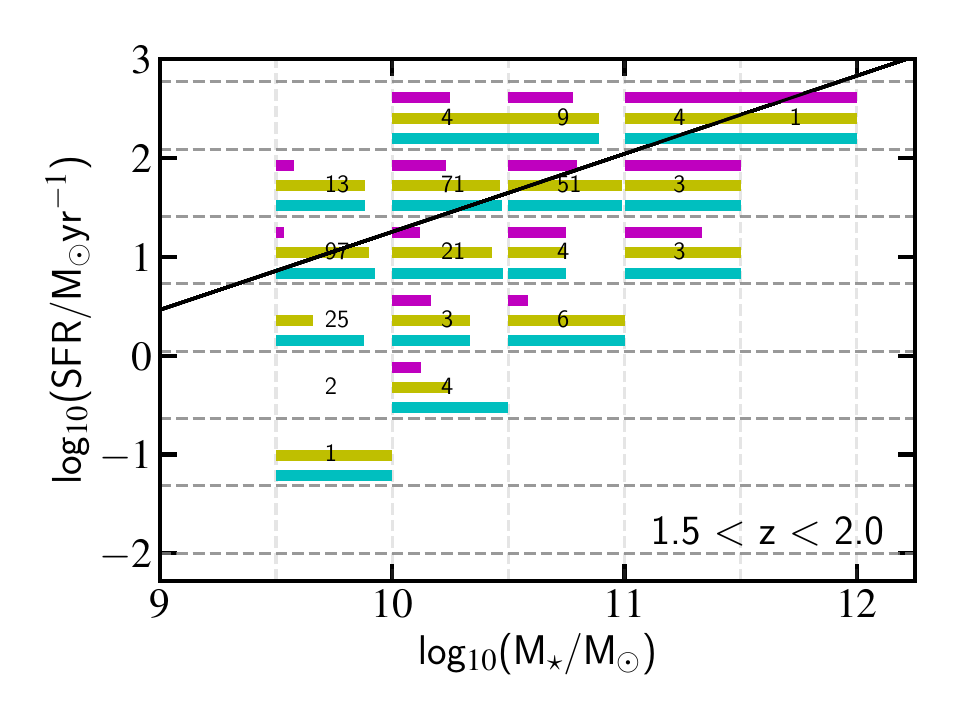}\hfill
\end{minipage}
\caption{As for Fig.~\ref{fig:discussion_MScompl_w_depth}, but for the additional redshift bins. Percentage completeness of each survey, shown in bins of stellar mass and SFR with respect to the Main Sequence (black solid line, \citealt{ref:M.Sargent2014}). The panels are also divided into bins of redshift, as indicated in the bottom right corner. Three bars are shown in each SFR-stellar mass bin, one for each survey (yellow: wide, magenta: deep, cyan: UD), where the length of the bar represents the percentage completeness. The number written in each bin is the number of input galaxies.}
\label{fig:discussion_MScompl_w_depth_appendix}
\end{figure*}

In Fig.~\ref{fig:discussion_MScompl_w_depth_appendix}, we show the completeness of each tier, defined as the number of detected galaxies divided by the number of input galaxies, as a function of SFR and stellar mass. This is the continuation of Fig.~\ref{fig:discussion_MScompl_w_depth}, in that we show the remaining redshift bins.

\section{Parameter recovery as a function of SFR and redshift}
\label{sec:appendixB}

\begin{figure*}
\centering
\begin{minipage}{\textwidth}
\includegraphics[clip, trim=0.cm 0.cm 0.cm 0.cm, width=0.5\textwidth]{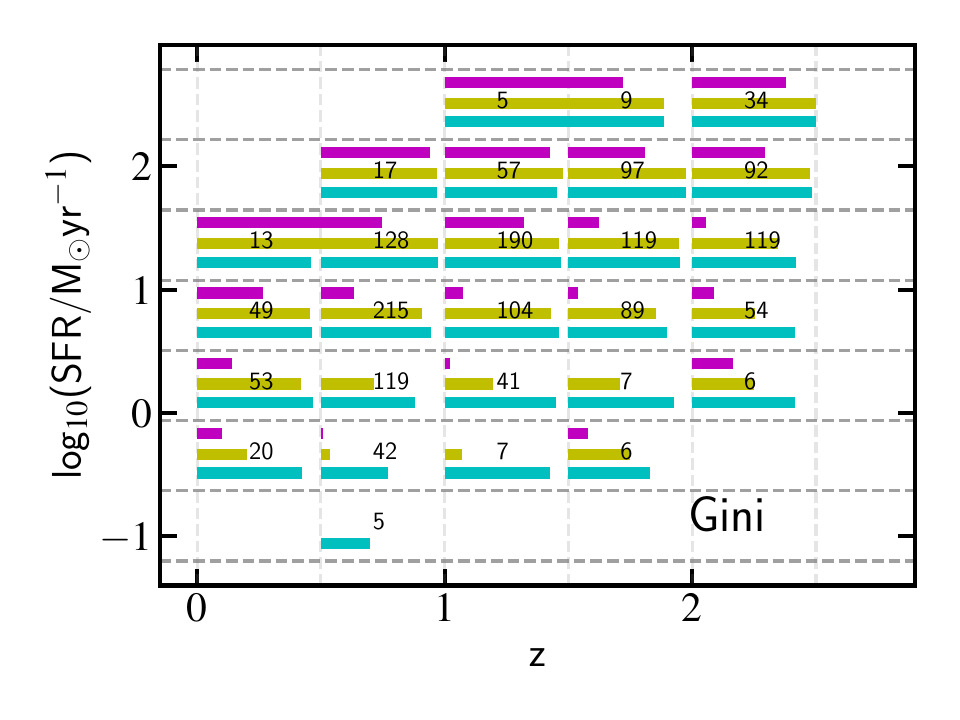}\hfill
\includegraphics[clip, trim=0.cm 0.cm 0.cm 0.cm, width=0.5\textwidth]{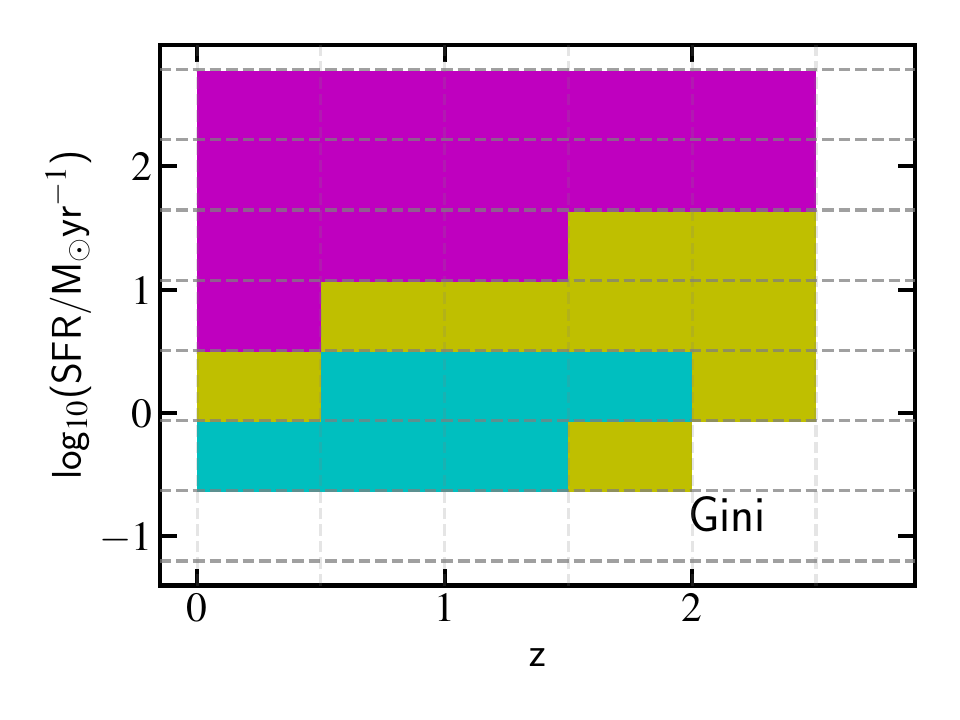}\hfill
\end{minipage}
\begin{minipage}{\textwidth}
\includegraphics[clip, trim=0.cm 0.cm 0.cm 0.cm, width=0.5\textwidth]{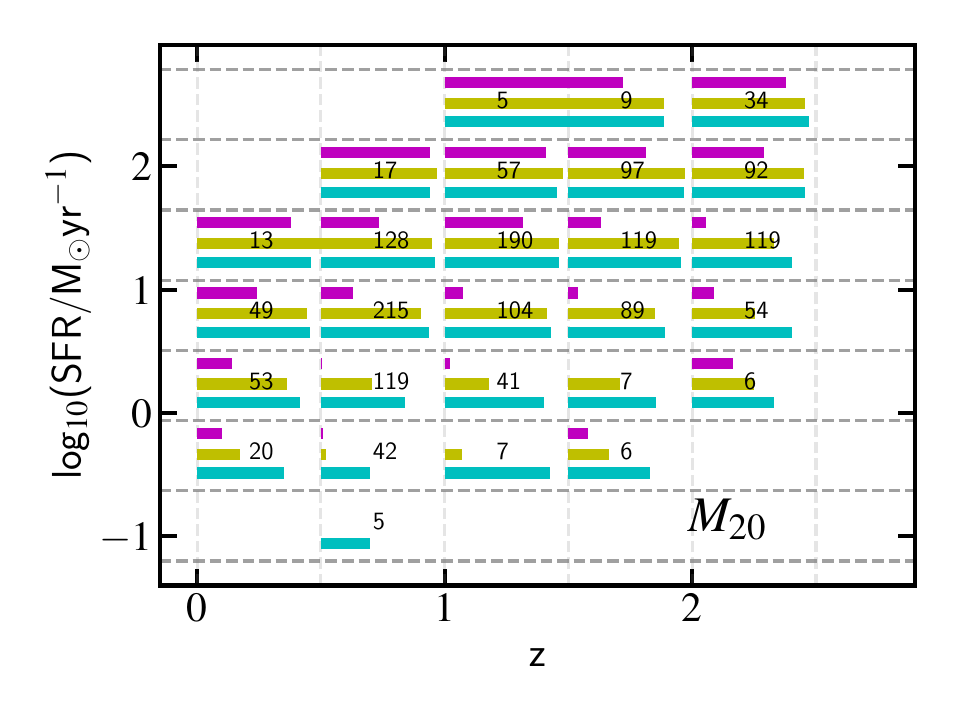}\hfill
\includegraphics[clip, trim=0.cm 0.cm 0.cm 0.cm, width=0.5\textwidth]{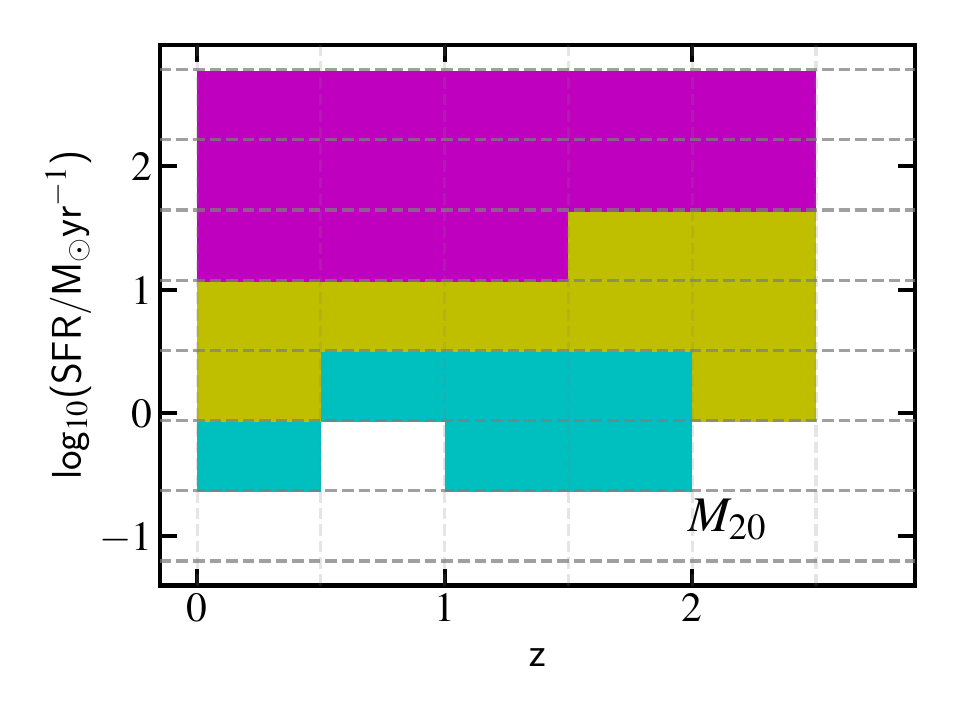}\hfill
\end{minipage}
\begin{minipage}{\textwidth}
\includegraphics[clip, trim=0.cm 0.cm 0.cm 0.cm, width=0.5\textwidth]{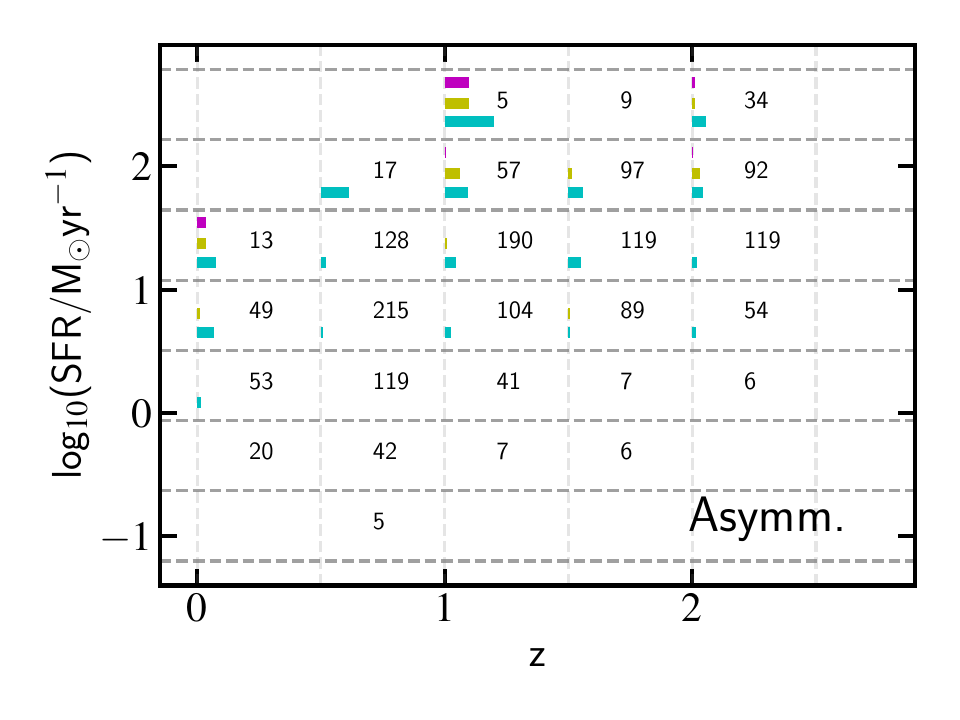}\hfill
\end{minipage}
\caption{As for Fig.~\ref{fig:param_recovery_fct_sfr}, but for Gini, $M_{20}$ and asymmetry. Parameter recovery as a function of SFR and redshift. Each row corresponds to one parameter (indicated in the bottom right), and our galaxy sample is divided into bins of SFR (y-axis) and redshift (x-axis). {\it Left}: In each bin, the length of the bars indicates the fraction of all input galaxies in that bin that had the given parameter recovered to within 50\% of the input value. This is shown separately for each tier (magenta: wide, yellow: deep, cyan: UD). The number of total input galaxies in each bin is indicated in black, and we only show bins that contain at least 5 galaxies. {\it Right}: The same SFR - z bins, colour-coded by the shallowest tier in which $\geq$50\% of the detected galaxies have the given parameter recovered to within 50\% of the input value. There is no right panel for asymmetry, as all bins were empty (meaning less than 50\% of galaxies had output asymmetries within 50\% of the input value for all bins).}
\label{fig:param_recovery_fct_sfr_appendix}
\end{figure*}

In Fig.~\ref{fig:param_recovery_fct_sfr_appendix}, we demonstrate parameter recovery in SFR-z space, for Gini, $M_{20}$ and asymmetry. As for Fig.~\ref{fig:param_recovery_fct_sfr}, we see a clear dependence on SFR and redshift, if we wish to recover parameter values to within 50\% of their input values for $>$50\% of the population. The bottom row of Fig.~\ref{fig:param_recovery_fct_sfr_appendix} provides further confirmation that asymmetry is very poorly recovered in all tiers.

\bsp	
\label{lastpage}
\end{document}